\documentclass[12pt]{elsarticle}
\date{}
\journal{Progress in Particle and Nuclear Physics}
\usepackage[a4paper,margin=2cm]{geometry}
\usepackage{newtxtext}
\usepackage[varvw,bigdelims]{newtxmath}

\usepackage{graphicx}
\usepackage[colorlinks=true,allcolors=blue]{hyperref}
\usepackage{amsmath}
\usepackage{amssymb}
\usepackage{bm}
\newcommand{\nuc}[2][]{{}^{#1}\mathrm{#2}}

\begin{document}

\title{Dynamics of clusters and fragments in heavy-ion collisions}
\author{Akira Ono}
\address{Department of Physics, Tohoku University, Sendai 980-8578, Japan}

\begin{abstract}
  A review is given on the studies of formation of light clusters and heavier fragments in heavy-ion collisions at incident energies from several tens of MeV/nucleon to several hundred MeV/nucleon, focusing on dynamical aspects and on microscopic theoretical descriptions.  Existing experimental data already clarify basic characteristics of expanding and fragmenting systems typically in central collisions, where cluster correlations cannot be ignored.  Cluster correlations appear almost everywhere in excited low-density nuclear many-body systems and nuclear matter in statistical equilibrium where the properties of a cluster may be influenced by the medium.  On the other hand, transport models to  solve the time evolution have been developed based on the single-nucleon distribution function.  Different types of transport models are reviewed putting emphasis both on theoretical features and practical performances in the description of fragmentation.  A key concept to distinguish different models is how to consistently handle single-nucleon motions in the mean field, fluctuation or branching induced by two-nucleon collisions, and localization of nucleons to form fragments and clusters.  Some transport codes have been extended to treat light clusters explicitly.  Results indicate that cluster correlations can have strong impacts on global collision dynamics and correlations between light clusters should also be taken into account.
\end{abstract}
\maketitle
\eject
\tableofcontents

\section{Introduction}

By the studies of heavy-ion collisions in the past decades, researchers have been trying to explore nuclear matter or nuclear many-body systems under various conditions of densities, excitation energies, isospin asymmetries and so on.  In collisions of two nuclei with similar mass numbers at the incident energy per nucleon between several tens of MeV and several hundred MeV in the laboratory frame, the two nuclei can overlap considerably in the early stage of the reaction so that the system may be compressed up to the density of about $2\rho_0$ depending on the incident energy, where $\rho_0\approx 0.16\ \text{fm}^{-3}$ is the nuclear saturation density.  The process reaching to this compressed system can be regarded as a violent phase in which each nucleon in the system experiences many collisions with other nucleons.  Therefore a large degree of thermalization or equilibration can be expected, though some limited number of particles may leave the rest of the system before reaching equilibrium.  However, the full equilibration will never be reached and we still see a quite dynamic evolution of the system.  The compressed system immediately starts to expand without any external force to keep the system at high density and high pressure.  It is likely a good picture that most of the degrees of the system are thermalized on top of the dynamic degrees of freedom of collective expansion.  As the system expands, the interactions between particles become less frequent, and eventually the particles will no longer interact with each other until they are detected by experimental apparatuses.  In this article, we will mainly focus on collisions in the incident energy region between several tens of MeV/nucleon and several hundred MeV/nucleon.  The typical time scale of the dynamics of compression and expansion is of the order of 10-100 fm/$c$ in these collisions.  If the incident energy is simply converted to the excitation energy, the whole system has an excitation energy of the order of 10-100 MeV per nucleon, which is much larger than the nuclear binding energy of about 8 MeV per nucleon.  However, this never means that the system would disintegrate into free nucleons.

It is very well established by heavy-ion experiments that the finally emitted particles are dominantly light clusters and heavier fragment nuclei even in these highly excited systems.  Here light clusters mean nuclei with mass numbers $A_{\text{c}}=2,3$ and $4$, which are deuterons, tritons, $\nuc[3]{He}$ nuclei and $\alpha$ particles.  For example, in $\nuc{Xe}+\nuc{Sn}$ central collisions at 50 MeV/nucleon, the INDRA data \cite{hudan2003} show that only 10\% of total protons in the system are emitted as free protons and all the other protons are bound in heavier fragments and clusters in the final state.  About 20\% of protons are bound in $\alpha$ particles, about 10\% are bound in clusters of $A_{\text{c}}=2$ and 3, and the other 60\% are bound in fragments.  Even at higher incident energies, 250 MeV/nucleon for example, the ratio of free protons is still about 20\% as shown by the FOPI data \cite{reisdorf1997,reisdorf2010} for the central $\nuc{Au}+\nuc{Au}$ collisions.  To the best of the author's knowledge, it has never been possible to break any heavy-ion collision system into free protons and neutrons without any other particles emitted, except for collisions of very light nuclei.

In the expanding system, therefore, clusters and fragments must exist when the particles stop interacting in the expanding system, i.e., at freeze-out.  One of the questions we want to cover in this article is how these clusters and fragments have emerged.  We regard the system as composed of nucleons, and therefore clusters and fragments are composite particles.  The existence or production of clusters and fragments should be understood as a specific kind of many-body correlations.  To describe a cluster or a fragment with mass number $A_{\text{c}}$, we need to consider at least $A_{\text{c}}$-body correlations that the $A_{\text{c}}$ nucleons are spatially close to each other and they have similar momenta.  Most likely we are exploring highly correlated nuclear many-body systems that are dynamically evolving in heavy-ion collisions, which requires microscopic and dynamical approaches of theory.  At the same time, such systems may be closely linked with the equilibrium properties of nuclear matter at finite temperatures below the saturation density $\rho_0$.  In fact, e.g., fragment observables have often been well explained by statistical models in many cases.  The relation of observables to nuclear liquid-gas phase transition has also been indicated based on experimental data (see Ref.~\cite{borderie2008} for a review).

The intermediate states in heavy-ion collisions and the nuclear matter at finite temperatures may be regarded as excited states of a nuclear many-body system.  At the lowest energy $E^*=0$ in a system composed of $A$ nucleons, we have a nucleus in its ground state.  As is well known in nuclear structure studies, with excitation energies of the order of $E^*\lesssim 10$ MeV, there are identified levels with collective and non-collective characteristics which can be understood by single-particle excitations.  In the same excitation energy region, however, it is also known that there are states which can be explained by the cluster degrees of freedom.  Typical cluster states appear around the threshold excitation energies for the separation of cluster(s) from the nucleus as represented by Ikeda diagram \cite{ikeda1968}.  It has been recently found that some states, such as the Hoyle state of the $\nuc[12]{C}$ nucleus, are rather simply described as independent motions of $\alpha$ clusters.  If the consideration is extended to higher excitation energies by assuming a virtual container with a fixed volume $V$, the states realized in heavy-ion collisions typically corresponds to $A/V$ between $\sim\frac{1}{2}\rho_0$ and $\sim\frac{1}{10}\rho_0$.  Even with high excitation energies $E^*\sim A\times 10$-100 MeV as mentioned above, the abundance of light clusters and heavier fragment nuclei observed in heavy-ion collisions suggests that many-body correlations are still very important.  It seems there is a huge region of excitation energies where cluster correlations play important roles in nuclear many-body systems.  Uncorrelated nucleon gas will be realized only in the limit of large $V$ which is not relevant due to finite reaction times.  The low-density nuclear matter is also important for astrophysical problems such as the core-collapse supernovae.  The equation of state (EOS) for astrophysical applications has been calculated recently with elaborated approaches allowing the existence of clusters at low densities.  The composition of clusters may be important for some phenomena in compact stars such as the neutrino transport in core-collapse supernovae.

It is a great challenge to try to solve the time evolution of heavy-ion collisions from an initial to the final state.  Many-body correlations are essential as already mentioned.  It is a quantum mechanical problem in many senses.  For the nuclei not only in the initial state but also in the final state, the fermionic nature of nucleons are fundamentally important.  Light clusters have only a single bound intrinsic state, which cannot be emulated by classical motions of nucleons.  Furthermore, even starting with the same initial state, different configurations of decomposition into clusters and fragments can be realized in the final state, which is the probablistic nature of quantum mechanics.  It is not possible to exactly solve such quantum many-body problem from the first principle with computational facilities available currently or in near future.  However, it has been very important to have practical models in order to extract valuable information of physics such as the nuclear matter properties from heavy-ion collisions.  Traditionally there are two types of transport models.  One is the mean-field models which are based on the distributions of nucleons moving independently in the mean field.  The other is the molecular dynamics models which is similar to the $A$-body classical dynamics but nucleons are represented by Gaussian wave packets.  Two-nucleon collisions are additionally considered in both types of models.  Naive statements about the defects of these models are that the mean-field models lack many-body correlations and that the molecular dynamics models are classical, though the actual situation is not so simple.  Some of the mean-field and molecular dynamics models have been improved e.g.\ by introducing a kind of fluctuations or by considering light clusters explicitly, which can be regarded as incorporating quantum mechanical features mentioned above in practical manners.  These attempts of transport models will be reviewed in this article with emphasis on the description of the formation of fragments and clusters.

Clusters and fragments are important not only for their existence and their production mechanism but also because they may carry valuable information on other properties of nuclear matter such as the density dependence of the symmetry energy.  An important question is whether the correlations to form clusters and fragments have only a small effect on top of the global dynamics of collisions, such as compression, expansion and collective flows, or they have large reverse impacts on the global dynamics.  The latter case is possible because of the released energy and the reduction of the degrees of freedom by the formation of clusters and fragments.  We will see some examples in theoretical results reviewed here.

This article is organized as follows.  In Sec.~\ref{sec:exp}, we will give an overview of representative experimental data which characterize expanding systems in which clusters and fragments are formed.  We will mainly focus on central heavy-ion collisions.  In Sec.~\ref{sec:matter}, we will theoretically explore the statistical properties of excited nuclear many-body systems, or warm nuclear matter, at low densities $\rho<\rho_0$, putting emphasis on cluster correlations in such environments.  We then review practical transport models for heavy-ion collisions in Secs.~\ref{sec:basicmodels}, \ref{sec:flctmodels} and \ref{sec:clstmodels}, where both formal characteristics and representative results are discussed.  Basic and standard transport models in Sec.~\ref{sec:basicmodels} do not necessarily describe clusters and fragments sufficiently.  Some of them are extended in Sec.~\ref{sec:flctmodels} by taking into account fluctuations in the mean-field dynamics.  There is another direction of extension as in Sec.~\ref{sec:clstmodels} to handle cluster correlations explicitly.  Section~\ref{sec:summary} is devoted for a summary and perspectives.

\section{\label{sec:exp}Basic observations in heavy-ion collisions}

Let us first review the experimental data on the basic characteristics of heavy-ion collisions in which clusters and fragments are produced, in the incident energy region from several tens of MeV/nucleon to several hundred MeV/nucleon.  Experimental information has been accumulated in the past decades in particular by employing multiple detector apparatuses which cover large solid angles close to $4\pi$.  Large part of the produced charged particles are detected on the event-by-event basis.  Without aiming at a complete review of the experimental data, we will focus on the most basic characteristics of the expanding and disintegrating systems.

There are two essential factors.  One is the global or collective dynamics such as from the initial phase of the collision to the late phase of the expanding system, which may be represented by the time evolution of the one-body distribution function without seeing many-body correlations.  The other is the production of clusters and fragments, which can be viewed theoretically as the appearance of many-body correlations in the evolving system.  It may be intuitively evident that the collective one-body dynamics influences the production of clusters and fragments, e.g., many clusters and fragments will be formed only if the system expands.  The opposite is less obvious but is possible theoretically, i.e., strong many-body correlations may influence the collective one-body dynamics through the energy conservation and the reduced number of degrees of freedom.  So it is a chicken or egg question which one of the factors we should see first.  In the following, we will first overview the decomposition of the system into clusters and fragments and then the collective characteristics of the collision dynamics.

Various phenomena occur in heavy-ion collisions not only depending on the incident energy but also depending on the impact parameter $b$ and on the combination of the projectile and target nuclei.  Here we will mainly discuss the central collisions of two nuclei with similar sizes ($b\approx0$ and $A_{\text{projectile}}\approx A_{\text{target}}$).

\subsection{Particle yields}
\begin{figure}
\centering
\includegraphics[scale=0.5]{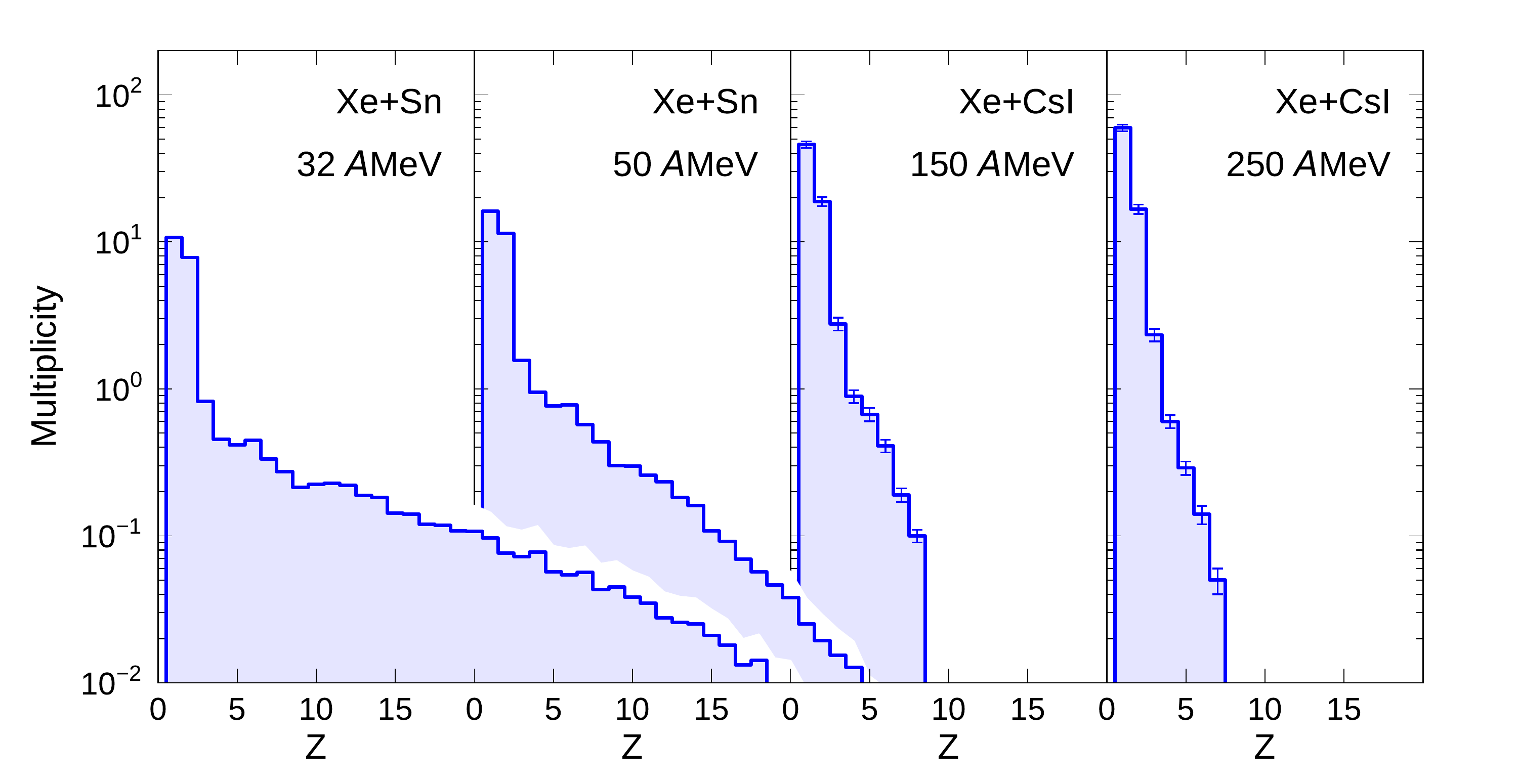}
\caption{\label{fig:zmult_exp}
Charge distribution of observed fragments in central heavy-ion collisions for $\nuc[129]{Xe}+\nuc[\text{nat}]{Sn}$ at 32 and 50 MeV/nucleon, and for $\nuc[129]{Xe}+\nuc{CsI}$ at 150 and 250 MeV/nucleon.  This figure was produced by plotting the INDRA data \cite{hudan2003} for $\nuc[129]{Xe}+\nuc[\text{nat}]{Sn}$ and the FOPI data \cite{reisdorf2010} for $\nuc[129]{Xe}+\nuc{CsI}$.}
\end{figure}

Figure \ref{fig:zmult_exp} shows the distributions of the charge $Z$ of 
fragments produced in central collisions, measured by the INDRA detector and the FOPI detector.  The $\nuc[129]{Xe}+\nuc[\text{nat}]{Sn}$ results at the lower two incident energies, 32 and 50 MeV/nucleon, are taken from Ref.~\cite{hudan2003} of the INDRA data and the $\nuc[129]{Xe}+\nuc{CsI}$ results at the higher two incident energies, 150 and 250 MeV/nucleon, are taken from Ref.~\cite{reisdorf2010} of the FOPI data.  In spite of the differences in the details of the analysis of these data, we can see the evolution of the fragmentation pattern as a function of the incident energy since $\nuc{Xe}+\nuc{Sn}$ and $\nuc{Xe}+\nuc{CsI}$ have similar system sizes.  In all the shown cases, the fragment multiplicity $Y(Z)$ is a decreasing function of $Z$.  At the high incident energies, $Y(Z)$ falls rapidly as exponential for $Z\ge 3$ fragments with an exception of $Z=4$ for which $\nuc[8]{Be}$ is unstable \cite{reisdorf1997}.  At lower incident energies, the distribution of $Y(Z)$ extends to heavier fragments but there is no residue component near $Z$ of the projectile and target nuclei.  The mean multiplicity of $Z\ge3$ fragments is about 7 at 50 MeV/nucleon and 5.5 at 32 MeV/nucleon \cite{leneindre1999thesis} in the $\nuc{Xe}+\nuc{Sn}$ system.  These are typical cases of multifragmentation in  which the system is disintegrating into many fragment nuclei together with light particles.

\begin{figure}
\centering
\includegraphics[scale=0.5]{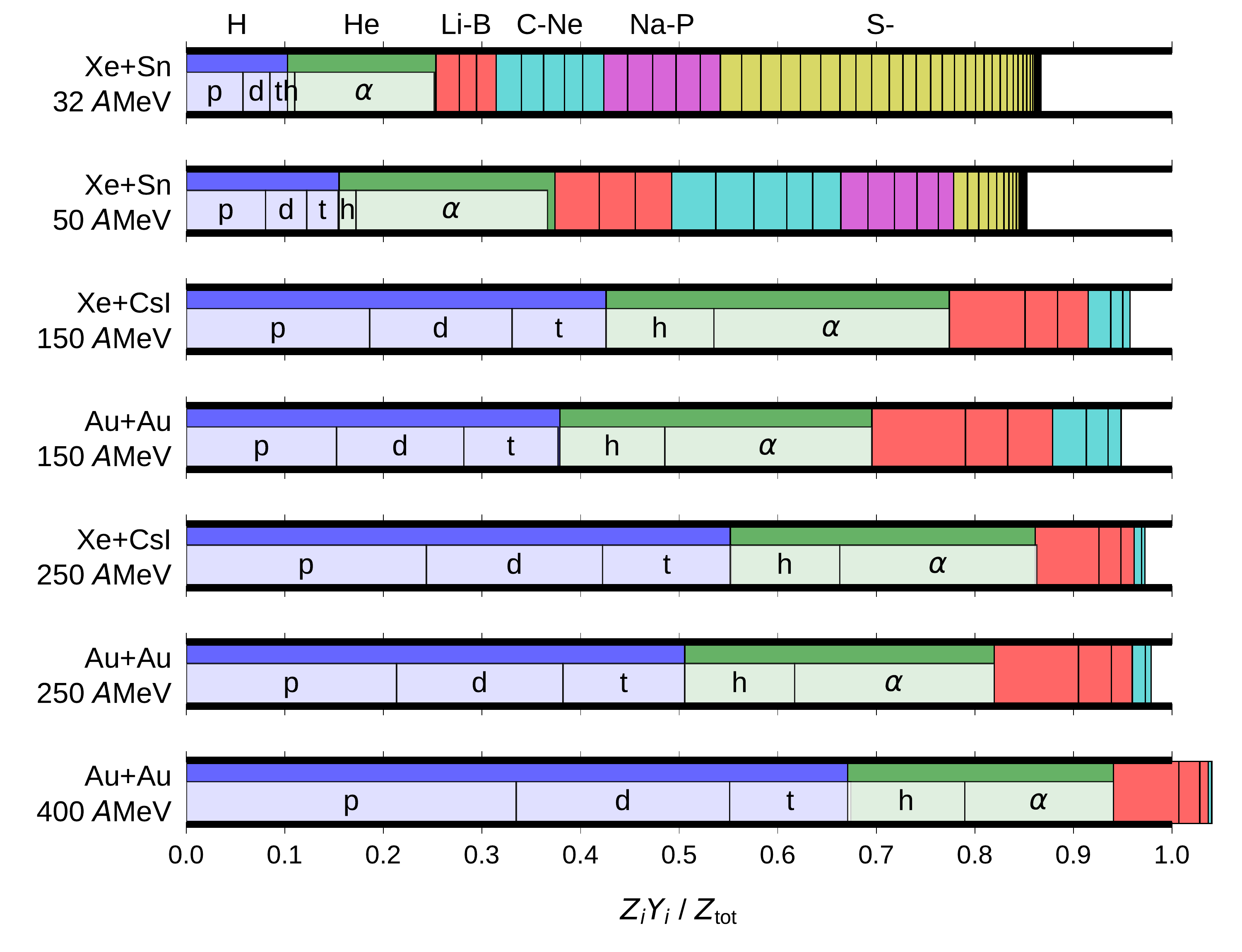}
\caption{\label{fig:zratiobars_exp}
Decomposition of protons into final products in central heavy-ion collisions of $\nuc[129]{Xe}+\nuc[\text{nat}]{Sn}$ (32 and 50 MeV/nucleon),  $\nuc[129]{Xe}+\nuc{CsI}$ (150 and 250 MeV/nucleon) and $\nuc[197]{Au}+\nuc[197]{Au}$ (150, 250 and 400 MeV/nucleon).  Fractions were obtained from the INDRA data \cite{hudan2003} for $\nuc[129]{Xe}+\nuc[\text{nat}]{Sn}$ and from the FOPI data \cite{reisdorf2010} for $\nuc[129]{Xe}+\nuc{CsI}$ and $\nuc[197]{Au}+\nuc[197]{Au}$ without taking into account the uncertainties in the experimental data.  Vertical lines separate fractions from different elements.  For H and He, the fractions are shown separately for $p$, $d$, $t$, $\nuc[3]{He}$ ($h$) and $\alpha$.}
\end{figure}

Already in Fig.~\ref{fig:zmult_exp}, we can carefully see that many light particles (H and He) are emitted.  The number of heavy fragments is small but it does not necessarily mean that they are unimportant because many nucleons are contained in a heavy fragment.  These points are clearer in the way of plotting as in Fig.~\ref{fig:zratiobars_exp} where the charge fractions $x_i=Z_iY_i/Z_{\text{tot}}$ of particle species $i$ are shown for different systems of central collisions, where $Z_{\text{tot}}=Z_{\text{projectile}}+Z_{\text{target}}$ is the total charge of the system.  The same INDRA and FOPI data are used as in Fig.~\ref{fig:zmult_exp}, with additional isotopic information on light charged particles, i.e., $p$, $d$, $t$, $\nuc[3]{He}$ (labeled as $h$) and $\alpha$.  For the INDRA data of $\nuc[129]{Xe}+\nuc[\text{nat}]{Sn}$ at 32 and 50 MeV/nucleon, the difference of the sum of $x_i$ from 1 corresponds to the particles which were not detected by the INDRA detector in each event.  The same differences in the FOPI data of $\nuc[129]{Xe}+\nuc{CsI}$ and $\nuc[197]{Au}+\nuc[197]{Au}$ are due to systematic uncertainties in the yields $Y_i$.  An important point of observation is that the fraction of protons is a relatively small part in all the cases shown here.  It is only about 10\% in $\nuc{Xe}+\nuc{Sn}$ at the low incident energy of 32 MeV/nucleon.  At the highest incident energy shown here, the fraction of protons is still 30-40\% in $\nuc{Au}+\nuc{Au}$ at 400 MeV/nucleon.  In all the cases, the charge fraction of $\alpha$ particles is as important as that of protons.  When the incident energy is raised, the fractions of deuterons, tritons and $\nuc[3]{He}$ increase and the fractions of heavier fragments decrease.  When the $\nuc{Xe}+\nuc{CsI}$ and $\nuc{Au}+\nuc{Au}$ systems are compared at the same incident energy, the $\nuc{Xe}+\nuc{CsI}$ system tends to disintegrate into smaller pieces.  This may be because of the different system sizes and/or because of the different neutron-to-proton ratio of the system.
In these experiments, neutrons were not measured and the fragment mass numbers were not measured except for light charged particles.  It will be interesting to study mass fractions in future.

It is sometimes important to know how the central events were selected in the data analysis.  Since the impact parameter is not a direct observable, one needs to choose an observable of each event that is believed to be strongly and monotonically correlated to the impact parameter.  In the INDRA data \cite{hudan2003} shown in Figs.~\ref{fig:zmult_exp} and \ref{fig:zratiobars_exp}, the flow angle $\theta_{\text{flow}}$ is defined for the kinetic energy tensor calculated from $Z\ge3$ fragments in the center-of-mass frame.  The events with $\theta_{\text{flow}}\ge 60^\circ$ (for 32 MeV/nucleon) or $\theta_{\text{flow}}\ge 45^\circ$ (for 50 MeV/nucleon) are regarded as central events with an additional precondition that the total detected charge is $\ge80\%$ of the total charge of the system.  In the FOPI data \cite{reisdorf2010} shown in Figs.~\ref{fig:zmult_exp} and \ref{fig:zratiobars_exp}, central events are selected by using the observable called ERAT which is the ratio of the total transverse to longitudinal kinetic energies in the center-of-mass frame.  These impact parameter observables should not be automatically correlated to the observables of our interest.

\begin{figure}
\centering
\includegraphics[width=0.6\textwidth]{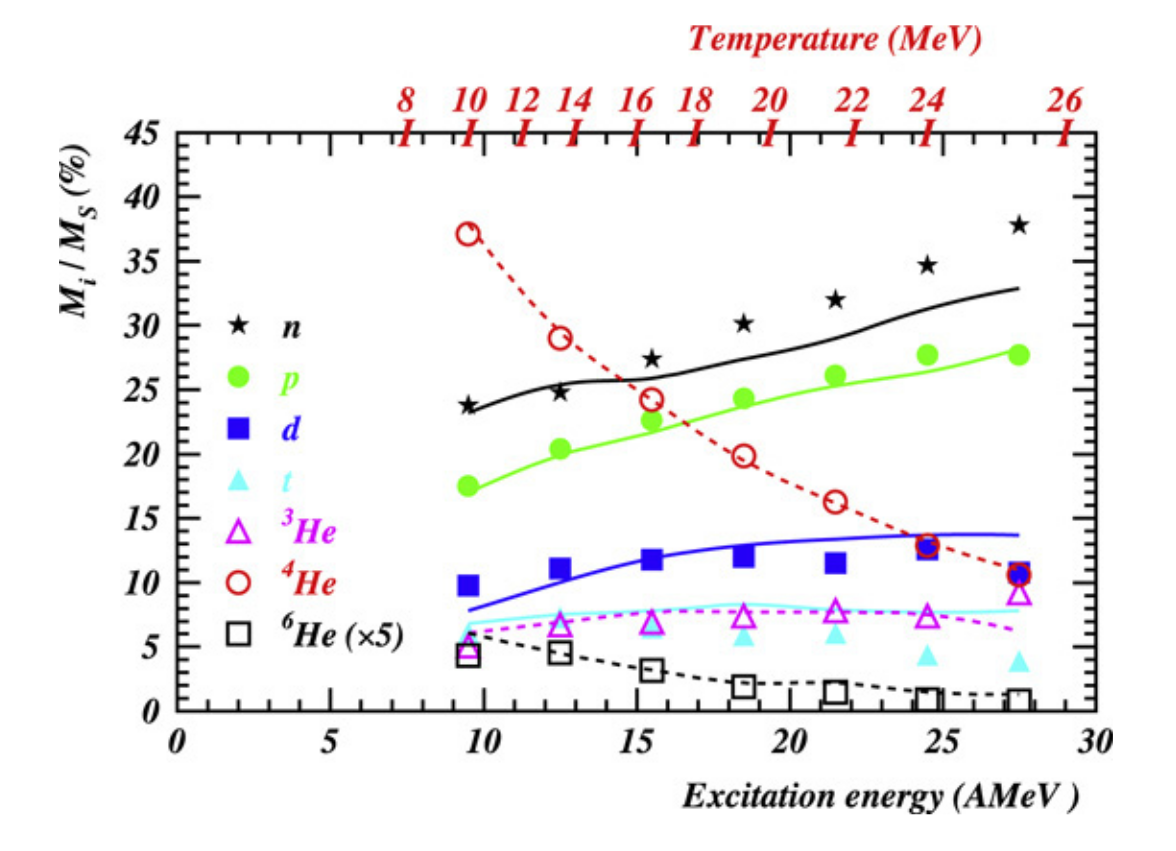}
\caption{\label{fig:borderie1999}
Mass fractions of vaporized quasi-projectiles formed in $\nuc[36]{Ar}+\nuc[58]{Ni}$ collisions at 95 MeV/nucleon, as functions of the excitation energy per nucleon.  Symbols are for data and lines are the results of a statistical model.  The temperatures in the model are also shown.  Taken from Ref.~\cite{borderie2008}.  Adapted from Ref.~\cite{borderie1999}.
}
\end{figure}
The importance of clusters in highly excited systems can also be found in the data by Borderie et al.~\cite{borderie1999} which show the composition of products from vaporized quasi-projectiles in $\nuc[36]{Ar}+\nuc[58]{Ni}$ collisions at 95 MeV/nucleon.  For the selected vaporized events in which all the products from the quasi-projectile have $Z<3$, the mass fractions of the products are shown in Fig.~\ref{fig:borderie1999} as functions of the excitation energy.

\subsection{\label{sec:exp-collective}Radial expansion and stopping}

\begin{figure}
\centering
\includegraphics{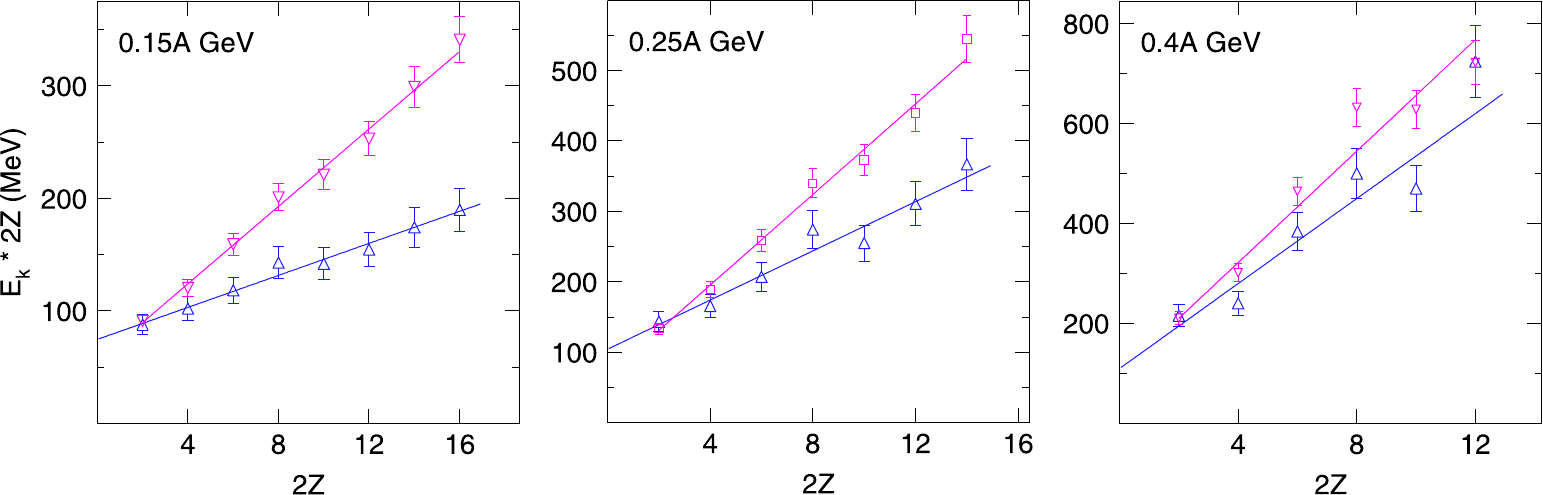}
\caption{\label{fig:ekinav_exp} Kinetic energies of fragments in central $\nuc{Au}+\nuc{Au}$ collisions ($b_0<0.15$) at three different incident energies.  The average kinetic energy per nucleon multiplied by $2Z$ is shown as a function of $2Z$ where $Z$ is the fragment charge.  Pink square points are for fragments emitted at angles in $5^\circ<\theta<45^\circ$ in the center-of-mass frame, while blue triangles are for $80^\circ<\theta<100^\circ$. Adapted from Ref.~\cite{reisdorf2010}.}
\end{figure}

Kinetic energies or momenta of emitted particles carry information strongly related to the collective dynamics such as isotropic or anisotropic radial expansion.  Experimental data suggesting the existence of radial flow have been published by the EOS collaboration \cite{lisa1995}, the FOPI collaboration (Ref.~\cite{reisdorf2010} and references therein) and the INDRA collaboration \cite{marie1997}.  Figure \ref{fig:ekinav_exp} shows the FOPI data \cite{reisdorf2010} for the average kinetic energies of fragments $\langle E'\rangle\equiv\langle E/A\rangle\times A'$ in the center-of-mass frame as a function of $A'=2Z$, where $A$ and $Z$ are the mass number and the charge of the fragment.  The shown quantity $\langle E'\rangle$ agrees with the kinetic energies of fragments as a function of the fragment size if $A=A'$ is a valid approximation.  For different incident energies for these central heavy-ion collisions and for forward and transverse emission directions ($5^\circ<\theta<45^\circ$ and $80^\circ<\theta<100^\circ$), the fragment kinetic energies are well fitted by linear functions $\langle E'\rangle\approx E_0+E_{\text{flow}}A'$.  This can be explained if the particle velocities can be expressed as a sum of two independent parts $\bm{v}=\bm{v}_{\text{flow}}+\bm{v}_{\text{th}}$, where the distribution of the flow velocity $\bm{v}_{\text{flow}}$ is common to all the particles and the thermal velocity $\bm{v}_{\text{th}}$ is for a temperature $T_0$ common to all the particle species, in the non-relativistic kinematics here for simplicity.  We see that radial flow is strong in the sense that the flow energy $E_{\text{flow}}A'$ is a large part of the total kinetic energy.

\begin{figure}
\centering
\includegraphics{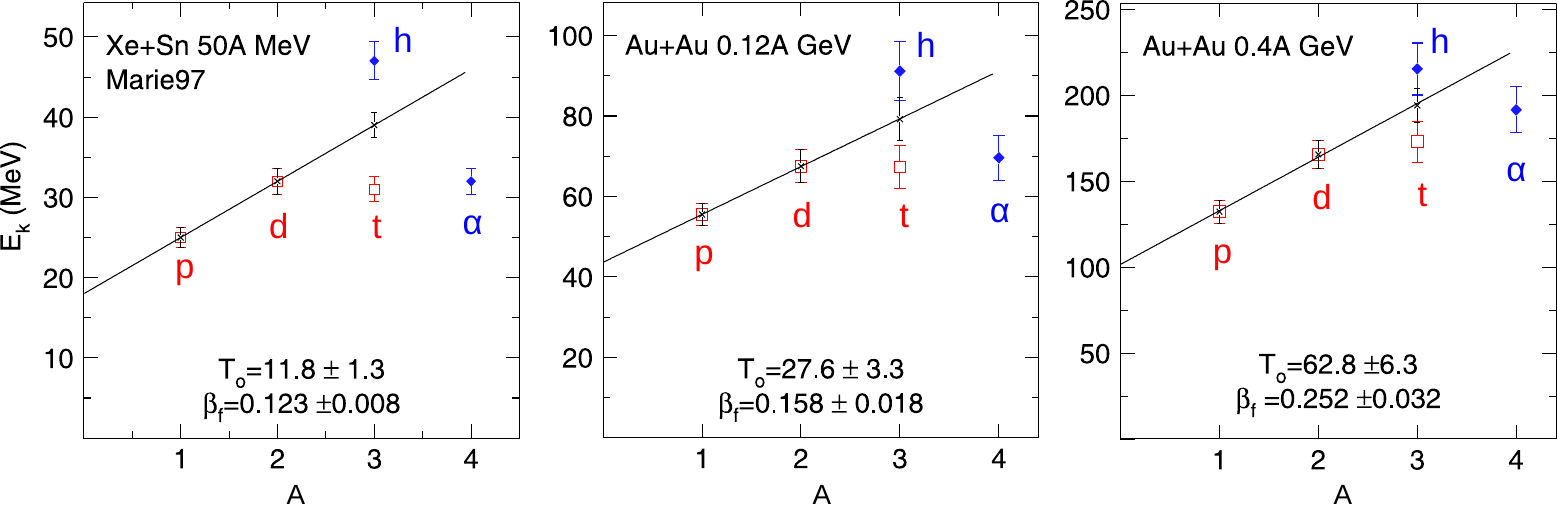}
\caption{\label{fig:ekinav_lpc_exp}
Average kinetic energies of H (red open squares) and He (blue filled diamonds) isotopes emitted at $90^\circ$ in the center-of-mass frame in $\nuc{Au}+\nuc{Au}$ central collisions ($b_0<0.25$) at 120 MeV/nucleon (middle panel) and 400 MeV/nucleon (right panel).  The left left panel is for the INDRA data \cite{marie1997} for $\nuc{Xe}+\nuc{Sn}$ at 50 MeV/nucleon.  The straight line is a linear fit up to mass 3.  The derived apparent temperatures $T_{\text{o}}$ and flow velocities $\beta_{\text{f}}$ are indicated.  Adapted from Ref.~\cite{reisdorf2010}.  
}
\end{figure}
The kinetic energies of light charged particles are also shown in a similar way in Fig.~\ref{fig:ekinav_lpc_exp}.  The fitting parameters $E_{\text{flow}}$ and $E_0$, for $1\le A\le 3$ in this case, may be related to the average flow velocity $\beta_{\text{flow}}$ and the temperature $T_0$, which are shown in each panel of the figure.  However, we see here that the linear dependence is not as clear as for heavier fragments, in particular at lower incident energies.  The difference between tritons and $\nuc[3]{He}$ can be explained only partly by the different Coulomb interactions.  For this puzzle of tritons and $\nuc[3]{He}$, no satisfactory solutions have been obtained by microscopic and dynamical approaches.

\begin{figure}
\centering
\includegraphics[scale=0.5]{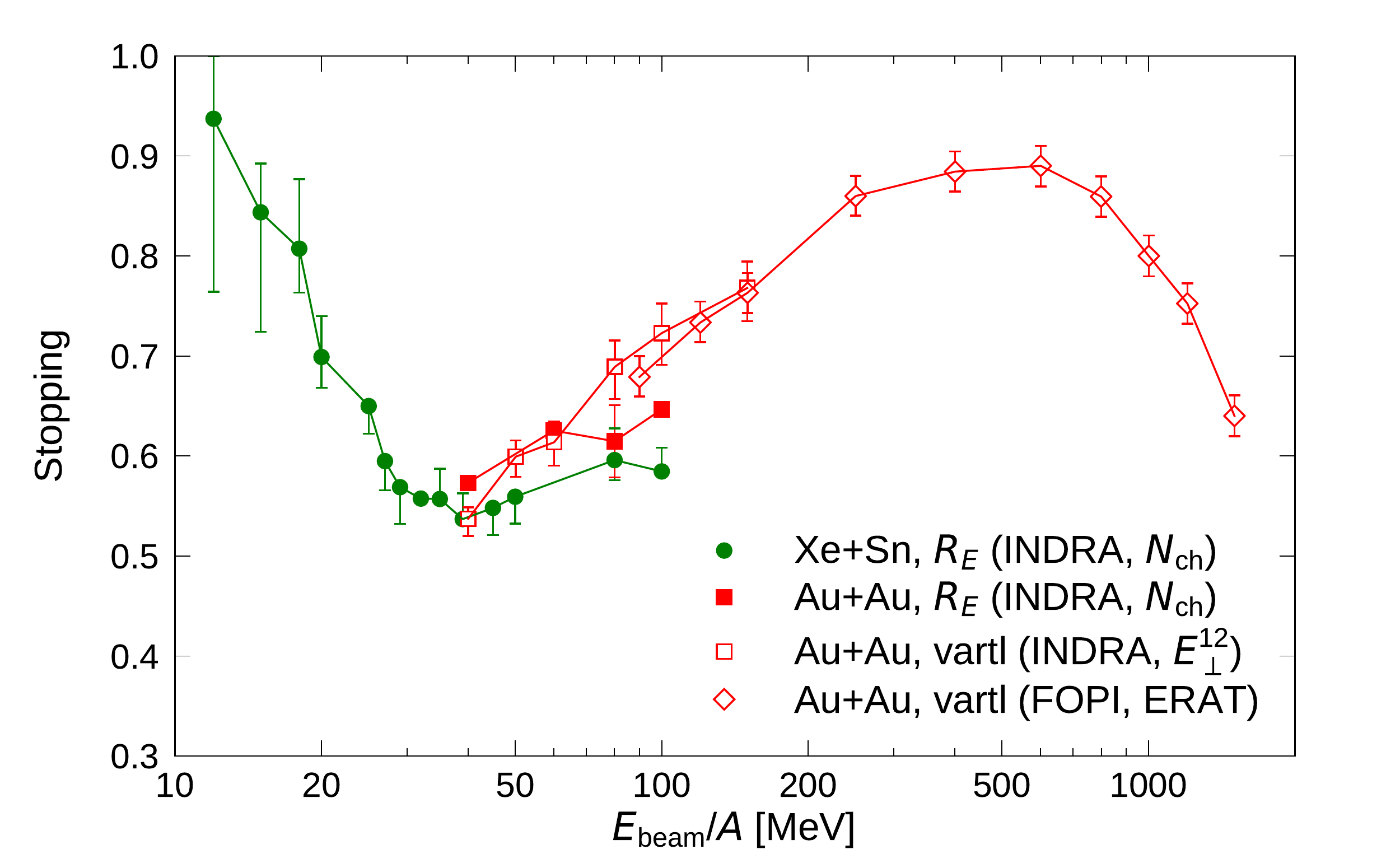}
\caption{\label{fig:stopping} The degree of stopping, $R_E$ and \textit{vartl}, as a function of the incident energy.  Filled green circles ($\nuc{Xe}+\nuc{Sn}$) and filled red squares ($\nuc{Au}+\nuc{Au}$) are INDRA data of $R_E$ averaged for central events selected by the total charged particle multiplicity $N_{\text{ch}}$ \cite{lehaut2010}.  Open red squares (INDRA data) and open red diamonds (FOPI data) show the value of \textit{vartl} for $\nuc{Au}+\nuc{Au}$ central events selected by the light-particle transverse energy $E_\perp^{12}$ (INDRA) or ERAT (FOPI) \cite{reisdorf2004,andronic2006}.  Data in Refs.~\cite{lehaut2010,andronic2006} are plotted.  }
\end{figure}

The difference of the kinetic energies in Fig.~\ref{fig:ekinav_exp} between forward and transverse angles shows that the expansion of the system is not isotropic but elongated in the beam direction to some degree.  This means that stopping is not complete, i.e., the memory of the direction of the initial momenta of the two colliding nuclei is not lost completely even in these violent central collisions.  The degree of stopping is often characterized by defining some quantities.  The quantity \textit{vartl} is defined in Ref.~\cite{reisdorf2004} as the ratio of the variances of the transverse to that of the longitudinal rapidity distributions, with the integration intervals limited between $-1$ and 1 in the scaled rapidities.  The autocorrelation of \textit{vartl} to the ERAT centrality variable is avoided by removing the particle of interest from ERAT.  The data from INDRA and FOPI measurements are combined in Ref.~\cite{andronic2006} for a wide region of the incident energy.  The values of $\textit{vartl}$ in $\nuc{Au}+\nuc{Au}$ collisions are plotted in Fig.~\ref{fig:stopping} with open symbols above 40 MeV/nucleon  (Filled symbols will be explained later).   This figure shows the global stopping variable which is defined for the $Z$-weighted sum of the rapidity distributions.  With this \textit{vartl} variable, from low to high incident energies, it seems that stopping becomes strong with a broad plateau extending from 200 to 800 MeV/nucleon and then it decreases at higher energies.

\begin{figure}
\centering
\includegraphics{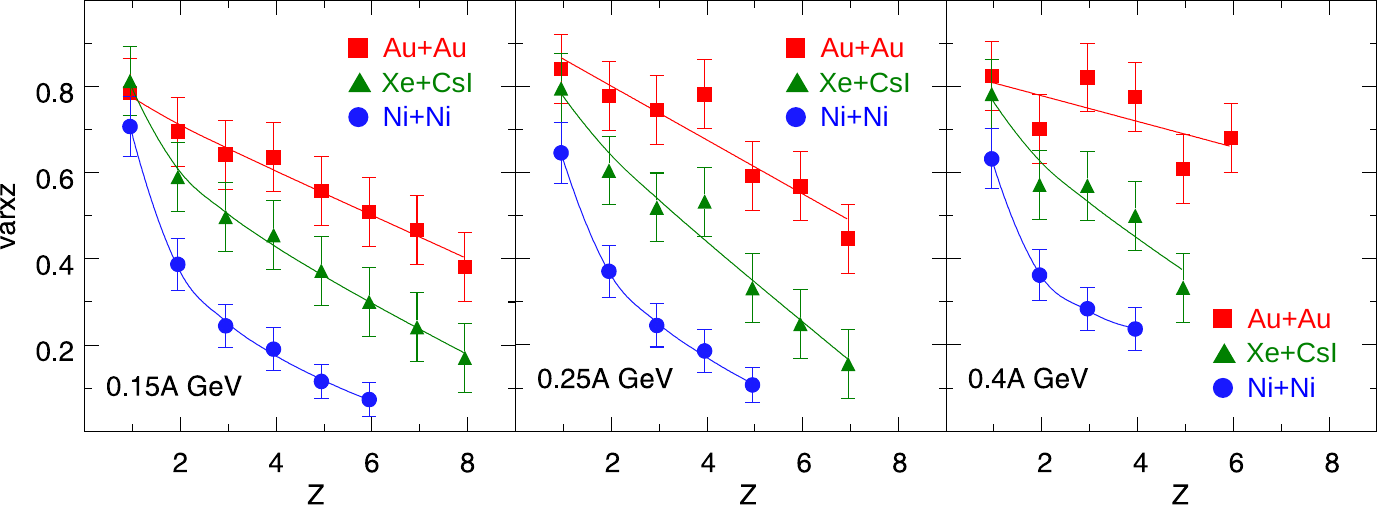}
\caption{\label{fig:stopping_fopi_sysz}
Dependence of the stopping observable \textit{varxz} on the system size ($\nuc{Au}+\nuc{Au}$, $\nuc{Xe}+\nuc{CsI}$ and $\nuc{Ni}+\nuc{Ni}$) and the fragment charge number $Z$ for central collisions ($b_0<0.15$) at the incident energy shown in each panel.  Lines are the fits to guide the eye.  Adapted from Ref.~\cite{reisdorf2010}.}
\end{figure}
Stopping depends not only on the incident energy but also on the size of the reaction system as shown in Fig.~\ref{fig:stopping_fopi_sysz} (taken from Ref.~\cite{reisdorf2010}).  This figure shows a stopping variable called \textit{varxz} which is similar to \textit{vartl} but obtained from rapidity distributions without a limit for the integration range.  Weaker stopping is observed for lighter system, which may be naturally understood that a light nucleus does not have enough thickness to stop the other nucleus.  Another important point here is that stopping appears weaker when the stopping variable is defined for heavier fragments with larger $Z$.  This is expected to some degree if the fragment velocities follow the form $\bm{v}=\bm{v}_{\text{flow}}+\bm{v}_{\text{th}}$ as introduced above, where $\bm{v}_{\text{flow}}$ is the flow velocity whose distribution is now anisotropic but common to all the particles while the distribution of the thermal velocity $\bm{v}_{\text{th}}$ is broader for lighter particles.  However, the strong $Z$-dependence in these experimental data suggests something more beyond this trivial effect.

For the INDRA data, analyses of stopping have been published in Refs.~\cite{lehaut2010,lopez2014}, where the stopping is characterized by the ratio of transverse and longitudinal kinetic energies
\begin{equation}
R_E=\frac{\sum E_\perp}{2\sum E_\parallel}.
\end{equation}
Here $E_\perp$ and $E_\parallel$ is the transverse and longitudinal kinetic energies of emitted particles in the center-of-mass frame, and the sum is taken over all the detected products (neutrons are not detected).  The full stopping with isotropic emissions corresponds to $R_E=1$.  Central events are selected using the total multiplicity $N_{\text{ch}}$ of detected charged products which is expected to be biased minimally to avoid autocorrelation.  The stopping observable $R_E$ can be defined on event-by-event basis.  For the selected central events, the mean value of $R_E$ is shown in Fig.~\ref{fig:stopping} with filled symbols as a function of the incident energy below 100 MeV/nucleon.  The filled red squares are for $\nuc{Au}+\nuc{Au}$ collisions and the filled green circles are for $\nuc{Xe}+\nuc{Sn}$ collisions.  The dependence on the reaction system shown here seems to suggest a similar system-size dependence of the stopping as also seen in Fig.~\ref{fig:stopping_fopi_sysz} at higher energies in the FOPI data.  The value of $R_E$ is about 0.5-0.7, suggesting partial stopping, in a wide range of the incident energy above 30 MeV/nucleon.  At lowest energies, the stopping is strong ($R_E\approx 1$) suggesting almost isotropic emission of particles.  The variance of the event-by-event fluctuation of $R_E$ is almost as large as the mean value of $R_E$ \cite{lehaut2010}.  In the energy region between 40 and 100 MeV/nucleon, we can compare the $R_E$ value averaged for events selected by $N_{\text{ch}}$ and the $\textit{vartl}$ value for events selected by the light-particle transverse energy $E_\perp^{12}$ \cite{andronic2006}.  Although it is not very clear, it may be the case that $R_E$ with the $N_{\text{ch}}$ selection has a weaker dependence on the incident energy than $\textit{vartl}$ with the $E_\perp^{12}$ selection.

In Ref.~\cite{lopez2014}, similar characteristics of stopping are shown for the quantity $R_E^p$ which is similar to $R_E$ but it is defined for the sum of the kinetic energy components of detected protons only.  Another difference is that the sums in the numerator and the denominator are first taken for the protons from all the selected central events and then the ratio is calculated.  It is mentioned in Ref.~\cite{lopez2014} that the latter procedure weakly lowers (5-10\%) the mean values for $R_E^p$ as compared to the event-by-event determination.

Here we have limited ourselves to the central collisions where the distributions are axially symmetric around the beam axis.  In semi-peripheral collisions with intermediate impact parameters, the azimuthal distributions have additional and important information, where the azimuthal angle $\phi$ is measured from the reaction plane and the distributions are usually characterized by the directed and elliptic flow variables.  Such flow observables are quite important in order to extract the information of high-density EOS of isospin-symmetric and asymmetric nuclear matter from heavy-ion collisions \cite{danielewicz2002,reisdorf2012,russotto2016}.  However, for a reasonable description of flow observables, we first need to well understand the stopping and expansion of the participant part of the system, i.e., the geometrically overlapping region of the projectile and the target.  Therefore we will mainly focus on the stopping and the radial expansion in this article.  The projectile-like part in high-energy collisions also undergoes multifragmentation \cite{schuttauf1996} with weaker expansion than in central collisions.  In the Fermi energy domain, fragments with intermediate velocities suggest formation of a low-density neck region between the projectile-like and target-like parts (see e.g.\ Ref.~\cite{defilippo2014}).  The projectile-like fragment is likely a deformed object in the dynamical stage \cite{hudan2014,jedele2017mcintosh}.

\subsection{Space-time information from correlations}
Even though the experimental information reviewed in the above two subsections certainly indicates a kind of collision dynamics and fragment/cluster formation mechanisms, the measured observables do not have truly direct space-time information on the collision dynamics during which particles are still strongly interacting.  However, more space-time information has been extracted from the momentum correlations between emitted particles.  See Ref.~\cite{verde2006} for a review.


The idea of getting spatial information of an object from the particle correlations was originally introduced in astronomy by Hanbury-Brown and Twiss \cite{brown1956hbt}, and the formulations convenient in heavy-ion collisions are found in Refs.~\cite{koonin1977,pratt1984,pratt1987}.  We consider the probability $P(\bm{p}_1,\bm{p}_2)$ that two particles, e.g.\ two protons, are emitted with momenta $\bm{p}_1$ and $\bm{p}_2$, both of which correspond to similar velocities $\bm{v}$, so that $\bm{p}_1=m_1\bm{v}+\bm{q}$ and $\bm{p}_2=m_2\bm{v}-\bm{q}$ with a small $\bm{q}$.  These two particles should have been interacting with the rest of the system during the violent collision dynamics before some space-time points $(\bm{r}_1,t_1)$ and $(\bm{r}_2,t_2)$, respectively, after which they no longer interact with the rest but they can still interact with each other.  An essential assumption made here is that these space-time points are not correlated so that they can be sampled independently from one-particle source functions $D_1(\bm{r}_1,\bm{p},t_1)$ and $D_2(\bm{r}_2,\bm{p},t_2)$.  Then, for the case of two-proton correlations, Koonin wrote \cite{koonin1977}
\begin{equation}
P(\bm{p}_1,\bm{p}_2)=\int dt_1 dt_2\int d\bm{r}_1d\bm{r}_2
D(\bm{r}_1,m\bm{v},t_1)D(\bm{r}_2,m\bm{v},t_2)
\Bigl(
\tfrac14|\psi_{\bm{q}}^{(S=0)}(\bm{r}_1'-\bm{r}_2)|^2
+\tfrac34|\psi_{\bm{q}}^{(S=1)}(\bm{r}_1'-\bm{r}_2)|^2
\Bigr)
\label{eq:Koonin}
\end{equation}
with $\bm{r}_1'=\bm{r}_1+\bm{v}(t_2-t_1)$.  This expression corresponds to taking a snapshot of the independently emitted particles with the two-particle scattering states which are the singlet state $\psi_{\bm{q}}^{(S=0)}$ and the triplet state $\psi_{\bm{q}}^{(S=1)}$ in this case (we will revisit here in Sec.~\ref{sec:clstmodels}).  With some assumptions for the source functions, one may write the correlation function in the form of so-called Koonin-Pratt equation \cite{pratt1987}
\begin{equation}
\frac{P(\bm{p}_1,\bm{p}_2)}{P_1(\bm{p}_1)P_2(\bm{p}_2)}
= C(\bm{q},\bm{v})
= 1+R(\bm{q},\bm{v})
= 1+ \int S(\bm{r},\bm{v})K(\bm{r},\bm{q})d\bm{r},
\label{eq:KooninPratt}
\end{equation}
which relates the correlation in the relative momentum $\bm{q}$ to the two-particle source function $S(\bm{r},\bm{v})$ that is the probability of the two particles being emitted with the relative distance $\bm{r}$.  The kernel $K(\bm{r},\bm{q})$ contains the information of the final-state interaction.  Different approaches have been taken in order to derive $S(\bm{r},\bm{v})$ from the measured $R(\bm{q},\bm{v})$ \cite{verde2002}.  Different values of $\bm{v}$ may be interpreted as different stages of particle emissions.  Information on the source lifetime is contained in the dependence of $R(\bm{q},\bm{v})$ on the direction of $\bm{q}$ relative to that of $\bm{v}$.  See Ref.~\cite{verde2006} and references therein.  For the FOPI data of central collisions at 400 MeV/nucleon, the effect of strong radial expansion of the source was taken into account to extract the space-time information of emission of protons and light clusters \cite{kotte1999spacetime}.

When fragment nuclei are emitted in heavy-ion collisions so that they no longer interact with the other part of the system, most of them are actually resonances which will decay by evaporating particle(s) in a long time scale.  Therefore the final fragment and  a particle originating from the same primary fragment are correlated in the velocity space.  The correlation between a fragment and light particles were analyzed to reconstruct the primary fragment \cite{marie1998,hudan2003} so that the excitation energies of primary fragments and the multiplicities of evaporated particles were extracted.  It may be possible to think that this kind of application is similar to Eq.~(\ref{eq:Koonin}) but the snapshots are taken with resonances or with continuum states containing resonance structures.  However, nucleons in a primary fragment should have been already correlated at freeze-out.

\begin{figure}
\centering
\includegraphics[scale=0.4]{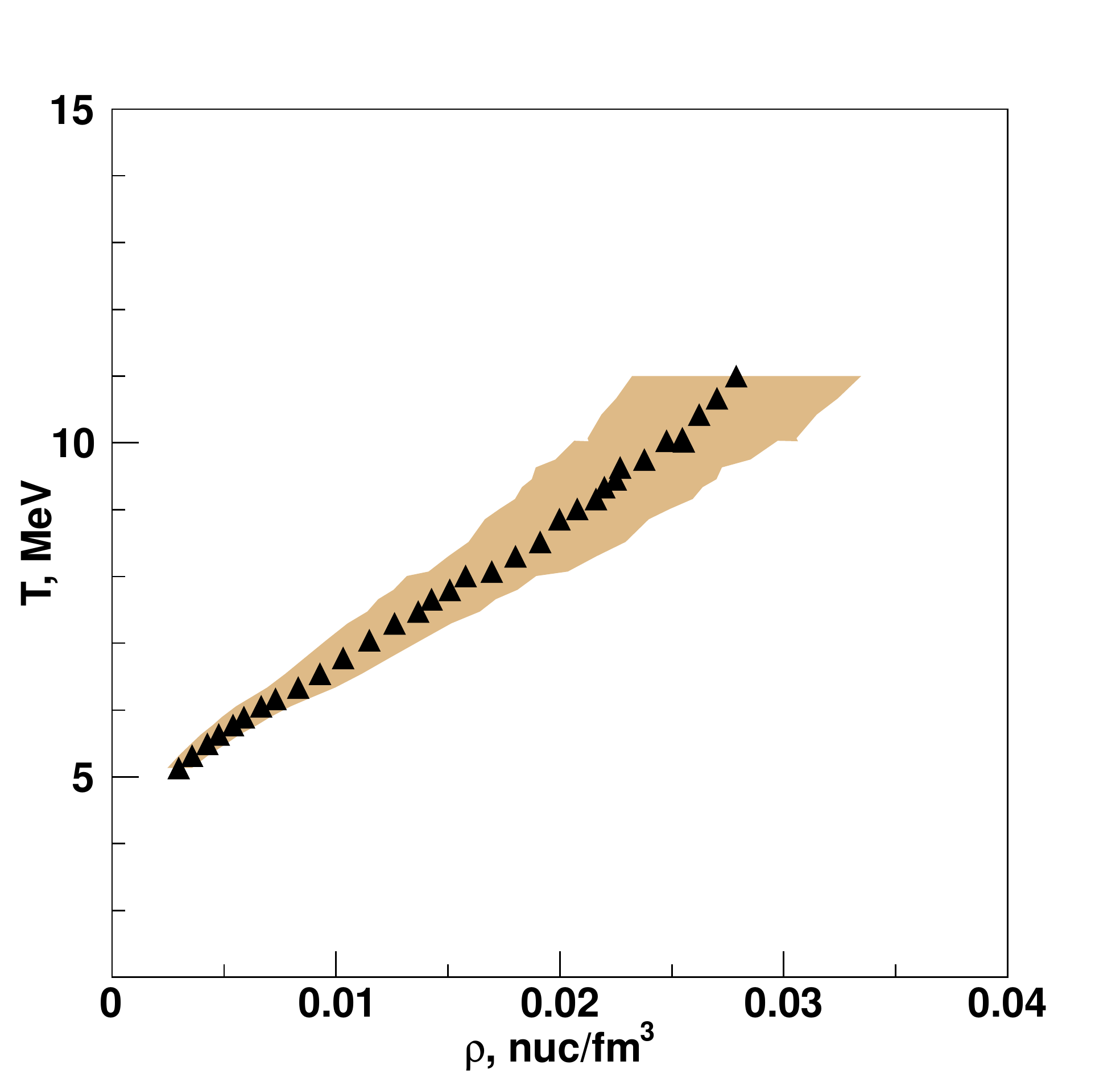}
\caption{\label{fig:qin2012texas}
Temperatures and densities sampled from the expanding intermediate velocity source.  The shaded band indicates the uncertainty in the density.  Taken from Ref.~\cite{qin2012texas}.
}
\end{figure}

Observing a cluster may also be considered as measuring the correlation between the particles in the cluster.  The difference from the momentum correlation is whether the internal state of the particles is a bound state or continuum states.  If the scattering wave functions in Eq.~(\ref{eq:Koonin}) were replaced by the cluster wave function, one will have a kind of coalescence model which relates the distribution function, e.g.~the momentum spectrum, of clusters to the $A_{\text{c}}$-th power of that of nucleons, where $A_{\text{c}}$ is the mass number of the cluster (see a related argument in Sec.~\ref{sec:clstmodels}).  However, the independent emission of nucleons is not a necessary condition for the coalescence relation in measured spectra.  In fact, as considered by Mekjian \cite{mekjian1978,mekjian1977}, in ideal Boltzmann gas of nucleons and clusters in thermal and chemical equilibrium under a (free) volume $V$, the numbers and the distributions of clusters are related to those of nucleons just as mentioned above.  In Refs.~\cite{qin2012texas,wada2012}, the data of $\nuc[40]{Ar}+\nuc[112,124]{Sn}$ and $\nuc[64]{Zn}+\nuc[112,124]{Sn}$ collisions at 47 MeV/nucleon measured by NIMROD multidetector at Texas A\&M University were analyzed to extract the evolution of the density and the temperature of the intermediate velocity source as shown in Fig.~\ref{fig:qin2012texas}.  For each velocity $v_{\text{surf}}$ of the emitted particle at the nuclear surface prior to Coulomb acceleration, the temperature is determined by employing double isotope yield ratios \cite{albergo1985,kolomiets1997temperature} and the density is determined by using the volume $V$ extracted with the Mekjian model \cite{mekjian1978} for the measured light particles.  The figure implies how the temperature and the density evolve as the intermediate velocity source expands.  The relevant densities are around $\sim\frac{1}{10}\rho_0$.

The fragment-fragment velocity correlations measured in $\nuc{Xe}+\nuc{Sn}$ central collisions were used in Ref.~\cite{lefevre2008} to extract the information of the deformed shape of the source.  In Ref.~\cite{piantelli2008}, the freeze-out properties, such as the neutron and proton numbers, the volume and the collective and internal excitation energies, were estimated from a simulation based on the experimental data for $\nuc{Xe}+\nuc{Sn}$ at 32-50 MeV/nucleon.

\subsection{Information from isospin degrees of freedom}

Heavy-ion collisions have been an important opportunity to study nuclear matter under various conditions in terrestrial experiments.  In particular, developments of RI beam facilities have enabled the studies of nuclear systems by varying the neutron and proton contents.  Nuclear EOS of asymmetric nuclear matter is not only relevant to heavy-ion collisions but also to nuclear structure problems and in astrophysical problems such as neutron stars, core-collapse supernovae and neutron star mergers.  Our knowledge has been increasing on the density dependence of the symmetry energy and other related properties such as the effective masses of neutrons and protons.  See Refs.~\cite{baran2005,bali2008,tsang2012,lattimer2013,horowitz2014,defilippo2014,baldo2016,bali2018} for reviews.  Transport model calculations are indispensable to relate dynamics in heavy-ion collisions to these nuclear matter properties.  Here, without going into this important subject, we will only make a few comments from a viewpoint of dynamics of central collisions and effects of clusters and fragments.

In Sec.~\ref{sec:exp-collective} we have reviewed how completely or incompletely the initial incident energy of colliding nuclei is distributed to all the directions so that the momentum distribution becomes equilibrated and isotropic (e.g.~$R_E\approx 1$) in the center-of-mass system.  In Refs.~\cite{rami2000,hong2002}, in order to see the degree of the mixing of the nucleons from the projectile and the target, the FOPI data of central collisions at 400 MeV/nucleon were analyzed for the four different combinations of ${}^{96}_{44}\nuc{Ru}$ and ${}^{96}_{40}\nuc{Zr}$ nuclei which have the same mass number but different neutron and proton contents.  If the nuclei are perfectly transparent, an isospin-related observable $X$ (e.g.~the proton multiplicity or $t/\nuc[3]{He}$ ratio) in a forward rapidity region should take the same value for ${}^{96}_{44}\nuc{Ru}+{}^{96}_{40}\nuc{Zr}$ and ${}^{96}_{44}\nuc{Ru}+{}^{96}_{44}\nuc{Ru}$.  On the other hand, in the case of complete mixing, $X$ for ${}^{96}_{44}\nuc{Ru}+{}^{96}_{40}\nuc{Zr}$ will be similar to the average of $X$ for ${}^{96}_{44}\nuc{Ru}+{}^{96}_{44}\nuc{Ru}$ and ${}^{96}_{40}\nuc{Zr}+{}^{96}_{40}\nuc{Zr}$. It is convenient to transform $X$ to an isospin transport ratio
\begin{equation}
\label{eq:isospin-transport-ratio}
R_{\text{i}}=\frac{2X-X({}^{96}_{44}\nuc{Ru}+{}^{96}_{44}\nuc{Ru})-X({}^{96}_{40}\nuc{Zr}+{}^{96}_{40}\nuc{Zr})}{X({}^{96}_{44}\nuc{Ru}+{}^{96}_{44}\nuc{Ru})-X({}^{96}_{40}\nuc{Zr}+{}^{96}_{40}\nuc{Zr})},
\end{equation}
so that $R_{\text{i}}=1$ for perfect transparency and $R_{\text{i}}=0$ for complete mixing.  The experimental results indicate incomplete mixing and partial transparency even in the most central collisions at 400 MeV/nucleon.

Similar isospin transport ratios were employed at lower energies \cite{tsang2004,tsang2009} in order to extract information on the density dependence of the symmetry energy from the experimental data taken at the National Superconducting Cyclotron Laboratory at Michigan State University.  For $\nuc[112,124]{Sn}+\nuc[112,124]{Sn}$ collisions at 35 MeV/nucleon, the impact parameter dependence of the transport ratio was studied in Ref.~\cite{sun2010} for the observable $X=\ln[Y(\nuc[7]{Li})/Y(\nuc[7]{Be})]$, which shows that the chemical equilibrium is not reached even in central collisions, $R_{\text{i}}=0.3$ - 0.4.  On the other hand, for the INDRA data of $\nuc[136,124]{Xe}+\nuc[124,112]{Sn}$ collisions at 32 MeV/nucleon, a high degree of chemical equilibration is reported in Ref.~\cite{bougault2018} for the light charged particles $p$, $d$, $t$ and $\alpha$, with an exceptional behavior of $\nuc[3]{He}$.

The spectra of neutrons and protons are predicted to be observables sensitive to the neutron-proton effective mass splitting in asymmetric nuclear matter by employing transport models (see Ref.~\cite{bali2018} and references therein).  Spectra of light particles, including clusters, have been measured for collisions of nuclei with various $N/Z$ ratios \cite{coupland2016,chajecki2014,liu2012}.  The data should be very useful not only to constrain the effective mass splitting or the density dependence of the symmetry energy but also to constrain the cluster production mechanisms.  So far, however, comparisons with transport models \cite{coupland2016,kong2015} have been done for `coalescence-invariant' spectra which are constructed for the sums of the emitted neutrons/protons and those bound in light clusters.

\subsection{Discussions}
As partly reviewed in this section, a lot of experimental data have been accumulated in the past decades, which suggest how the nucleons, light clusters and heavier fragments are emitted from the system that may be dynamically evolving.  However, the experimental observables are inevitably on the final products that left the rest of the system long time ago, and therefore some model assumption is always necessary if one tries to relate the observables to the information on the intermediate states of collisions.

Many observations such as incomplete stopping and partial isospin equilibration suggest that heavy-ion collisions should be understood as a dynamical phenomenon, starting with two boosted initial nuclei, without assuming local equilibrium.  One may have in mind transport models which describe the time evolution of the one-body distribution of nucleons.  However, we also see that, at least in the final state, the many-body correlations are often very strong so that many clusters and fragments are formed.  The states near the end of the dynamical stage may be close to local equilibrium in which nucleons are strongly correlated to form clusters and fragments.  Nevertheless, dynamical approaches are still necessary to describe the collective dynamics, the emission mechanism of particles, and any deviation from local equilibrium.

Many-body correlations and one-body dynamics are interrelated in general.  In other words, these correlations should not be treated as perturbation on top of the one-body dynamics.  For example, Reisdorf discussed in Ref.~\cite{reisdorf2010} by comparing the experimental data and a transport calculation as follows.  In a transport model without sufficient cluster correlations, too many non-clustered protons, together with the total energy conservation, lead to too small average kinetic energy of protons, and thus the two observables, cluster yeilds and radial flow, are interrelated.

\section{\label{sec:matter}Clusters in nuclear many-body systems}

\subsection{Nuclear system at various excitation energies}

The lowest-energy state of a many-nucleon system is a nucleus in its ground state.  It is usually described well in the mean-field or shell-model picture based on independent motions of nucleons (with the short-range two-nucleon correlations renormalized in effective interactions and two-nucleon collisions mostly blocked by the Pauli principle).  As nuclear structure problems, many kinds of nuclear excited states with collective or non-collective characteristics can be understood based on single-particle excitations such as the one-particle one-hole excitations.  There are, however, another kind of excited states in which cluster correlations are important.  See e.g.\ Refs.~\cite{horiuchi2012,kanada2012,freer2017} for recent reviews.  Developed cluster structures appear near the excitation energies corresponding to the cluster-decay thresholds as predicted by the Ikeda diagram \cite{ikeda1968}.  Some clusterized states, such as the Hoyle state in the $\nuc[12]{C}$ nucleus, can be interpreted by gas-like independent motions of $\alpha$ clusters as in the Bose-Einstein condensate \cite{tohsaki2001,freer2014,tohsaki2017}.  Above the threshold energy, further cluster structures are known such as molecular resonances.

\begin{figure}
\centering
\includegraphics{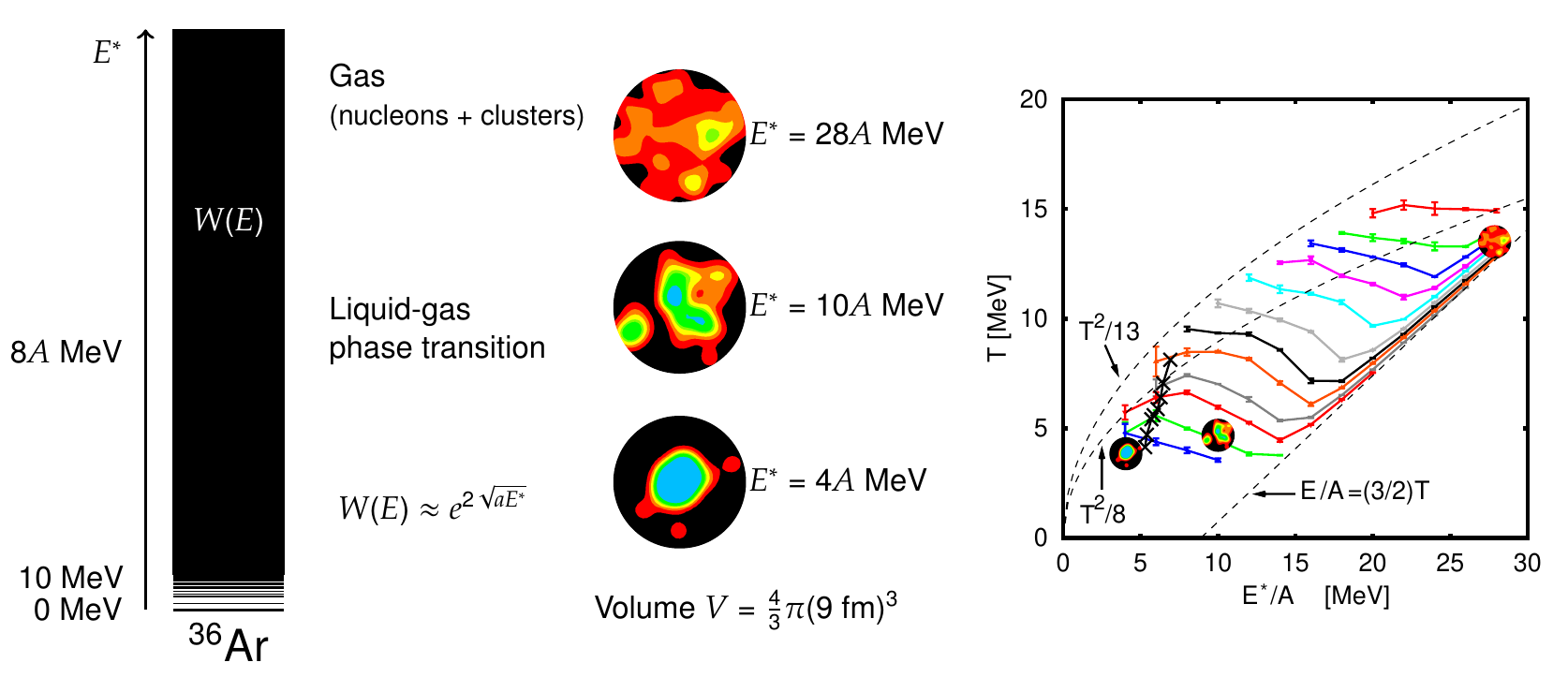}
\caption{\label{fig:exnucleus}
Excited states of a many-nucleon system with $N = Z = 18$ confined in a virtual spherical container with a radius of 9 fm. The density distribution, in the middle part, at each excitation energy ($E^*/A = 4$, 10, and 28 MeV) shows a snapshot taken from the AMD equilibrium calculation of Ref.~\cite{furuta2009}.  The right figure shows the constant pressure caloric curves calculated with AMD for the $A = 36$ system \cite{furuta2009}.  The points corresponding to the conditions of $E^*/A=4$, 10 and 28 MeV with the container radius of 9 fm are indicated by small images of density distributions.  The cross symbols indicate the equilibrium states corresponding to the reaction system from $t = 80$ to 300 fm/$c$ in $\nuc[40]{Ca} + \nuc[40]{Ca}$ central collisions at 35 MeV/nucleon.
}
\end{figure}

We can continue this exploration toward high excitation energies by confining the system in a virtual container of a volume $V$ which should be larger than the volume of the nucleus (Fig.~\ref{fig:exnucleus}).  This may be a similar situation to heavy-ion collisions in which most particles are distributed and interacting within a finite volume at a finite reaction time.  As mentioned in previous sections, in heavy-ion collisions at the incident energies from several tens of MeV/nucleon to several hundred MeV/nucleon, the excitation energy measured from the ground state nucleus is similar to or higher than the binding energy of a nucleus, $E^*/A\gtrsim 8$ MeV, which is of course much higher than the particle-decay thresholds.  As discussed in Sec.~\ref{sec:exp}, the experimental data of heavy-ion collisions suggest that light clusters and heavier nuclei still exist in such highly excited states.  Only a fraction of the protons in the system exist as unbound particles.  It seems there is a huge region of the excitation energy in which cluster correlations play essential roles in low-density systems.  The middle part of Fig.~\ref{fig:exnucleus} shows snapshots of the density distribution taken from microcanonical statistical ensembles at three different excitation energies for a system with 36 nucleons in a container of radius 9 fm.  The ensembles were generated by solving the time evolution for a long time with the antisymmetrized molecular dynamics (AMD) with wave-packet splitting (see Sec.~\ref{sec:amd-wpsplit}) \cite{furuta2006,furuta2009}.  At $E^*/A=4$ MeV, there is a single nucleus surrounded by a few nucleons in typical states, which may be regarded as a liquid droplet surrounded by a gas.  At $E^*/A=28$ MeV, nucleons are rapidly moving around but clusters are still found, so that it is like a gas composed of nucleons and clusters.  At an intermediate excitation energy $E^*/A=10$ MeV, we often find several nuclei, which may be regarded as a mixture of liquid and gas.

Appearance of clusters and fragments can be naturally understood as a consequence of nuclear saturation property, i.e., the fact that it does not cost much energy to separate a nucleus into two or more nuclei, while it costs much energy to break a nucleus into free nucleons.  At the same time, however, at high excitation energies in particular, we also need to consider the entropy or the phase space of the translational motions of clusters and nucleons, which favors a larger number of constituents.  The equilibrium composition of clusters and fragments may be roughly understood as the competition of these effects.

One of the aims of the study of multifragmentation was to identify the nuclear liquid-gas phase transition which is expected from the similarity of the nuclear EOS to the van der Waals EOS.  The concept of phase transition in finite many-body systems, as in heavy-ion collisions, has been advanced \cite{gross1997,gross2002,chomaz2000}.  Phase transition is clearly defined in finite systems by considering microcanonical ensembles as a backbending of caloric curves or a negative heat capacity, i.e., $dT/dE < 0$ for the temperature $T(E)$ as a function of the energy of the system.  Statistical models for multifragmentation can be employed to calculate caloric curves for a finite system.  Microcanonical caloric curves at constant pressure in Ref.~\cite{raduta2001} actually shows backbending if the effect of Coulomb interaction is weak.  Different versions of molecular dynamics models were also employed to study caloric curves and phase transition \cite{ohnishi1993,ono1996mflct,ono1996nazotoki,schnack1997,sugawa1999,sugawa2001,furuta2006,furuta2009}.  Constant-pressure caloric curves have been obtained with AMD in Refs.~\cite{furuta2006,furuta2009}, where the phase transition is clearly identified in the caloric curves as shown in the right part of Fig.~\ref{fig:exnucleus}.  Another signal of the first-order liquid-gas phase transition is the presence of abnormally large kinetic energy fluctuations which corresponds to the presence of negative heat capacity \cite{chomaz2000}.  Representative experimental data on caloric curves and heat capacity are found in Refs.~\cite{pochodzalla1995,natowitz2002,dagostino2004,borderie2013}.  See also Ref.~\cite{borderie2008} for a review.

It is a fundamental question whether thermal and chemical equilibrium is realized in heavy-ion collisions, i.e., whether all the microscopic states in the volume $V$ and with a given energy $E^*$ are equally realized without biases.  It is rather easy to derive a negative answer because, e.g., just before particles stop interacting in heavy-ion collisions, there is a collective expansion so that the states in which some particles are moving inwards are rarely realized.  However, it is empirically known that most degrees of freedom, except for such collective ones, can be explained by equilibrium.  Many fragmentation data have been reproduced reasonably well by statistical models for multifragmentation \cite{gross1990,bondorf1995,dagostino1996,raduta2002,botvina2010} which assume equilibrium of nuclei under an assumed volume and an excitation energy before they suddenly stop interacting to be emitted.  If Ref.~\cite{furuta2009}, the same AMD model was applied both to the equilibrium system confined in a container and to a heavy-ion collision system.  The calculations show that there exists an equilibrium ensemble which well reproduces the reaction ensemble at each reaction time $t$ for the investigated period $80 \le t \le 300$ fm/$c$ in $\nuc[40]{Ca} + \nuc[40]{Ca}$ central collisions at 35 MeV/nucleon, as far as fragment observables, such as fragment yields and excitation energies, are concerned.  In Fig.~\ref{fig:exnucleus}, the path of the collision is shown in the $E^*$-$T$ plane by cross symbols.  In Refs.~\cite{ono2003,ono2004}, it was shown that the fragment yields in the AMD calculation for the $\nuc{Ca}+\nuc{Ca}$ systems satisfy the isoscaling relation \cite{xu2000}, which also suggests relevance of equilibrium and the low-density symmetry energy in the isotopic yields of fragments.

The system confined in a volume $V$ may be regarded as a piece of infinite nuclear matter, as in compact stars such as neutron stars and core-collapse supernovae, specified by a density $\rho=A/V<\rho_0$.  Although leptons do not take part in the equilibrium in heavy-ion collisions and in isolated nuclei, physics of hadronic degrees of freedom should be common.  In general, heavy-ion collisions are supposed to be an important opportunity to study nuclear EOS at various densities in terrestrial experiments.  Here we focus on a region at low densities $\rho<\rho_0$ and at finite temperatures which has a close link to fragmenting systems in heavy-ion collisions.  Equations of state which cover this domain have been developed and improved by many researchers \cite{lattimer1991, shen1998, ishizuka2003, botvina2010,buyukcizmeci2014, raduta2010, typel2010, sumiyoshi2008, furusawa2013, hempel2010, shen2011horowitz, togashi2017}.  See Refs.~\cite{oertel2017,fischer2014} for recent reviews.  Some of them have been developed based on statistical models for multifragmentation which are also applied to heavy-ion collisions.  Different approximations have been employed to handle the existence and interaction of light clusters and heavier nuclei in low-density matter. A comparison of different models can be found in Ref.~\cite{buyukcizmeci2013} with detailed analyses of nuclear composition.

\subsection{\label{sec:clusterinmed}Clusters and correlations in medium}

The concept of cluster is most safely understood as a kind of spatial many-nucleon correlations in a nuclear many-body system.  When we treat a full many-body wave function as in microscopic nuclear structure calculations, we do not need to ask in principle what is meant by cluster.  On the other hand, in statistical calculations for nuclear EOS and for nuclear fragmentation, one usually treats clusters and nuclei as different species of particles for which thermal and chemical equilibrium is considered.  This is called nuclear statistical ensemble.  This is probably a good picture if the particles are separated from one anther most of the time except that collisions and/or chemical reactions occur between particles from time to time to achieve equilibrium.  However, when the density is high so that particles frequently overlap, we cannot forget the fact that clusters and nuclei are composed of nucleons.

The compositeness and interaction of clusters and nuclei are partly taken into account in statistical models in many different ways.  In the heuristic excluded-volume approach, the volumes of the finite-size nuclei are subtracted from the total volume of the system to obtain the available volume for the translational motions.  The excited states of each nucleus are usually taken into account in the form of the explicit nuclear levels or a simplified function of the level density $D(\epsilon)$ (or the free energy) for the internal degrees of freedom of a nucleus.  In statistical models for multifragmentation, a formula such as a Fermi-gas formula $D(\epsilon)\sim e^{2\sqrt{a\epsilon}}$ with a level density parameter $a$ is usually employed without setting an upper limit of the excitation energy.  In some models, however, a cut-off is introduced in the level density (see Ref.~\cite{raduta2010} for example).  It sounds reasonable to take into account the bound excited states and long-lived resonances of a nucleus in nuclear statistical ensemble.
The other excited states counted by a Fermi-gas formula for a nucleus will quickly decay by emitting one or more particles, so they may be more suitably described as scattering continuum states of two or more particles.  The long-lived resonances can also be treated in the scattering states, in principle.  Then it is required to treat inter-particle interactions and correlations in nuclear statistical ensemble.  In some EOS calculations \cite{horowitz2006,shen2011horowitz}, the interactions between particles are taken into account in the form of the virial expansion which is valid only in the low-density limit.

\begin{figure}
\centering
\includegraphics[scale=0.5]{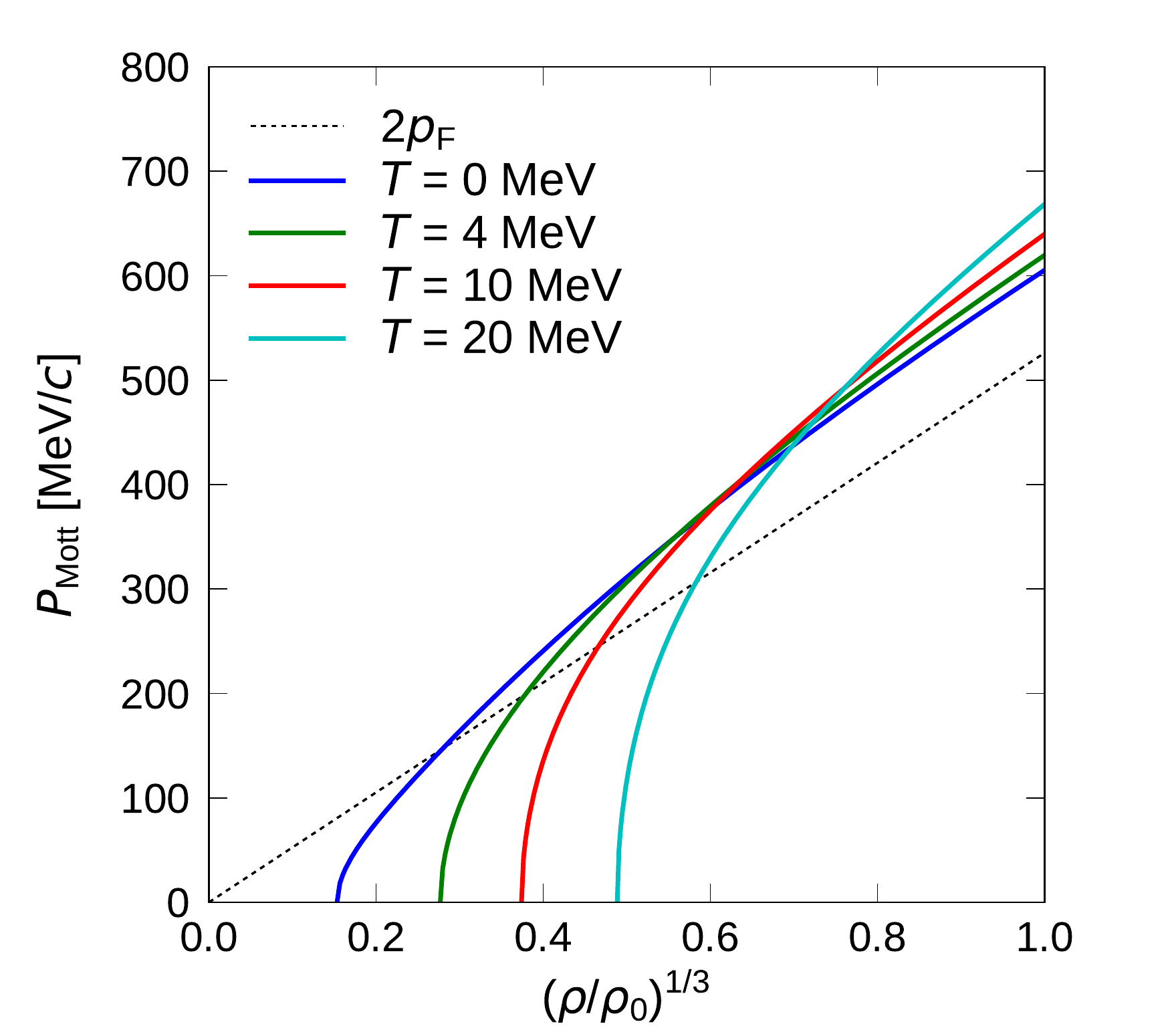}
\caption{\label{fig:roepke2011}
The Mott momentum $P_{\text{Mott}}$ for a deuteron is shown as a function of the cubic root of the density, $(\rho/\rho_0)^{1/3}$ with $\rho_0=0.16\ \text{fm}^{-3}$, for different temperatures $T$ of the symmetric nuclear matter.  A deuteron bound state exists in the region $P>P_{\text{Mott}}$ of the deuteron momentum $P$.  The solid curves show the parametrized formula by R\"opke \cite{roepke2011} that fits the solution of Eq.~(\ref{eq:deuteron-inmed}).  The dotted line shows twice the Fermi momentum at $T=0$.
}
\end{figure}

A more fundamental framework has been developed by R\"opke and his collaborators based on the quantum statistics description with thermodynamic Green's functions.  See Refs.~\cite{roepke2009,roepke2015} and references therein.  Using the cluster decomposition of the single-nucleon self-energy, the total neutron and proton densities, which are functions of the temperature and the chemical potentials, are expressed as sums of contributions of different quasiparticles specified by $NZ\nu$,
\begin{equation}\label{eq:qs}
\begin{split}
\rho_n(T,\mu_n,\mu_p)&=\int\frac{d\bm{P}}{(2\pi\hbar)^3}
\sum_{NZ\nu}N\frac{1}{\exp[E_{NZ\nu}(\bm{P},T,\mu_n,\mu_p)-N\mu_n-Z\mu_p]-(-1)^{N+Z}}
\\
\rho_p(T,\mu_n,\mu_p)&=\int\frac{d\bm{P}}{(2\pi\hbar)^3}
\sum_{NZ\nu}Z\frac{1}{\exp[E_{NZ\nu}(\bm{P},T,\mu_n,\mu_p)-N\mu_n-Z\mu_p]-(-1)^{N+Z}}
\end{split}
\end{equation}
in the case of infinite matter.  The index $\nu$ is for internal (discrete) states of a quasiparticle composed of $N$ neutrons and $Z$ protons.  The quasiparticle energy $E_{NZ\nu}(\bm{P},T,\mu_n,\mu_p)$, which can depend on the momentum $\bm{P}$ of the quasiparticle, may be calculated from the in-medium Schr\"odinger equation.  In the case of the deuteron, it is \cite{danielewicz1991}
\begin{multline}\label{eq:deuteron-inmed}
[E_n(\tfrac12\bm{P}+\bm{p})+E_p(\tfrac12\bm{P}-\bm{p})]\psi_{\bm{P}}(\bm{p})
\\
+[1-f_n(\tfrac12\bm{P}+\bm{p})-f_p(\tfrac12\bm{P}-\bm{p})]
\int\frac{d\bm{p}'}{(2\pi\hbar)^3}\langle\bm{p}|v|\bm{p}'\rangle\psi_{\bm{P}}(\bm{p}')
=E_{11}(\bm{P})\psi_{\bm{P}}(\bm{p}),
\end{multline}
where the dependence on $(T,\mu_n,\mu_p)$ should be understood implicitly.  This is similar to the usual Schr\"odinger equation for a deuteron in the momentum representation except that the nucleon kinetic energies are replaced by the nucleon quasiparticle energies $E_n(\bm{p}_1)$ and $E_p(\bm{p}_2)$, and that a factor $1-f_n(\bm{p}_1)-f_p(\bm{p}_2)$ takes account of the Pauli principle in the interaction term.  Thus the quasiparticle properties are modified by medium effects.  It should be noted that the effects depend also on the momentum $\bm{P}$ of the quasiparticle.  Above a certain density, the deuteron bound state exists only at high momenta $|\bm{P}|>P_{\text{Mott}}$, where the lower bound is called the Mott momentum.  Figure \ref{fig:roepke2011} shows the Mott momentum for a deuteron as a function of $(\rho/\rho_0)^{1/3}$ for various temperatures $T$, using a parametrized formula given by R\"opke \cite{roepke2011} to fit the solution of Eq.~(\ref{eq:deuteron-inmed}).  At low densities, the bound state exists at all values of $|\bm{P}|$, but at higher densities a deuteron cannot exist as a bound state if its momentum is placed inside or close to the Fermi sphere $|\bm{P}|\lesssim 2p_{\text{F}}$.  A similar calculation by Danielewicz et al.~\cite{danielewicz1991} showed that there is another branch of solutions near $\bm{P}\approx 0$ at $T=0$, corrsponding to a Cooper pair.  The states with $|\bm{P}|\gtrsim 2p_{\text{F}}$ may be similar to the usual deuteron with some modifications by the Pauli principle.  Similar considerations and calculations should be extended to heavier quasiparticles in principle.  In practical calculations by Typel et al.~\cite{typel2010}, the clusters with $A\le 4$ are taken into account as explicit particle degrees of freedom with medium modifications and heavier nuclei are treated in a generalized relativistic mean-field approach.

When the bound state of a cluster has disappeared by the medium effects, the correlation may still exist in the scattering continuum states.  As also mentioned above, the long-lived resonances in the excited states of a nucleus should contribute almost as bound states in the equilibrium.  A transparent picture is given by the generalized Beth-Uhlenbeck approach \cite{schmidt1990roepke,voskresenskaya2012typel,roepke2013,roepke2015} which handles bound and scattering states consistently.  For the contribution of the deuteron channel in Eq.~(\ref{eq:qs}), the bound and continuum states are more explicitly shown by  \begin{equation}
\sum_{\nu}(\cdots)\quad\rightarrow\quad
\int d\epsilon\ D(\epsilon)(\cdots),
\end{equation}
where the difference of the density of states from that of the non-interacting system is written for the deuteron channel as
\begin{equation}
\label{eq:den-of-states-deuteron}
D(\epsilon)=3\delta(\epsilon-E_{d})
+3\sum\frac{1}{\pi} 2\sin^2\delta(\epsilon)\frac{d\delta(\epsilon)}{d\epsilon},
\end{equation}
where the dependence on $\bm{P}$ and $(T,\mu_n,\mu_p)$ must be implicitly understood for $D(\epsilon)$, the bound state energy $E_d$ and the scattering phase shift $\delta(\epsilon)$.  The delta-functional bound-state contribution $3\delta(\epsilon-E_d)$ does not exist when the deuteron is not bound under the condition.  The factor $2\sin^2\delta(\epsilon)$, which does not exist in the standard Beth-Uhlenbeck formula, is to take into account that part of interaction effects is already contained in the quasiparticle energies.
Similar approach can be taken for two-nucleon correlations in the isospin-triplet channel with no bound state, which may be treated as a temperature-dependent effective resonance as in Ref.~\cite{voskresenskaya2012typel}.  Evidently more considerations are necessary to handle heavier clusters and nuclei with many possible breakup channels.  See Ref.~\cite{roepke2015} for more discussions.

Although the realized state is a mixture of various kinds of quasiparticles, i.e., nucleons, clusters and possibly heavier nuclei, the occupation probabilities $f_n(\bm{p}_1)$ and $f_p(\bm{p}_2)$ in the in-medium Schr\"odinger equation (\ref{eq:deuteron-inmed}) are for uncorrelated nucleons.  The same is true for the nucleon quasiparticle energies $E_n(\bm{p}_1)$ and $E_p(\bm{p}_2)$.  A part of correlations may be treated as an effective change of the temperature parameter in the occupation probabilities as adopted in Ref.~\cite{roepke2015}.  However, spatial correlations does not seem taken into account in these approaches.  For example, when many $\alpha$ clusters have been formed in moderately low-density system, there are more open spaces, that are not occupied by the nucleons in $\alpha$ clusters, compared to the case in which the same number of nucleons are distributed randomly or uniformly in the space.  If there are spatial or phase-space correlations in this way, it may be possible that a deuteron is formed in an open space because it is not blocked by the Pauli principle even though the total density of the system is not very low.

\begin{figure}
\centering
\includegraphics{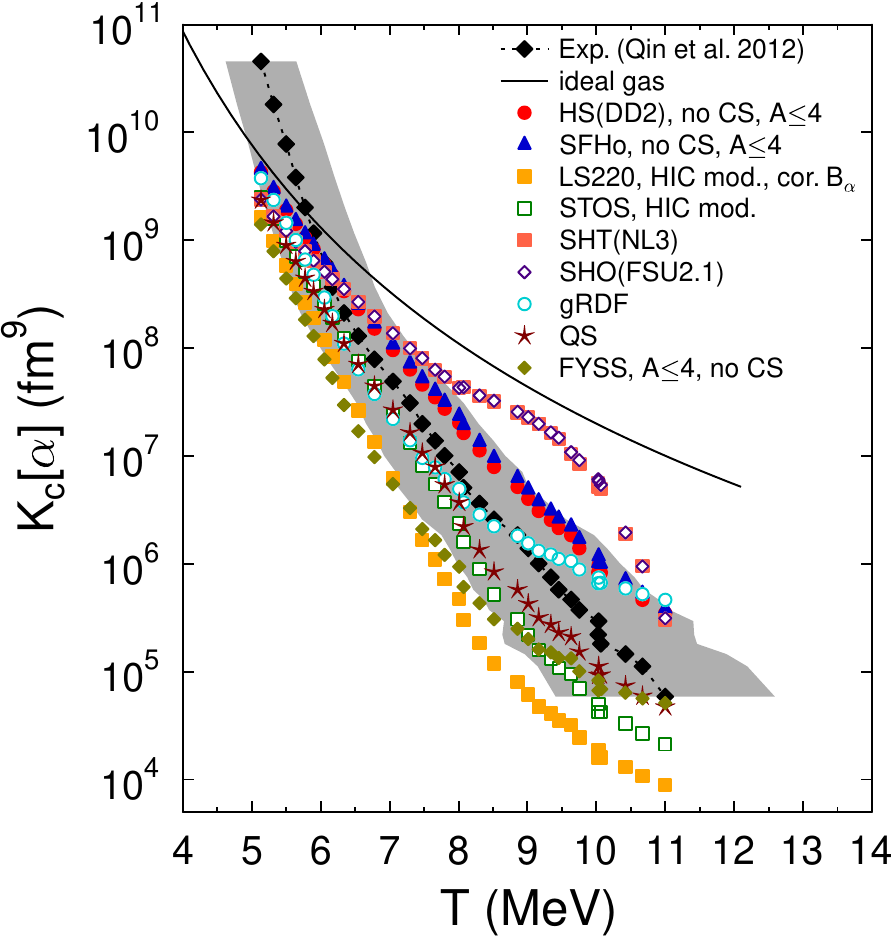}
\caption{\label{fig:hempel2015} Equilibrium constants for $\alpha$ particles extracted from heavy-ion experiments \cite{qin2012texas} (black diamonds) in comparison with those of various theoretical models, which are all adapted for the conditions in heavy-ion collisions, as far as possible \cite{hempel2015}.  The gray band is the experimental uncertainty in the temperature determination. The black line shows the equilibrium constant of the ideal gas model.  Taken from Ref.~\cite{oertel2017}.  }
\end{figure}

The cluster compositions of different EOS models were compared in Ref.~\cite{hempel2015} for the chemical equilibrium constants defined by
\begin{equation}
K_{\text{c}}[i] = \frac{\rho(N_i,Z_i)}{\rho(1,0)^{N_i}\rho(0,1)^{Z_i}}
\end{equation}
for the cluster species $i$, where $\rho(N,Z)$ is the number density of the nucleus $(N,Z)$.  The equilibrium constants may depend on the temperatures and the chemical potentials in general but it depends only on the temperature for the statistical ensemble of ideal Boltzmann gas.  Depending on the model, the considered species of clusters and nuclei are limited, which affects the cluster abundances.  In equilibrium constants, the differences originating from this reason are reduced, and therefore the comparison allows us to study the effects of other ingredients in different models.  For $\alpha$ clusters, the equilibrium constants for the calculated cluster abundances are shown in Fig.~\ref{fig:hempel2015} as functions of the temperature $T$ for the proton fraction $Y_p=0.41$.  The baryon number density is related to $T$ by the path of evolution in heavy-ion collisions shown in Fig.~\ref{fig:qin2012texas}.  All the calculated results approach to the ideal gas relation shown by a thin curve in the low-temperature (and therefore low-density) limit.  At higher temperatures (and therefore higher densities), the cluster abundance is reduced compared to what is expected for the ideal gas of clusters, which is in general consistent with the information from heavy-ion collisions shown by the black diamonds with the gray band of uncertainties.  By the comparison among the calculations and with the experimental data, it seems that important ingredients are the consideration of all relevant particle species, mean-field effects of the unbound nucleons, and a suppression mechanism for bound clusters at high densities.  It may be desirable to better understand the low-temperature and low-density behavior of the experimental data in relation to the emission mechanisms of clusters in heavy-ion collisions.

The existence of clusters is not a small effect.  It can strongly influence the EOS and the symmetry energy at low densities \cite{kowalski2007,natowitz2010,wada2012}.

\section{\label{sec:basicmodels}Basic transport models}

Dynamics of heavy-ion collisions is a complicated many-body problem.  Furthermore, many kinds of quantum mechanical aspects are expected to play important roles in particular for the formation of light clusters and heavier fragments.  A light cluster typically has only a single bound state which cannot be emulated by classical motions of constituent nucleons.  For a heavier nucleus, the fermionic nature of nucleons is important not only for the ground state but also for excited states.  Because of the density of states of a degenerate fermionic system, the excitation energy of a nucleus is related to the temperature by $E^*\approx aT^2$ with a level density parameter $a$, which essentially determines the characteristics of equilibrium.  To understand the evolution from the early phase of compression to the late phase of collective expansion and fragment formation, we need to solve the time evolution of a quantum system in some way.  We should prepare for the emergence of so many reaction channels, each of which corresponds to a configuration of decomposing the whole system into fragments.

Even if we may know the initial many-body state $|\Psi(t=0)\rangle$ completely, we can solve the many-body Shr\"odinger equation only approximately at best, to obtain an approximate solution $|\Psi_{\text{approx}}(t)\rangle$.  Even if approximations introduced in the equation of the time evolution appear to be reasonable, the small errors will accumulate during solving the equation for a long time from $t=0$ to e.g.\ $t_{\text{end}}=100$-300 fm/$c$, so that the final solution will be far from the exact one $|\langle\Psi_{\text{exact}}(t_{\text{end}})|\Psi_{\text{approx}}(t_{\text{end}})\rangle|^2 \ll 1$.  However, we still expect that the approximate solution may be able to reproduce important properties of the exact solution.  How is it possible?  One of the answers is to ensure the consistency of the equilibrium properties associated with the approximate time evolution with the exact ones.

Traditionally several types of transport models have been proposed and practically applied to heavy-ion collisions.  These calculations are actually important and useful to understand collision dynamics and to extract physical information from heavy-ion collision data such the nuclear EOS at various densities.  We will review these transport models below.  We will not try to rigorously derive or justify these models here but will put more emphasis on their achievements, advantages and disadvantages, which is useful in improving models and also in understanding important physics in heavy-ion collisions.  Although most of practical transport models are classical or semiclassical, the link with quantum mechanics will be still essential.

\subsection{Reduction to single-particle dynamics}

Practically all the transport models try to solve the time evolution of the one-body density operator $\hat{\rho}$ (or its Wigner transform) for an $A$-nucleon system which is defined
as
\begin{equation}
\hat{\rho} = A\mathop{\mathrm{Tr}}_{2,3,\ldots,A}|\Psi\rangle\langle\Psi|
\end{equation}
by taking the trace of the many-body density operator $|\Psi\rangle\langle\Psi|$ with respect to all the particles excluding one.  We assume the state $|\Psi\rangle$ is normalized.  Knowing the one-body density operator $\hat{\rho}(t)$ at a time $t$ allows us to calculate the expectation value of any one-body observable $O=\sum_{i=1}^Ao_i$ as $\langle\Psi(t)|O|\Psi(t)\rangle=\mathop{\mathrm{Tr}}[o\hat{\rho}(t)]$.  The aim is to have a closed equation of the time evolution for $\hat{\rho}(t)$.  Generally speaking, however, one should not assume that it is possible.  If there are two-nucleon interactions $\sum_{i<j}v_{ij}$, the equation for $\frac{d}{dt}\hat{\rho}$ can be written as
\begin{equation}
\label{eq:drhodt-general}
i\hbar\frac{d}{dt}\hat{\rho}=
[\frac{1}{2M}\bm{p}^2,\ \hat{\rho}]
+\mathop{\mathrm{Tr}}_2[v,\ \hat{\rho}^{(2)}]
\end{equation}
by using the two-body density operator which is defined by
\begin{equation}
\hat{\rho}^{(2)} = A(A-1)\mathop{\mathrm{Tr}}_{3,\ldots,A}|\Psi\rangle\langle\Psi|,
\end{equation}
and the equation for $\frac{d}{dt}\hat{\rho}^{(2)}$ contains the three-body 
density operator, and so on (see e.g.\ Ref.~\cite{wang1985}).  The equations will not be closed until the full many-body state $|\Psi(t)\rangle$ is included.

The time-dependent Hartree-Fock (TDHF) theory assumes a single Slater determinant as the many-body state $|\Psi\rangle$.  An idempotent one-body density matrix that satisfies $\hat{\rho}^2=\hat{\rho}$ with $\mathop{\mathrm{Tr}}\hat{\rho}=A$ uniquely corresponds to a many-body state which is a Slater determinant, and therefore the two-body density operator is written with $\hat{\rho}$ as
\begin{equation}
\label{eq:rho2-hf}
\hat{\rho}^{(2)}_{12}=\hat{A}_{12}\hat{\rho}_1\hat{\rho}_2,
\end{equation}
where $\hat{A}_{12}$ is the antisymmetrization operator acting on a two-body state.  The subscript 1 or 2 attached to a one- or two-body operator indicates on which particle it operates in a two-body state.  More explicitly, $\hat{\rho}^{(2)}(|\psi\rangle\otimes|\phi\rangle) = \hat{\rho}|\psi\rangle\otimes\hat{\rho}|\phi\rangle - \hat{\rho}|\phi\rangle\otimes\hat{\rho}|\psi\rangle$.
Then Eq.~(\ref{eq:drhodt-general}) is rewritten as
\begin{equation}
\label{eq:drhodt-tdhf}
i\hbar\frac{d}{dt}\hat{\rho}=
\Bigl[\frac{1}{2M}\bm{p}^2+U[\hat{\rho}],\ \hat{\rho}\Bigr]
\end{equation}
with the one-body mean-field potential which is defined depending on $\hat{\rho}$ as
\begin{equation}
\label{eq:meanfield-of-rho}
U_1[\hat{\rho}]=\mathop{\mathrm{Tr}}_2\hat{A}_{12}v_{12}\hat{\rho}_2,
\end{equation}
so that the equation for $\hat{\rho}$ is now closed.  This TDHF equation  (\ref{eq:drhodt-tdhf}) conserves the idempotency $\hat{\rho}^2=\hat{\rho}$ as well as the normalization $\mathop{\mathrm{Tr}}\hat{\rho}=A$.

The TDHF equation for $\hat{\rho}$ may be more general than the assumption of a Slater determinant because the assumption of the non-correlated two-body density operator by Eq.~(\ref{eq:rho2-hf}) can be introduced for a general $\hat{\rho}$ that does not correspond to a Slater determinant, i.e., in the case of $\hat{\rho}^2\ne\hat{\rho}$.  However, a benefit of restricting to the case of $\hat{\rho}^2=\hat{\rho}$ is that we then have a uniquely corresponding many-body wave function which is easily constructed from $\hat{\rho}$.  We will come back to this point later.

\begin{figure}
\centering
\includegraphics{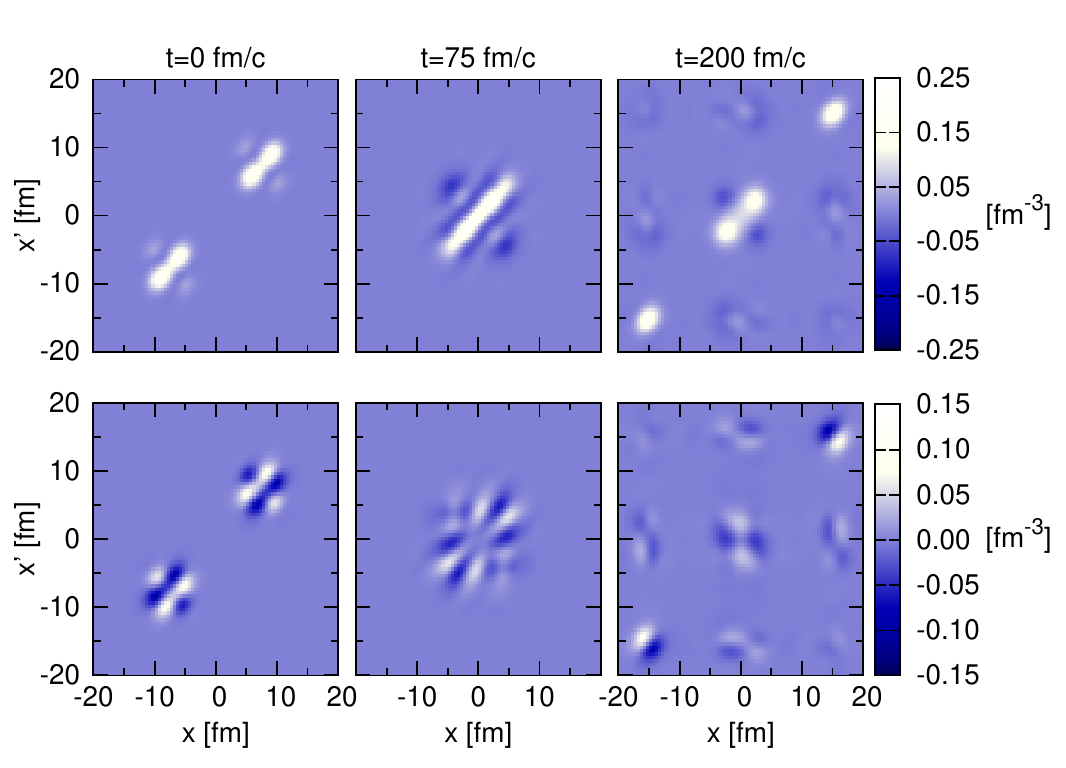}
\caption{\label{fig:rios2011}
Intensity plots representing the real (upper panels) and the imaginary part (lower panels) of the density matrix $\langle x|\hat{\rho}(t)|x'\rangle$ for a collision of two slabs at $E_{\text{cm}}/A=4$ MeV in one dimension.  A scaling has been applied so that the values along the diagonal coincide with the 3D density.
Taken from Ref.~\cite{rios2011}.
}
\end{figure}

In TDHF, the density matrix can be always diagonalized as $\hat{\rho}(t)=\sum_{i=1}^{A}|\psi_i(t)\rangle\langle\psi_i(t)|$, and therefore one only needs to consider the time evolution of the single-particle states $|\psi_i(t)\rangle$ ($i=1,2,\ldots,A$).  However, for further improvements and extensions of the approach, it is useful to directly handle the matrix elements of $\hat{\rho}(t)$, or more generally Green's functions.  In the coordinate representation, a diagonal element $\rho(\bm{r})=\langle\bm{r}|\hat{\rho}|\bm{r}\rangle$ is the density.  Rios et al.\ made a study in Ref.~\cite{rios2011} on the off-diagonal elements in collisions of two slabs in one dimension.  Figure \ref{fig:rios2011} represents the time evolution of the matrix elements $\langle x|\hat{\rho}(t)|x'\rangle$ in a collision where three fragments are produced.  It is seen at $t=200$ fm/$c$ that the three fragments are located at $x\approx -15$, 0, and 15 fm.  Even though fragments are well separated in space, nonzero off-diagonal elements $\langle x|\hat{\rho}|x'\rangle$ exist for $x$ and $x'$ in different fragments.  This means that the states of different fragments are correlated and entangled in the TDHF solution.  However, the details of the entanglement most likely contains uncontrollable errors due to the limitation of TDHF, i.e., many-body correlations will affect the entanglement in reality.  Do off-diagonal elements, if any, influence anything we are interested in?  In Ref.~\cite{rios2011}, it is demonstrated that the off-diagonal elements with large $|x-x'|$ are not important, i.e., a suitable suppression of these elements does not significantly affect the evolution the elements close to the diagonal.  This is an example to show how decoherence can be introduced to the fully coherent single-particle dynamics that is likely an artifact of the model assumption.

For an intuitive understanding as in classical mechanics, the Wigner transform of the one-body density operator is introduced by
\begin{equation}
f(\bm{r},\bm{p})=\int e^{-(i/\hbar)\bm{p}\cdot\bm{s}}\langle\bm{r}+\tfrac12\bm{s}|\hat{\rho}|\bm{r}-\tfrac12\bm{s}\rangle d\bm{s},
\end{equation}
with which the normalization and the expectation value of any one-body operator $O=\sum_{i=1}^A o_i$ are expressed as
\begin{equation}
\label{eq:f-normalization}
\iint f(\bm{r},\bm{p})\frac{d\bm{r}d\bm{p}}{(2\pi\hbar)^3}=A,
\qquad
\langle O\rangle=
\iint o(\bm{r},\bm{p})f(\bm{r},\bm{p})\frac{d\bm{r}d\bm{p}}{(2\pi\hbar)^3},
\end{equation}
which implies that $f(\bm{r},\bm{p})$ can be interpreted as a phase-space distribution function.  The Wigner transform of the one-body operator has been introduced by
\begin{equation}
o(\bm{r},\bm{p})=\int e^{-(i/\hbar)\bm{p}\cdot\bm{s}}\langle\bm{r}+\tfrac12\bm{s}|o|\bm{r}-\tfrac12\bm{s}\rangle d\bm{s}.
\end{equation}
Wigner transforms, as well as $f(\bm{r},\bm{p})$, are still matrices in the spin and isospin space, which must be understood implicitly in equations.  For example, in Eq.~(\ref{eq:f-normalization}), the trace should be taken for the spin and isospin space.  For the product of two one-body operators $C=\sum_{i=1}^Ac_i=\sum_{i=1}^Aa_ib_i$, the Wigner transform is not a simple product of the Wigner transforms of individual operators, but it includes higher order terms of $\hbar$ as
\begin{equation}
c(\bm{r},\bm{p})
=a(\bm{r},\bm{p})\exp(\tfrac12i\hbar\Lambda)b(\bm{r},\bm{p})
\end{equation}
with
\begin{equation}
\Lambda = \frac{\overleftarrow\partial}{\partial\bm{r}}
\cdot\frac{\overrightarrow\partial}{\partial\bm{p}}
-\frac{\overleftarrow\partial}{\partial\bm{p}}
\cdot\frac{\overrightarrow\partial}{\partial\bm{r}},
\end{equation}
where the direction of an arrow represents whether the differential is operated on the function located on the left or the right of the operator.
By performing the Wigner transform of the both sides of the TDHF equation  (\ref{eq:drhodt-tdhf}) and ignoring the $O(\hbar^3)$ terms, we have the Vlasov equation
\begin{equation}
\label{eq:vlasov}
\frac{\partial f(\bm{r},\bm{p},t)}{\partial t}
+\biggl(\frac{\bm{p}}{M}+\frac{\partial U[f]}{\partial\bm{p}}\biggr)
\cdot\frac{\partial f}{\partial\bm{r}}
-\frac{\partial U[f]}{\partial\bm{r}}
\cdot\frac{\partial f}{\partial\bm{p}}
=0,
\end{equation}
where $U[f](\bm{r},\bm{p})$ is the Wigner transform of the mean-field potential 
 (\ref{eq:meanfield-of-rho}).  The most intuitive way to interpret this equation may be to represent $f(\bm{r},\bm{p},t)$ by many test particles as
\begin{equation}
\label{eq:testp-delta}
f(\bm{r},\bm{p},t)=\frac{1}{N_{\text{tp}}}\sum_{k=1}^{N_{\text{tp}}A}
(2\pi\hbar)^3\delta(\bm{r}-\bm{r}_k(t))\delta(\bm{p}-\bm{p}_k(t))
\end{equation}
with a sufficiently large $N_{\text{tp}}$ which represents the number of test particles per nucleon.  Then the Vlasov equation is equivalent to the classical motions of test particles in the mean field \cite{wong1982},
\begin{equation}
\label{eq:testp-eq-of-motion}
\frac{d}{dt}\bm{r}_k(t)
 = \frac{\bm{p}_k}{M}+\frac{\partial U[f](\bm{r}_k,\bm{p}_k)}{\partial\bm{p}_k}
,\qquad
\frac{d}{dt}\bm{p}_k(t)
=-\frac{\partial U[f](\bm{r}_k,\bm{p}_k)}{\partial\bm{r}_k}.
\end{equation}
This is also a practical numerical method to solve transport equations \cite{bertsch1988}.

The Slater-determinant condition  is
\begin{equation}
\hat{\rho}^2=\hat{\rho}\qquad\Leftrightarrow\qquad
f(\bm{r},\bm{p})\cos(\tfrac12\hbar\Lambda)f(\bm{r},\bm{p})=f(\bm{r},\bm{p})
\end{equation}
in the phase-space distribution function.  In the classical limit of $\hbar\rightarrow0$, it implies that $f$ takes only 1 or 0 at each phase-space point.  However, at the same time, $f$ should be smoothly connected, in the scale of $\hbar$, on the boundary between $f=1$ and $f=0$ regions in order that the higher order terms in $\hbar$ are negligible.  If there is a phase-space region in which $f$ takes a constant value but $f\ne1$ and $f\ne0$, this distribution function does not correspond to any Slater determinant.

\subsection{BUU models}

The Boltzmann-Uehling-Uhlenbeck (BUU) theory can be derived from the Born-Bogoliubov-Green-Kirkwood-Yvon (BBGKY) hierarchy, or by using the real-time Green's-function formalism to arrive at the Kadanoff-Baym equations \cite{danielewicz1984,botermans1990,buss2012}.  With the quasiparticle approximation, an approximation for the self-energy, and a semiclassical approximation, one arrives at the BUU equation for the one-body Wigner distribution function.  A necessary approximation is the separation of macroscopic and microscopic scales, i.e., the spatial and time scales of global collision dynamics are assumed to be large and slowly-changing compared to those of short-range interaction between two nucleons.  Then two nucleons collide locally and instantaneously in the macroscopic scale, as an energy-conserving scattering of two nucleons of definite incoming momenta.  There are similar but different ways of arriving at similar conclusions, e.g.\ Refs.~\cite{reinhard1992stdhf,aichelin1991,kawai1992}.

In the case without momentum dependence of the mean field and with the non-relativistic kinematics, we have the BUU equation
\begin{equation}
\label{eq:buu}
\frac{\partial f(\bm{r},\bm{p},t)}{\partial t}
+\frac{\bm{p}}{M}\cdot\frac{\partial f}{\partial\bm{r}}
-\frac{\partial U[f]}{\partial\bm{r}}
\cdot\frac{\partial f}{\partial\bm{p}}
=I_{\text{coll}}[f](\bm{r},\bm{p}),
\end{equation}
with a collision term
\begin{equation}
\label{eq:icoll-buu}
I_{\text{coll}}[f](\bm{r},\bm{p})
=\bar{I}_{\text{coll}}[f](\bm{r},\bm{p})
=\int\frac{d\bm{p}_1}{(2\pi\hbar)^3}d\Omega\
v_{\text{rel}}\frac{d\sigma}{d\Omega}
[f'f_1'(1-f)(1-f_1)-ff_1(1-f')(1-f_1')].
\end{equation}
In the above integral, the abbreviations mean $f=f(\bm{r},\bm{p},t)$, $f_1=f(\bm{r},\bm{p}_1,t)$, $f'=f(\bm{r},\bm{p}',t)$ and $f_1'=f(\bm{r},\bm{p}_1',t)$, with $\bm{p}'$ and $\bm{p}'_1$ related to $\bm{p}$ and $\bm{p}_1$ by the scattering angle $\Omega$ and the energy and momentum conservation.  The relative velocity is $v_{\text{rel}}=|\bm{p}-\bm{p}_1|/M=|\bm{p}'-\bm{p}_1'|/M$.  The first and the second terms in the collision term (\ref{eq:icoll-buu}) can be interpreted as the loss and the gain terms, respectively, for the phase-space point $(\bm{r},\bm{p})$ by two-nucleon collisions with Pauli
 blocking factors of the form $(1-f)$.  The mean field $U[f]$ and the in-medium two-nucleon collision cross section $d\sigma/d\Omega$ are consistently related to the self-energy in principle.  However, in actual applications, they are usually treated as independent inputs to the calculations.  The same equation is sometimes called by other names, such as Vlasov-Uehling-Uhlenbeck (VUU), Boltzmann-Nordheim-Vlasov (BNV), and Landau-Vlasov (LV) equation.

In the case of a general momentum-dependent mean field, we should treat the integration for the relative final momentum together with the delta function for the energy conservation, in principle.  With a suitable normalization of the matrix element $\mathcal{M}$, the cross sections in the loss and gain terms are related to $\mathcal{M}$ by
\begin{gather}
\label{eq:vsigma}
v_{\text{rel}}\sigma(\bm{p}\bm{p}_1\rightarrow\bm{p}'\bm{p}_1')
=\frac{2\pi}{\hbar}
\int |\mathcal{M}_{\bm{p}\bm{p}_1,\bm{p}'\bm{p}_1'}|^2\delta(E_{\bm{p}\bm{p}_1}-E_{\bm{p}'\bm{p}_1'})p_{\text{rel}}^{\prime\,2}dp_{\text{rel}}'
=\frac{2\pi}{\hbar}
|\mathcal{M}_{\bm{p}\bm{p}_1,\bm{p}'\bm{p}_1'}|^2\frac{p_{\text{rel}}^{\prime\,2}}{v_{\text{rel}}'},
\\
v_{\text{rel}}'\sigma(\bm{p}'\bm{p}_1'\rightarrow\bm{p}\bm{p}_1)
=\frac{2\pi}{\hbar}
\int |\mathcal{M}_{\bm{p}'\bm{p}_1',\bm{p}\bm{p}_1}|^2\delta(E_{\bm{p}\bm{p}_1}-E_{\bm{p}'\bm{p}_1'})p_{\text{rel}}^{2}dp_{\text{rel}}
=\frac{2\pi}{\hbar}
|\mathcal{M}_{\bm{p}\bm{p}_1,\bm{p}'\bm{p}_1'}|^2\frac{p_{\text{rel}}^{2}}{v_{\text{rel}}},
\end{gather}
where the time-reversal symmetry of $\mathcal{M}$ has been employed and the relative velocities are $v_{\text{rel}}=dE_{\bm{p}\bm{p}_1}/dp_{\text{rel}}$ and $v_{\text{rel}}'=dE_{\bm{p}'\bm{p}_1'}/dp_{\text{rel}}'$.  It is, however, not very clear from the literature how the energy conservation and the above condition for the detailed balance are taken into account in individual BUU codes.  In an IBUU04 calculation \cite{guo2014ibuu04}, a parametrization for the in-medium cross section $\sigma/\sigma_{\text{free}}=(\mu^*/\frac12M)^2$ was adopted, which corresponds to using the matrix element $|\mathcal{M}|^2$ in the free space with an assumption of $p_{\text{rel}}/v_{\text{rel}}=p_{\text{rel}}'/v_{\text{rel}}'$ ($=\mu^*$).

In most of the BUU calculations currently applied to heavy-ion collisions, the distribution functions are considered for neutrons and protons separately but  the nucleon spins are assumed to be saturated, i.e., the phase-space distribution function, which is a matrix in the spin and isospin space, is assumed to have a form $\langle\sigma\tau|f(\bm{r},\bm{p})|\sigma'\tau'\rangle=\delta_{\sigma\sigma'}\delta_{\tau\tau'}f_\tau(\bm{r},\bm{p})$ with $\sigma$ ($\sigma'$) and $\tau$ ($\tau'$) being the spin and the isospin coordinates, respectively.  However, there are some attempts to treat spins more explicitly \cite{xia2016spin}.

The BUU equation is numerically solved by introducing test particles as in Eq.~(\ref{eq:testp-delta}) to simulate heavy-ion collisions.  For the collision integral $I_{\text{coll}}$, a detailed description of the prescriptions employed by different codes is found in Ref.~\cite{yxzhang2018} where comparisons were made under a controlled condition in a periodic box among codes that are currently used and/or developed actively for heavy-ion collisions.  For the collision attempt rate that corresponds to the factor $v_{\text{rel}}\sigma ff_1$ in the collision integral Eq.~(\ref{eq:icoll-buu}), the prescription described by Bertsch and Das Gupta \cite{bertsch1988} is employed with little modification by many codes.  In this prescription, a pair of two test particles attempts a collision in a time step interval $\Delta t$ (e.g.~$\Delta t=0.5$ or 1 fm/$c$) based on a geometrical consideration when the two test particles approach closest to each other during the current time step interval and the closest distance is less than $\sqrt{(\sigma/N_{\text{tp}})/\pi}$.  In some BUU codes, the test particles are divided into $N_{\text{tp}}$ groups and the collision attempts are tested only within the same group, to save the computational time.  In this parallel ensemble method, the maximum distance for a collision is $\sqrt{\sigma/\pi}$.  There are other codes which more directly sample pairs of test particles from the probability $v_{\text{rel}}\sigma ff_1$.  Many codes treat the relativistic kinematics more precisely than Eq.~(\ref{eq:icoll-buu}) that is for the non-relativistic case.  Contrary to a naive expectation, some disagreements were found in the calculated collision attempt rates by different BUU codes in the comparison of Ref.~\cite{yxzhang2018}.  A part of the observed differences was well understood as originating from the different choices of the above-mentioned prescriptions.

For the Pauli blocking factor $(1-f)$ in $I_{\text{coll}}$, a more fundamental issue of the test particle method was discussed clearly in the comparison of Ref.~\cite{yxzhang2018}.   Here one needs to know the distribution $f$ as a smooth function of the phase-space point.  Let $f$ represent the true distribution function which hypothetically exists but cannot be exactly represented by test particles as in Eq.~(\ref{eq:testp-delta}) with the singularities of delta functions.  The test particles can be regarded as samples randomly taken from the true distribution $f$, and it is impossible to reconstruct $f$ from the finite number of samples.  Therefore, BUU codes need to introduce some kind of smearing by counting the number of test particles in phase-space cells or by replacing the delta functions in Eq.~(\ref{eq:testp-delta}) by smooth functions, e.g.~Gaussian functions, centered at individual test particles.  The obtained $\tilde{f}$ with smearing is different from the original $f$ for two main reasons.  The distribution $\tilde{f}$ is broader than $f$ due to the smearing, and $\tilde{f}$ fluctuates due to the random sampling of a finite number of test particles.  In particular, when $\tilde{f}$ is greater than 1 due to fluctuations, it must be replaced by 1 to be used as the Pauli blocking probability.  As demonstrated by the benchmark calculations in Ref.~\cite{yxzhang2018}, the distribution gradually deviates from the initially prepared Fermi-Dirac distribution toward a Boltzmann distribution, the time scale of which depends on the prescriptions.  However, in principle, the problem can be avoided by choosing a very large number of test particles ($N_{\text{tp}}\rightarrow\infty$) and a very narrow width for the smearing.  A smooth distribution function is also necessary for the evaluation of the mean-field term in the BUU equation, i.e.~the forces acting on the test particles.

\begin{figure}
\centering
\includegraphics{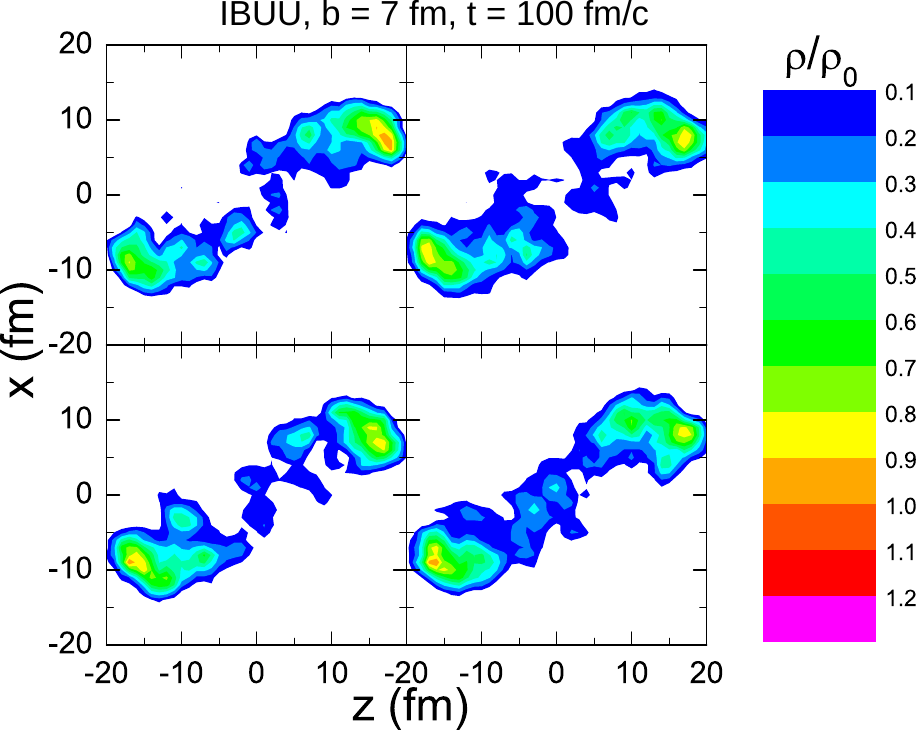}
\caption{\label{fig:xu2016-ibuu}
Density distributions in the reaction plane at $t=100$ fm/$c$ from four runs of a BUU model (IBUU \cite{bali2008,bali1997,bali1998,lwchen2014}) with $N_{\text{tp}}=100$ test particles per nucleon, for $\nuc{Au}+\nuc{Au}$ at 100 MeV/nucleon with the impact parameter $b=7$ fm.  Adapted from Ref.~\cite{xu2016}.
}
\end{figure}
An example of BUU calculations is shown in Fig.~\ref{fig:xu2016-ibuu}, where the density distribution at $t=100$ fm/$c$ calculated by a BUU model (IBUU \cite{bali2008,bali1997,bali1998,lwchen2014}) is plotted in each panel for $\nuc{Au}+\nuc{Au}$ collisions at 100 MeV/nucleon with the impact parameter $b=7$ fm.  The four panels show different runs of the simulation with $N_{\text{tp}}=100$.  There are differences among the four panels in the details, in particular in the neck region between the projectile-like and target-like parts, which is due to the finite number of test particles.  Only in the limit of $N_{\text{tp}}\rightarrow\infty$, one would get the deterministic solution of the BUU equation.  However, we will see later that the difference among these four panels with $N_{\text{tp}}=100$ is small compared to the fluctuations among different events in the QMD results shown in Fig.~\ref{fig:xu2016-imqmd}.

In the BUU equation, a part of two-body correlations is taken into account by the collision term.  However, only the average effect of two-nucleon collisions can be reflected in the deterministic time evolution of $f$.  A two-nucleon collision does not induce any subsequent correlations as is seen in the $ff_1$ factor in $I_{\text{coll}}$ which assumes that two nucleons always collide as if they were not correlated before they met to collide.  The mean field $U[f]$ is also calculated for $f$ as if there were no correlations.  Nevertheless, if the many-body correlations have only perturbative countereffects on the one-body distribution, one may expect that the BUU equation can predict one-body observables in heavy-ion collisions.  Formation of clusters and fragments, which requires many-body correlations, is beyond the scope of the BUU equation, though density fluctiations may occur in calculations by the amplification of numerical noise due to the spinodal instability in uniform low-density matter \cite{chomaz2004,colonna1993}.

Except for the possibility of spinodal instability, the problem in the description of clusters and fragments can be understood as follows.  In a late stage of a heavy-ion collision where the system has expanded, the distribution $f$ may be a widely-spread smooth function in the coordinate space.  If nucleons are independently distributed in the phase space following $f$, it can happen that, e.g., four nucleons are close to one another in the phase space so that they are likely forming an $\alpha$ particle.  Even in this case, the mean filed $U[f]$, calculated for the widely-spread smooth function $f$, is a shallow and smooth potential which cannot bind the four nucleons to move together as an $\alpha$ particle.  This problem, of course, originates from the omission of many-body correlations which are, however, difficult to handle directly.  A usual strategy to go beyond the BUU equation is to treat a part of many-body correlations as stochastic fluctuations on $f$ or $U[f]$.  We will come to this point later.

\subsection{\label{sec:qmd}QMD models}

Besides the BUU models, there are quantum molecular dynamics (QMD) models which are also actively employed in the studies of heavy-ion collisions.  Regardless of its name and its historical origin \cite{aichelin1991}, the QMD approach can be understood as a method to avoid the problems of the BUU equation discussed at the end of the previous subsection, i.e., to handle the propagation of correlations induced by two-nucleon collisions and to avoid too smooth mean-field potential, in order to enable the formation of clusters and fragments.

In QMD, the one-body distribution function is restricted to the sum of Gaussian wave packets as
\begin{equation}
\label{eq:f-qmd}
f_{\text{QMD}}(\bm{r},\bm{p},t)
=\sum_{k=1}^A \Bigl(\frac{2\nu}{\pi}\Bigr)^{3/2}e^{-2\nu(\bm{r}-\bm{R}_k(t))^2}
(2\pi\hbar)^3\delta(\bm{p}-\bm{P}_k(t)),
\end{equation}
where $\nu$ or $\Delta x=(4\nu)^{-1/2}$ is a fixed parameter representing the width of the wave packet of each nucleon.  The deterministic part of the equation of motion for the phase-space centroids $(\bm{R}_k, \bm{P}_k)$ is
\begin{equation}
\label{eq:qmd-eq-of-motion}
\frac{d}{dt}\bm{R}_k = \frac{\partial\mathcal{H}}{\partial\bm{P}_k},\qquad
\frac{d}{dt}\bm{P}_k = -\frac{\partial\mathcal{H}}{\partial\bm{R}_k},
\end{equation}
where $\mathcal{H}$ stands for the energy calculated for Eq.~(\ref{eq:f-qmd}).  This equation may be derived by averaging the equation of motion of the test particles (\ref{eq:testp-eq-of-motion}) for each Gaussian wave packet in Eq.~(\ref{eq:f-qmd}).  Thus this QMD equation of motion gives an approximate solution of the Vlasov equation (\ref{eq:vlasov}) or the left hand side of the BUU equation (\ref{eq:buu}), under the constraint of the form of Gaussian wave packets.

In the literature of QMD, e.g.~Ref.~\cite{aichelin1991}, it is often stated that the many-body state is approximated by a non-antisymmetrized product of Gaussian wave packets $e^{-\nu(\bm{r}-\bm{R}_k)^2}e^{(i/\hbar)\bm{P}_k\cdot\bm{r}}$ as the single-particle wave functions.  This corresponds to the one-body distribution function
\begin{equation}
\label{eq:f-qmd-w}
f_{\text{W}}(\bm{r},\bm{p})
=8\sum_{k=1}^A e^{-2\nu(\bm{r}-\bm{R}_k)^2}e^{-(\bm{p}-\bm{P}_k)^2/2\hbar^2\nu}
\end{equation}
with Gaussian distributions also in the momentum space, and the same equation of motion (\ref{eq:qmd-eq-of-motion}) can be derived from the time-dependent variational principle.  The difference of Eq.~(\ref{eq:f-qmd}) and Eq.~(\ref{eq:f-qmd-w}) appears in the value of the energy $\mathcal{H}$.  For the same $\bm{R}_k$ and $\bm{P}_k$, the value of the kinetic energy for $f_{\text{W}}$ of Eq.~(\ref{eq:f-qmd-w}) is larger by $3\hbar^2\nu/2M\times A$ than for $f_{\text{QMD}}$ of Eq.~(\ref{eq:f-qmd}).  This affects how an initial state nucleus should be prepared, for example.  The difference of the kinetic energy is quite large, e.g., it is about $10A$ MeV for $\nu=1/(2.5\ \text{fm})^2$.  In practically all the QMD codes, this additional kinetic energy from the momentum width is not counted, which means that these models assume Eq.~(\ref{eq:f-qmd}) as an representation of the many-body state.  In the benchmark comparison in Ref.~\cite{yxzhang2018}, all the QMD codes initialized the state so that $\bm{P}_k$, and therefore $f_{\text{QMD}}$ of Eq.~(\ref{eq:f-qmd}), follow the specified momentum distribution.

In addition to the deterministic part of the equation of motion given by Eq.~(\ref{eq:qmd-eq-of-motion}), two-nucleon collisions are introduced as stochastic processes in QMD models, based on in-medium cross sections $d\sigma/d\Omega$.  A typical procedure is as follows.  A pair of two nucleons, represented by $(\bm{R}_i, \bm{P}_i)$ and $(\bm{R}_j, \bm{P}_j)$ will collide when the distance $|\bm{R}_i-\bm{R}_j|$ becomes minimum during the current time step interval $\Delta t$ and the closest distance is less than $\sqrt{\sigma/\pi}$, just in the same way as the geometrical prescription by Bertsch and Das Gupta \cite{bertsch1988}.  It should be remarked that the distance between the wave-packet centroids is directly used for this collision prescription and therefore the Gaussian widths in the phase-space distribution $f_{\text{QMD}}$ of Eq.~(\ref{eq:f-qmd}) are not taken into account, as far as for the QMD codes participating in Ref.~\cite{yxzhang2018}.  A scattering angle is randomly selected according to the angular distribution of $d\sigma/d\Omega$, to update the momenta to $\bm{P}_i'$ and $\bm{P}_j'$.  Due to the Pauli blocking effect, the collision is assumed to occur with a probability $P_{\text{block}}$ which should be similar to the factor $(1-f)(1-f_1)$ in the BUU collision term.  Here one cannot use a singular function $f_{\text{QMD}}$ defined by Eq.~(\ref{eq:f-qmd}) as the probability.  Many QMD codes use a smeared distribution function given by Eq.~(\ref{eq:f-qmd-w}) but the self-contribution from the scattered nucleon $i$ or $j$ is excluded.  See Ref.~\cite{yxzhang2018} for various methods of Pauli blocking adopted by different QMD codes, which should be regarded as a real difference of the model assumptions in these codes.  Thus, on average, the effect of a collision is similar to the BUU collision term.  The essential difference is that a collision is now treated as a stochastic process by choosing the scattering angle randomly.

The stochastic process of a two-nucleon collision may be thought as a quantum-mechanical transition which is assume to happen instantaneously.  It is a finite jump from a classical path, which follows Eq.~(\ref{eq:qmd-eq-of-motion}), to many other possible paths corresponding to different scattering angles.  To be specific, let us write the one-body distribution as $\hat{\rho}(\Omega)$ for the final state of a two-nucleon collision with the scattering angle $\Omega$, then the two-body distribution after a collision is assumed to be
\begin{equation}
\hat{\rho}_{12}^{(2)}=\int \hat{\rho}_1(\Omega)\hat{\rho}_2(\Omega) w(\Omega)d\Omega
\end{equation}
with a suitable weight function $w(\Omega)$.  This includes correlations in contrast to the assumption of Eq.~(\ref{eq:rho2-hf}) for the mean-field models.  These correlations are propagated in QMD calculations by choosing a path $\hat{\rho}_1(\Omega)\hat{\rho}_2(\Omega)$ randomly and by solving the subsequent time evolution independently of other paths.  By many two-nucleon collisions in a heavy-ion collision, multiple branching of paths will occur, so that many different final states can be realized even starting with the same initial state.  This is an important difference of QMD from the BUU equation.  Another difference is that, in each path going to a final state, the nucleons are still represented by wave packets with a fixed width and therefore the interactions between some nucleons may be strong enough to form a fragment in principle.

\begin{figure}
\centering
\includegraphics{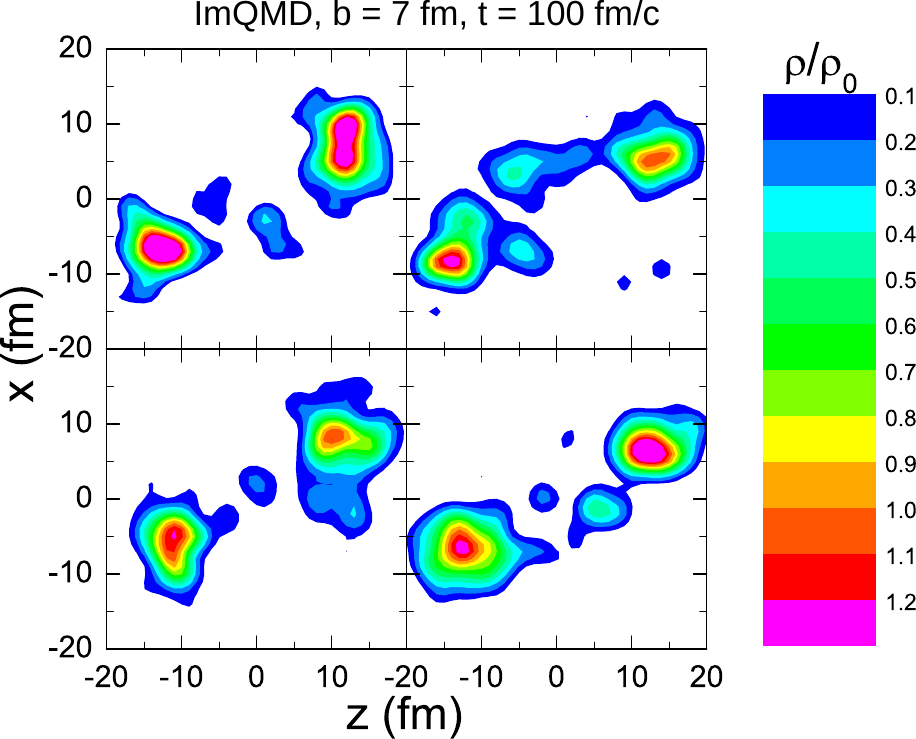}
\caption{\label{fig:xu2016-imqmd}
Density distributions in the reaction plane at $t=100$ fm/$c$ of four events calculated with a QMD model (ImQMD \cite{yxzhang2006,yxzhang2007stopping,yxzhang2008}), for $\nuc{Au}+\nuc{Au}$ at 100 MeV/nucleon with the impact parameter $b=7$ fm.  Adapted from Ref.~\cite{xu2016}.
}
\end{figure}
As an example, Fig.~\ref{fig:xu2016-imqmd} shows the density distributions at $t=100$ fm/$c$ of four different events calculated by a QMD model (ImQMD \cite{yxzhang2006,yxzhang2007stopping,yxzhang2008}), for $\nuc{Au}+\nuc{Au}$ collisions at 100 MeV/nucleon with the impact parameter $b=7$ fm.  Due to the stochastic nature of QMD, different events are generated with eventually different fragmentation configurations.  Each calculated event corresponds to a real event.  In this example, we can clearly see that fragments are produced in each event in the neck region between the projectile-like and target-like parts.  The event-by-event fluctuation in the QMD result here is much larger than the numerical fluctuation in the BUU result with $N_{\text{tp}}=100$ shown in Fig.~\ref{fig:xu2016-ibuu}.

There are, however, some drawbacks in QMD.  The one-body distribution $f_{\text{QMD}}$ of Eq.~(\ref{eq:f-qmd}) or $f_{\text{W}}$ of Eq.~(\ref{eq:f-qmd-w}) is not consistent with the Pauli principle, i.e., it does not correspond to a one-body density operator $\hat{\rho}$ with all the eigenvalues between 0 and 1.  Furthermore, Heisenberg's uncertainty principle is not respected by $f_{\text{QMD}}$.  For these reasons, the preparation of a ground state nucleus is not straightforward.  In the benchmark calculation in a box \cite{yxzhang2018}, due to the violation of the Pauli principle and the smearing in the blocking factor $f_{\text{W}}$, the initially prepared low temperature Fermi-Dirac distribution evolves toward a Boltzmann distribution more quickly than in the BUU models.  This may be also responsible to the unphysical time evolution of a ground state nucleus seen in the other benchmark comparison of Ref.~\cite{xu2016}.  It also implies that the decay of a hot nucleus is not within the applicability of QMD, because the fermionic nature, like the relation $E^*=aT^2$ between the excitation energy and the temperature, is essential for such a problem.

\begin{figure}
\centering
\includegraphics[scale=1.4]{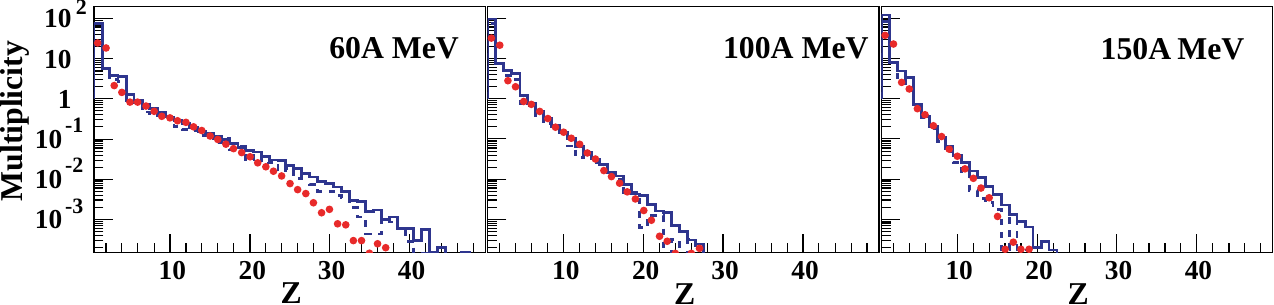}
\caption{\label{fig:zbiri2007}
Fragment charge distribution in central $\nuc{Au}+\nuc{Au}$ collisions at 60, 100 and 150 MeV/nucleon (different panels).  Central events of both the QMD results (blue histograms) and the INDRA data (red points) have been selected using the total transverse energy of light charged particles with $Z=1$ and 2.  The solid (dashed) histograms show the QMD results with (without) applying the INDRA filter.  Adapted from Ref.~\cite{zbiri2007}.
}
\end{figure}

QMD models have been employed to study dynamical characteristics of fragmentation \cite{tsang1993,hagel1994,maruyama1997toshiki,nebauer1999,zbiri2007,yxzhang2005,reisdorf2010,yxzhang2012}.  Let us limit our discussion to the points that are commonly seen in most of the calculations, avoiding the difficulties in comparing the results for different reaction systems calculated by different codes with different input parameters.  When applied to central collisions, QMD models generally explain, at least qualitatively, the copious production of intermediate mass fragments (IMF), e.g.~$3\le Z\le30$, at various incident energies ranging from 35 MeV/nucleon for $\nuc[40]{Ca}+\nuc[40]{Ca}$ \cite{hagel1994,yxzhang2005} to 100-400 MeV/nucleon for $\nuc[197]{Au}+\nuc[197]{Au}$ \cite{tsang1993,maruyama1997toshiki,zbiri2007}.  Good reproductions of the experimental data were often obtained \cite{maruyama1997toshiki,zbiri2007} for the large IMF multiplicity $N_{\text{IMF}}\approx 10$ and/or the IMF charge distribution in central $\nuc{Au}+\nuc{Au}$ collisions, as shown e.g.~in Fig.~\ref{fig:zbiri2007}.  An effective interaction corresponding to a soft EOS is usually favored to explain fragmentation.  However, these are for the primary fragments recognized in QMD at e.g.~$t=200$ fm/$c$.  Some authors directly compare the primary fragments with experimental data assuming that they are already very cold, while other authors consider the statistical decay of primary fragments, by which the calculated IMF multiplicity becomes too small in some cases \cite{tsang1993,maruyama1997toshiki}.  For the light charged particles, probably without any exception, QMD models overestimate hydrogen ($Z=1$) emission and underestimate helium ($Z=2$) emission.  In the published results of QMD e.g.~in Refs.~\cite{hagel1994,nebauer1999,yxzhang2005,zbiri2007,yxzhang2012}, and as also seen in Fig.~\ref{fig:zbiri2007}, the calculated the hydrogen yield is larger than the helium yield by about an order of magnitude, $N(\nuc{H})/N(\nuc{He})\sim10$ in various central reactions, while it is of the order of $\sim2$ in experimental data.  This is most likely because the classical motions of wave packets cannot describe the production of light clusters, such as $\alpha$ particles, which are quantum objects with only one or two bound states.  In peripheral collisions, the fragmentation of the projectile-like part seems more difficult to describe for QMD, because there is no strong collective expansion in this case.  In the results of Refs.~\cite{tsang1993,maruyama1997toshiki}, there is a strong tendency that a large projectile-like residue is produced with emissions of light particles, without undergoing multifragmentation.  A similar problem was seen in $\alpha+\nuc{Au}$ collisions at 5 GeV/nucleon \cite{maruyama1997tomoyuki}.  It is sometimes argued that this is because the effective interaction range is too large in QMD because of the spatial smearing by the wave packets.  Although a tuning of the Gaussian width parameter may improve the results, this is nothing more than a technical solution.  A more fundamental origin of the problem is probably a non-fermionic classical statistics and/or an underestimation of the degeneracy pressure in the hot residue, due to the violation of the Pauli principle as mentioned in the previous paragraph.

Fragments are usually recognized in QMD models after solving the dynamics sufficiently long, e.g.~up to $t\sim 200$ fm/$c$, by connecting two nucleons $i$ and $j$ if $|\bm{R}_i-\bm{R}_j|<R_0$, where $R_0$ is a distance parameter. This is often called the minimum spanning tree (MST) method.   In some cases, a condition in the momentum space may be also taken into account.  In principle, if the time evolution is solved for a sufficiently long time, different fragments should be automatically separated well in the coordinate space, and therefore the result of MST should not depend much on the distance parameter $R_0$ as long as it is not too small.  A traditional value seems to be $R_0\sim 3$ fm.  However, this may be too small for a neutron-rich fragment with halo or skin neutrons which may be misidentified as isolated.  It is proposed to use $R_0= 6$ fm for neutrons in Ref.~\cite{yxzhang2012clusterrecog}.

Dorso and Randrup \cite{dorso1993} proposed a method to recognize fragments at a relatively early time in molecular dynamics simulations, by finding a fragmentation configuration that maximizes the sum of the internal binding energies of individual fragments.  This method can recognize fragments before they are spatially separated, and it may be employed in order to go around the problems in the late fragmentation stage of the QMD time evolution.  In fact, Gossiaux et al.~\cite{gossiaux1997} were able to reproduce the observed multiplicity in peripheral collisions by recognizing fragments at an early time.  A recent development is ongoing \cite{lefevre2016}, where symmetry energy and structure information, such as shell and pairing effects, are taken into account in calculating the binding energies and the secondary decay of fragments are also considered.  This new method seems to explain the large yield of $\alpha$ particles.

\begin{figure}
\centering
\includegraphics[scale=1.25]{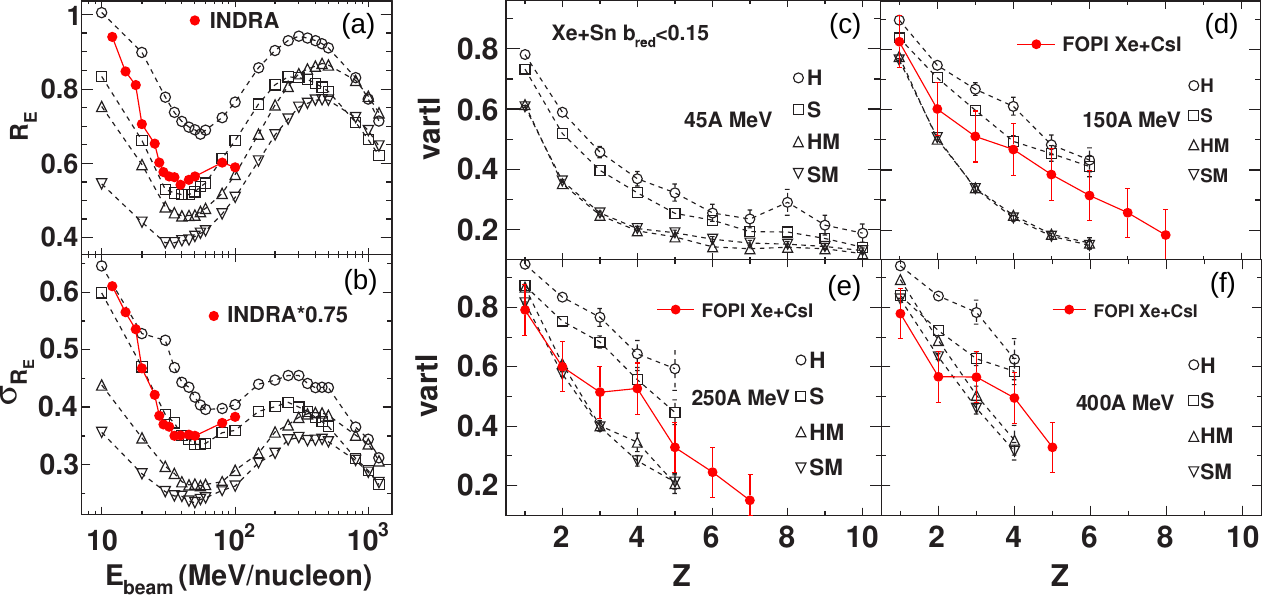}
\caption{\label{fig:gqzhang2011}
Stopping observables ($R_E$, $\sigma_{R_E}$ and \textit{varxz}) calculated by the IQMD model for $\nuc[129]{Xe}+\nuc[120]{Sn}$ collisions with small impact parameters ($b_{\text{red}}<0.15$) \cite{gqzhang2011}. In all panels, INDRA or FOPI data \cite{lehaut2010,reisdorf2010} are shown by red filled circles (The INDRA data of $\sigma_{R_E}$ have been scaled by a factor 0.75).  Circles and squares represent calculations with a hard and a soft EOS, respectively, without momentum dependence of the mean field.  Up-triangles and down-triangles are with a hard and a soft EOS, respectively, with momentum dependence.  Panels (a) and (b) show the stopping observable $R_E$ and the width of its distribution, respectively, as a function of the incident energy.  Panels (c) - (f) show the stopping observable \textit{varxz}, for different incident energies, as a function of the fragment charge $Z$ for which \textit{varxz} is calculated.  Adapted from Ref.~\cite{gqzhang2011}.}
\end{figure}

Stopping in central collisions has been studied with QMD models \cite{liu2001stopping,yxzhang2007stopping,reisdorf2010,gqzhang2011,su2016,xing2017stopping}.  Even though stopping is a global characteristics of collision dynamics, it is not a simple one-body observable because it is defined with the translational motions of emitted nucleons, light clusters and heavier fragments.  Therefore it should be studied with models that can describe fragmentation.  As naturally expected, the stopping is sensitive to the in-medium two-nucleon cross sections as shown by some QMD calculations \cite{liu2001stopping,yxzhang2007stopping,su2016,xing2017stopping}.  The EOS dependence was seen in Refs.~\cite{reisdorf2010,gqzhang2011}, where stopping is stronger with hard EOS.  Furthermore, a large effect of the momentum dependence of the mean field was found in Ref.~\cite{gqzhang2011}, as shown in Fig.~\ref{fig:gqzhang2011}.  The momentum dependence makes it easy for the projectile and the target to penetrate each other, and therefore it makes stopping weak, in particular when it is observed for IMF.  In this calculation, the stopping is too weak compared with data in $\nuc{Au}+\nuc{Au}$ collisions at 150 MeV/nucleon if a momentum-dependent interaction is adopted.  The calculation in Ref.~\cite{zbiri2007} also shows for the same reaction system that stopping is too weak when observed with IMF.  This problem may be related to the calculated result of Ref.~\cite{nebauer1999} for $\nuc{Xe}+\nuc{Sn}$ at 50 MeV/nucleon in which large fragments are too often emitted in forward and backward directions in the center-of-mass frame.  The hierarchy of stopping, i.e.~the dependence of the stopping on the size of clusters and fragments, seems to be an important clue to extract information on EOS and the momentum dependence.  Then, however, the production of clusters and fragments must be simultaneously understood together with the stopping observables.

Let us here comment on the energy conservation in two-nucleon collisions, though this issue is not limited to QMD models but applies to all the transport models with two-nucleon collisions.  When a momentum-dependent mean field is adopted, the energy conservation is not straightforward.  In the center-of-mass frame of the colliding two nucleons, the momenta $\bm{p}_{\text{rel}}$ and $-\bm{p}_{\text{rel}}$ are changed to $\bm{p}_{\text{rel}}'$ and $-\bm{p}_{\text{rel}}'$ by a collision.  In many transport codes, only the direction of $\bm{p}_{\text{rel}}$ is changed by keeping its length $|\bm{p}_{\text{rel}}|=|\bm{p}_{\text{rel}}'|$, even with the momentum-dependent mean field.  Then it does not conserve the energy.  Cozma \cite{cozma2016} studied the influence of this energy violation in a QMD model.  He tried three cases; to conserve $|\bm{p}_{\text{rel}}|$, to conserve the sum of the kinetic and potential energies of the colliding two nucleons, and to conserve the total energy of the system.  The investigations of the flow and pion observables in Ref.~\cite{cozma2016} suggest that the energy conservation should be seriously treated in order to achieve a sufficient accuracy e.g.\ to constrain the high-density symmetry energy.  It may be important to point out that the integration for $p_{\text{rel}}=|\bm{p}_{\text{rel}}|$ with the energy-conservation delta function in Eq.~(\ref{eq:vsigma}) results in a change of the phase-space factor to $p_{\text{rel}}^{\prime\,2}/v_{\text{rel}}'$, with $v_{\text{rel}}'=dE/dp_{\text{rel}}'$, which is influenced by the momentum dependence of the mean field.  This does not seem to have been taken into account even by Ref.~\cite{cozma2016}.  However, the detailed balance relation is generally important to ensure the correct equilibrium.

\subsection{\label{sec:amd-basic}Basic version of AMD}

As already mentioned above several times, fermionic nature of nucleons most probably plays important roles in fragmentation reactions, as well as in many other problems in nuclear physics.

The antisymmetrized molecular dynamics (AMD) approach was first applied to a realistic case of heavy-ion collisions in Refs.~\cite{ono1992,ono1992prl}.  In AMD, the $A$-nucleon system is described by a Slater determinant of Gaussian wave packets,
\begin{equation}
\label{eq:amd-wf}
|\Phi_{\text{AMD}}(Z)\rangle
=\hat{A}\prod_{i=1}^{A}\varphi_i(i)
\end{equation}
with the full antisymmetrization operator $\hat{A}$.  Each single-particle state is a product of a Gaussian function and a spin-isospin state
\begin{equation}
\langle\bm{r}|\varphi_i\rangle =
\exp\Bigl[-\nu\Bigl(\bm{r}-\frac{\bm{Z}_i}{\sqrt{\nu}}\Bigr)^2\Bigr]
\otimes\chi_{\alpha_i}.
\end{equation}
The spin and isospin of each nucleon are fixed, $\alpha_i=p\uparrow,p\downarrow,n\uparrow$ or $n\downarrow$.  The Gaussian width parameter $\nu$ is almost always chosen to be $\nu=1/(2.5\ \text{fm})^2$.  The wave functions are usually not normalized in AMD and the single-particle states $\varphi_i$ are not orthogonal to each other.  However, the non-normalized Slater determinant  (\ref{eq:amd-wf}) is a proper many-body state of fermions as long as the single-particle states are linearly independent.  Thus the many-body state $|\Phi_{\text{AMD}}(Z)\rangle$ is parametrized by the Gaussian centroids $Z=\{\bm{Z}_1,\bm{Z}_2,\ldots,\bm{Z}_A\}$ which are all complex vectors.  The time-evolution of these variables may be determined by the time-dependent variational principle, from which we obtain an equation of motion
\begin{equation}
\label{eq:amd-eq-of-motion}
\frac{d}{dt}\bm{Z}_i=\{\bm{Z}_i,\mathcal{H}\}_{\text{PB}},
\end{equation}
where the Poisson bracket should be suitably defined for the non-canonical variables $Z$ \cite{ono2002,ono2004ppnp}.  The Hamiltonian $\mathcal{H}$ is the expectation value of the Hamiltonian operator $H$ with the subtraction of spurious zero-point kinetic energies of fragment center-of-mass motions \cite{ono1992,ono1993}.  An effective interaction is employed here, such as the Gogny force and the Skyrme force.  The equation of motion represents the motion of wave packets in the mean field.

Before the first application of AMD, Feldmeier independently proposed the fermionic molecular dynamics (FMD) \cite{feldmeier1990}, in which the width parameters $\nu_i$ for individual wave packets are also treated as time-dependent parameters which take complex values.  This allows the exact evolution of free particles and the spreading of single-particle states in expanding systems.  In AMD, on the other hand, the width parameter $\nu$ is always a fixed parameter, for the same reason mentioned in the previous subsection for QMD, in order to describe fragment formation.  However, when the AMD wave function is applied to nuclear structure problems, the wave function is extended by treating $\nu$ as a variational parameter, by allowing deformation of the wave packets, by superposing many wave functions, and so on.  See Ref.~\cite{kanada2012} for a review.

The one-body distribution function corresponding to the AMD wave function $|\Phi_{\text{AMD}}(Z)\rangle$ of Eq.~(\ref{eq:amd-wf}) is exactly expressed as
\begin{equation}
\label{eq:f-amd}
f_{\text{AMD}}(\bm{r},\bm{p})
=8\sum_{i=1}^A\sum_{j=1}^Ae^{-2\nu(\bm{r}-\bm{R}_{ij})^2} e^{-(\bm{p}-\bm{P}_{ij})^2/2\hbar^2\nu}
B_{ij}B^{-1}_{ji},
\end{equation}
where $B_{ij}=\langle\varphi_i|\varphi_j\rangle$ is the overlap matrix of non-orthogonal single-particle states.  The parameters that look like Gaussian centroids take complex values, $\bm{R}_{ij}=(\bm{Z}_i^*+\bm{Z}_j)/\sqrt{\nu}$ and $\bm{P}_{ij}=2i\hbar\sqrt{\nu}(\bm{Z}_i^*-\bm{Z}_j)$.  It is evident heare that the wave-packet centroids $Z$ do not have very intuitive meaning due to the antisymmetrization.  Therefore, for some purpose, an approximate physical coordinates $(\bm{R}_k,\bm{P}_k)$ ($k=1,2,\ldots, A$) were introduced by a certain non-linear transformation from $Z$ \cite{ono1992}, so that the one-body distribution function may be approximately represented by $f_{\text{W}}$ of Eq.~(\ref{eq:f-qmd-w}).  By using these physical coordinates, the two-nucleon collision process was introduced as a stochastic process in a similar way to the method in QMD, though there are some differences.  See Refs.~\cite{ono1992,ono1993} for the details.  It should be commented that the approximation of Eq.~(\ref{eq:f-qmd-w}) with physical coordinates may be sufficient for the two-nucleon collision process but it is far from satisfactory for the evaluation of the Hamiltonian $\mathcal{H}$, for example.

Important differences between QMD and AMD are commented here.  AMD has a well-defined antisymmetrized wave function and therefore the ground-state nucleus is well described as the minimum-energy state within the model space.  In particular, the ground state is completely stable when its time evolution is calculated.  Because of this character, AMD can never abandon Heisenberg's uncertainty principle.  The one-body distribution function $f_{\text{QMD}}$ of Eq.~(\ref{eq:f-qmd}) is not acceptable in AMD not because it is not antisymmetrized but because it violates the relation $\Delta x\Delta p\ge\frac12\hbar$.  In AMD, the momentum width $\Delta p$ is contained in a wave packet as Eq.~(\ref{eq:f-amd}) or Eq.~(\ref{eq:f-qmd-w}).  This is not a problem as long as the wave function with a momentum width is properly interpreted.  However, this is a problem if the momentum width is not properly reflected in the time evolution of the system, which of course depends on how the time evolution is approximated in the model.  This will be an essential point in the next section.  In QMD, the momentum distribution is represented by randomly sampled $\bm{P}_k$, and therefore different momentum components may be naturally reflected in the dynamics.

After some successful applications \cite{ono1992,takemoto1996,takemoto1998} of the basic version of AMD for fragmentation of light nuclei such as $\nuc[12]{C}$, it turned out that multifragmentation in collisions of heavier nuclei such as $\nuc[40]{Ca}$ is not well described with AMD in its basic version, a solution of which was given in Ref.~\cite{ono1996amdv} to be reviewed in the next subsection.  The problem of the original version was that the two colliding nuclei tend to penetrate each other even in central collisions and two large residue nuclei are left in $\nuc[40]{Ca}+\nuc[40]{Ca}$ collisions at 35 MeV/nucleon.  Even though they are highly excited by $E^*/A\sim 5$ MeV, they do not undergo multifragmentation.  The situation is similar in central $\nuc[129]{Xe}+\nuc{Sn}$ collisions at 50 MeV/nucleon, as we will see later in Fig.~\ref{fig:ono2013nn-denev}(a) together with improved results with extended versions of AMD.  The problem can be understood as related to the momentum width of wave packets that is frozen in the time evolution of the basic version of AMD.

As mentioned above, FMD proposed to use the width parameter of each wave packet as a dynamical variable \cite{feldmeier1990}.  There were several other works based on molecular dynamics with dynamical wave-packet widths.  It was found that the inclusion of a dynamical width improves the agreement with data in some cases, such as fusion cross section above the Coulomb barrier \cite{maruyama1996}.  Kiderlen et al.~\cite{kiderlen1997} studied the fragmentation of excited systems.  In response to the initial pressure, the excited system begins to expand but clusters were not produced even though Gaussian wave packets with many-body correlations were employed.  When the excited system expands, the widths of wave packets grow and then, in turn, the interaction between the packets weakens.  The mean field for such a configuration is very shallow and smooth and there is then little chance for clusters to appear.  Similarly, studies of spinodal instability by Colonna et al.~\cite{colonna1998md} showed that the zero sound is significantly affected when the width grows large and this spreading of the nucleon wave packet then inhibits cluster formation.

\section{\label{sec:flctmodels}Extended models with fluctuations}

If the many-body time-dependent Schr\"odinger equation could be solved starting with an initial state of two colliding nuclei, the intermediate and final states will be a superposition of so many channel wave functions each of which represents a specific configuration of fragmentation.  It is far from possible practically to handle all the coherence of the many-body state.  Furthermore, most of the interferences between different channels are, at the end, of no importance and of no interest.  Therefore, an acceptable approximate picture is to represent the many-body state at a given time $t$ as an statistical ensemble as
\begin{equation}
\label{eq:stdhf}
|\Psi(t)\rangle\langle\Psi(t)|\approx \sum_c w_c(t)|\Phi_c(t)\rangle\langle\Phi_c(t)|,
\end{equation}
where the interference terms in the right hand side are ignored.  Each channel wave function $|\Phi_c(t)\rangle$, which may be chosen to be time dependent, should be sufficiently simple and tractable.  The time evolution of the weights $w_c(t)$ may be described by the transitions between channels.  The stochastic TDHF by Reinhard and Suraud \cite{reinhard1992stdhf} is conceptually transparent.  Each channel wave function $|\Phi_c(t)\rangle$ is a Slater determinant and the corresponding one-body density operator $\hat{\rho}_c(t)$ follows the TDHF equation
\begin{equation}
\label{eq:drhodt-tdhf-c}
i\hbar\frac{d}{dt}\hat{\rho}_c=
\Bigl[\frac{1}{2M}\bm{p}^2+U[\hat{\rho}_c],\ \hat{\rho}_c\Bigr],
\end{equation}
where the mean field $U[\hat{\rho}_c]$ is defined in each channel $c$.  The transition rates between channels are given by Fermi's golden rule
\begin{equation}
\label{eq:ccrate}
W(c\rightarrow c')=\frac{2\pi}{\hbar}|\langle\Phi_{c'}|V|\Phi_c\rangle|^2\delta(E_{c'}-E_c)
\end{equation}
with a residual interaction $V$, under assumptions such as that the interaction occurs in a short time scale compared to that of the mean-field evolution (see Ref.~\cite{reinhard1992stdhf} for more rigorous arguments).  Although there is nothing stochastic in the formalism, one may employ a Monte-Carlo procedure to treat the transitions which induce stochastic branching into different channels.  The time evolution of a branch can be solved independently of other branches.  Different fragmentation, for example, will occur depending on the chosen branch.

In all the practical transport models reviewed in this article, transitions are assumed to occur instantaneously and conserving the energy, which can be justified formally only in limiting cases.  To go beyond this assumption, one may need off-shell transport (see Ref.~\cite{buss2012} for a review), which is, however, beyond the scope of this article.

\subsection{\label{sec:buu-ext}Mean-field models with fluctuation}

Evidently the BUU equation [Eq.~(\ref{eq:buu}) with Eq.~(\ref{eq:icoll-buu})] does not agree with this picture.  Even though the two-nucleon collision rates like Eq.~(\ref{eq:ccrate}) are considered, only the average effect is reflected in the time evolution of the one-body distribution or $\hat{\rho}(t)$ which is deterministic without branching into channels.  Bauer et al.~\cite{bauer1987} already had an idea that a two-nucleon collision should be treated as a quantum jump from a Slater determinant to another.  In the BUU calculation with $N_{\text{tp}}$ test particles per nucleon, they treated a two-nucleon collision not as a scattering of two test particles but as a scattering of two groups of test particles.  Each group is chosen to be composed of $N_{\text{tp}}$ test particles which are neighboring in the phase space, so that a collision changes the momenta of the states of two entire nucleons.  Pauli blocking is checked for the scattering of the initially chosen two test particles.  When applied to heavy-ion collisions, the calculation showed fragmentation qualitatively \cite{bauer1987}, though a good treatment of Pauli blocking requires more investigations as in Ref.~\cite{chapelle1992}.

The fluctuation (or branching) induced by two-nucleon collisions was formulated as the Boltzmann-Langevin equation by Ayik and Gregoire \cite{ayik1990}, which is the same as the BUU eqaution (\ref{eq:buu}) but with a fluctuating collision term
\begin{equation}
\label{eq:icoll-bl}
I_{\text{coll}}[f](\mathbf{r},\mathbf{p})
=\bar{I}_{\text{coll}}[f](\mathbf{r},\mathbf{p})
+\delta I_{\text{coll}}[f](\mathbf{r},\mathbf{p}),
\end{equation}
where $\bar{I}_{\text{coll}}$ is the BUU collision term (\ref{eq:icoll-buu}) and $\delta I_{\text{coll}}$ is a stochastic term which is zero on average.  The equivalent Fokker-Plank equation was derived by Randrap and Remaud \cite{randrup1990}, and a more formal derivation using Green's function techniques is found in Ref.~\cite{reinhard1992boltzmann}.  The variances and the correlations of $\delta I_{\text{coll}}$ are related to the two-nucleon collision rates and therefore to the average term $\bar{I}_{\text{coll}}$.  This relation can be understood as the fluctuation-dissipation theorem.  It seems usually assumed that $\delta I_{\text{coll}}$ are Gaussian fluctuations, which allow gradual bifurcation of the trajectory of $f$.

The Boltzmann-Langevin equation has not been directly applied to simulate realistic heavy-ion collisions, but it has been playing a role as a theoretical background of more phenomenological approaches and approximations.  For example, Colonna et al.~\cite{colonna1993} explored a possibility of simulating the stochastic part $\delta I_{\text{coll}}$ by a numerical noise associated with the finite number of test particles $N_{\text{tp}}$ in the usual BUU calculation.  In two-dimensional matter, it was shown that $N_{\text{tp}}$ can be adjusted so that the BUU calculation yields a good reproduction of the spontaneous clusterization in the Boltzmann-Langevin solution.
Chomaz et al.~\cite{chomaz1994bob} introduced a powerful method by approximating $\delta I_{\text{coll}}$ by a suitable stochastic one-body force, $\delta I_{\text{coll}}\rightarrow -\delta\bm{F}\cdot(\partial f/\partial\bm{p})$, with the Brownian force tuned at each point in space and time to ensure the dynamics of important collective modes emulates the results of the Boltzmann-Langevin equation for nuclear matter.  This Brownian one-body dynamics model was applied in Ref.~\cite{guarnera1997multifragmentation} to the study of an initially compressed nucleus which expanded into a hollow and unstable configuration, resulting in several intermediate mass fragments.

The stochastic term $\delta I_{\text{coll}}$ in the Boltzmann-Langevin equation will generate an ensemble of the one-body distribution function $f(\bm{r},\bm{p})$.  We consider here the occupation probabilities $f_\alpha$ of individual phase-space cells, e.g., square cells of the volume $(2\pi\hbar)^3$.  If a local equilibrium is reached during the time evolution, the fluctuation term $\delta I_{\text{coll}}$ should have induced the fluctuation on $f_\alpha$ with the variance $\overline{\Delta f_\alpha^2}=\bar{f}_\alpha(1-\bar{f}_\alpha)$, where $\bar{f}_\alpha$ is the mean value.  Colonna et al.~\cite{colonna1998} introduced a stochastic mean-field (SMF) model based on the BUU equation and extending it by giving fluctuations to $f_\alpha$ or spatial density.  The basic idea is to give fluctuations to $f_\alpha$ by randomly changing $f_\alpha$ to $f_\alpha+\Delta f_\alpha$ with the above-mentioned variances, from time to time after the local thermal equilibrium is reached.  Correlations between different cells are assumed to be small, but minimal correlations are introduced for the conservation laws.  However, in actual SMF calculations, the fluctuations are projected and implemented only onto the coordinate space, agitating the spatial density profile.

\begin{figure}
\centering
\includegraphics[width=\textwidth]{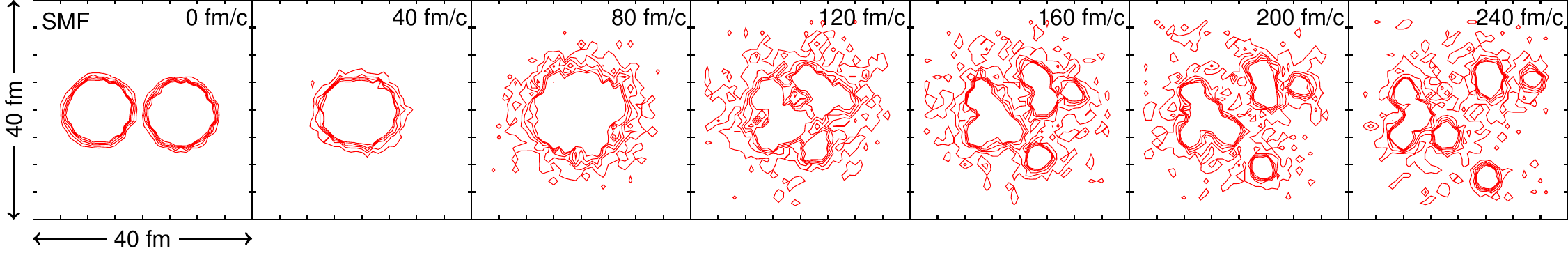}
\caption{\label{fig:denev-smf}
SMF result of the time evolution of the density distribution projected onto the reaction plane in an central event of $\nuc[112]{Sn}+\nuc[112]{Sn}$ at 50 MeV/nucleon.  Adapted from Ref.~\cite{colonna2010}.
}
\end{figure}

Fragmentation described by SMF is close to the picture of spinodal decomposition \cite{colonna2004,chomaz2004}.  When the compressed system is expanding, the density distribution may be almost uniform at first.  However, when the system is expanding toward low densities, it may enter the spinodal region in the density-temperature plane where the uniform matter is unstable against density fluctuations.  In the SMF calculation with fluctuations, the density inhomogeneity, which may be small at first, will be amplified by the mean field dynamics so that the system is eventually broken into fragments, as shown in Fig.~\ref{fig:denev-smf} for a $\nuc{Sn}+\nuc{Sn}$ central collision at 50 MeV/nucleon.  The size of fragments is mainly governed by the unstable modes of density fluctuation.  It is also seen in the figure that many nucleons are continuously emitted into free space.  In Sec.~\ref{sec:amd-wpsplit} and in Fig.~\ref{fig:denev-amd}, we will review the comparison of this SMF result with AMD.

\begin{figure}
\centering
\includegraphics[scale=1.1]{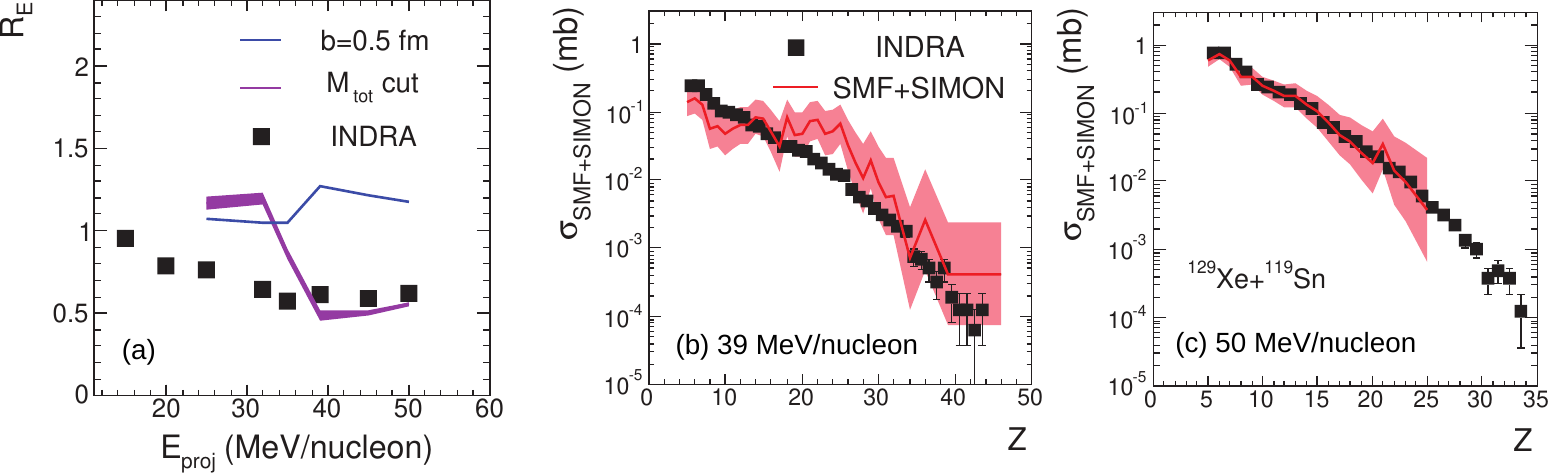}
\caption{\label{fig:bonnet2014} (a) Stopping observable $R_E$ calculated by SMF with SIMON secondary decay for $\nuc[129]{Xe}+\nuc[119]{Sn}$ collisions.  The evolution with beam energies of the mean value of $R_E$ is shown for $b=0.5$ fm (blue line) and for events selected using a total charged particle multiplicity ($M_{\text{tot}}$) cut (violet line). (b) Fragment charge distribution at $E/A=39$ MeV calculated by SMF with SIMON secondary decay for the events selected by $M_{\text{tot}}$ cut (red line). (c) The same as (b) but at $E/A=50$ MeV.  In all panels, the INDRA data are shown by black filled squares.  Adapted from Ref.~\cite{bonnet2014}.}
\end{figure}

The results of SMF, with the SKM* effective interaction and the free two-nucleon cross sections, were compared with the INDRA data by Bonnet et al.~\cite{bonnet2014} for $\nuc{Xe}+\nuc{Sn}$ collisions in Fermi energy domain.  Above the incident energy of 39 MeV/nucleon, SMF shows multifragmentation in the dynamical stage and reproduces the data of IMF charge distributions well.  Below this energy, density fluctuations still occur and lead to formation of prefragments, but they recombine to produce an evaporation residue or fission fragments, which is not consistent with data.  This problem may be interrelated to the underestimation of fragment velocities which suggests too weak radial expansion in the calculation.  For central collisions with $b<0.5$ fm, SMF shows full stopping with $R_E\gtrsim1$ at all the calculated incident energies between 25 to 50 MeV/nucleon, as shown by the blue line in Fig.~\ref{fig:bonnet2014}(a), which might appear inconsistent with the data.  However, if central events are selected by using a total charged particle multiplicity in a similar way to the experimental data, weaker stopping $R_E\approx 0.5$-0.7, which is consistent with the data, is obtained above 39 MeV/nucleon as shown by the violet line.  With the same event selection, a good agreement between SMF and data is also seen for the fragment charge distribution above 39 MeV/nucleon [panels (b) and (c) of Fig.~\ref{fig:bonnet2014}].

The distribution of $f_\alpha$ with the variance $\overline{\Delta f_\alpha^2}=\bar{f}_\alpha(1-\bar{f}_\alpha)$ cannot be a Gaussian distribution with a small width.  If the value is limited in the range $0\le f_\alpha\le 1$, the only possible distribution with that variance and the mean value $\bar{f}_\alpha$ is the Bernoulli distribution, i.e., $f_\alpha=1$ with the probability $\bar{f}_\alpha$ and $f_\alpha=0$ with the probability $1-\bar{f}_\alpha$.  Therefore the effect of the fluctuation term is to make each phase-space cell completely occupied or vacant.  This condition, $f_\alpha=1$ or 0, is a classical limit of the condition $\hat{\rho}^2=\hat{\rho}$ for a Slater determinant.  We then arrive at the original idea by Bauer et al.~\cite{bauer1987} to change two entire nucleons by a two-nucleon collision, which corresponds to a jump from a Slater determinant to another.

\begin{figure}
\centering
\includegraphics[scale=0.4]{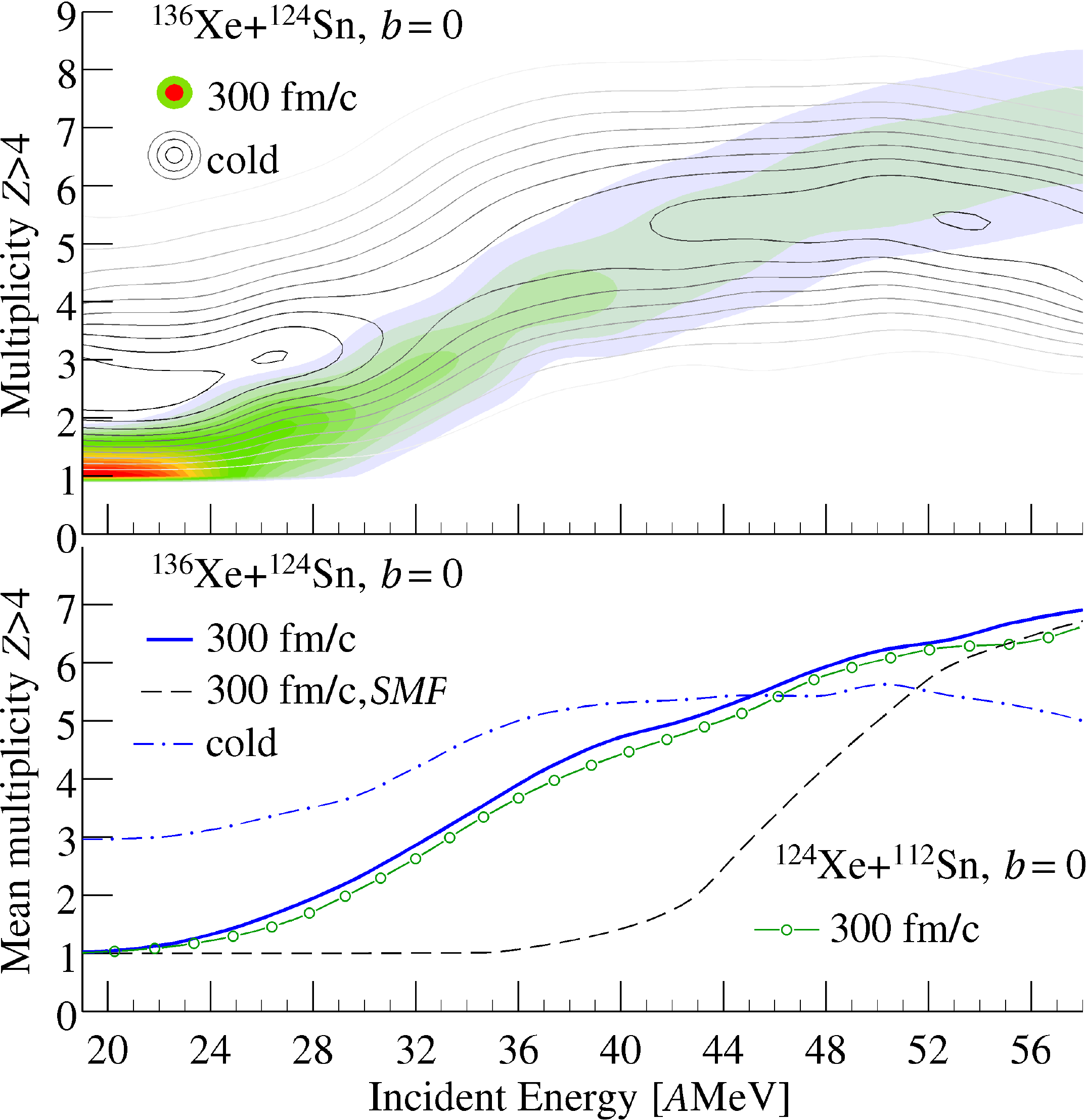}
\caption{\label{fig:napolitani2013} Top: Results of the BLOB model for the multiplicity distribution of fragments with $Z>4$ as a function of the incident energy in central $\nuc[136]{Xe}+\nuc[124]{Sn}$ collisions at $t=300$ fm/$c$ (filled contours) and after secondary decay (contour lines).  Bottom: Corresponding mean values.  The result of the SMF model at $t=300$ fm/$c$ is shown by a dashed line.  Taken from Ref.~\cite{napolitani2013}.  }
\end{figure}

Based on the original idea of Ref.~\cite{bauer1987}, an improved procedure of Pauli blocking was introduced and tested in idealized systems in a periodic box \cite{rizzo2008new}.  Napolitani and Colonna \cite{napolitani2012,napolitani2013,napolitani2015} introduced this improved method in a transport code, Boltzmann-Langevin one-body (BLOB) model.  In the BLOB model, a two-nucleon collision is treated as a scattering of two groups $\mathrm{A}$ and $\mathrm{B}$ each of which consists of $N_{\text{tp}}$ test particles, so that a group corresponds to a nucleon.  The groups of test particles after the collision, $\mathrm{A}'$ and $\mathrm{B}'$, will be determined by changing the momenta of test particles for the chosen scattering angle.  The groups are chosen to be compact in the coordinate space and in the momentum space but under the condition that the Pauli principle $f\le 1$ is not violated in the initial and final states.  Thus BLOB takes into account the full fluctuations in the phase space while the fluctuations were projected onto the coordinate space in SMF.  In fact, as shown in Fig.~\ref{fig:napolitani2013} taken from Ref.~\cite{napolitani2013}, the BLOB calculation for $\nuc[136]{Xe}+\nuc[124]{Sn}$ central collisions in the Fermi energy domain shows the onset of fragmentation at relatively low energies compared to the SMF case.  Fragmentation or spallation in light-ion induced reactions in the 1 GeV region was also studied by BLOB in Ref.~\cite{napolitani2015}.

\subsection{\label{sec:amd-wpsplit}AMD with wave-packet splitting}

AMD uses Gaussian wave packets with a fixed width as single-particle states.  One would naturally expect that one-body dynamics is better described by mean-field models such as TDHF and BUU models that allow unrestricted time evolution of single-particle states or distribution function.  For example, the problem of producing too large highly-excited residues, which we have seen in Sec.~\ref{sec:amd-basic}, may be related to the insufficient description of single-particle dynamics in AMD in its basic version.

Then, is there any advantage in using wave packets rather than more general wave functions?  An illustrative argument to understand this point can be found in Ref.~\cite{ono2004ppnp} for the case of channels with and without a nucleon emission.  Here, let us consider a similar but different example in which the system with $(A_1+A_2+1)$ nucleons has disintegrated (or is going to disintegrate) into two fragments, which is described by a single Slater determinant in a mean-field model as
\begin{equation}
|\Psi\rangle = |\psi\Phi_1\Phi_2\rangle
\qquad\mbox{with}\quad \psi=\frac{1}{\sqrt2}(\phi_1+\phi_2).
\end{equation}
We assume that $\Phi_1$ and $\Phi_2$ are the states of two nuclei with mass numbers $A_1$ and $A_2$, respectively, which are already well separated in space.  The antisymmetrization and normalization should be implicitly understood.  There exists another nucleon whose state $\psi$ has two components $\phi_i$ ($i=1,2$) that are localized in the space of the first and the second nuclei, respectively.  When the time evolution is further solved by the mean-field model, the mean field $U[\hat{\rho}]$ is calculated for the one-body density operator $\hat{\rho}$ of $|\Psi\rangle$.  In the regions of individual nuclei, $U[\hat{\rho}]$ is essentially made by $(A_1+0.5)$ and $(A_2+0.5)$ nucleons, respectively.  On the other hand, the same state is also written as
\begin{equation}
|\Psi\rangle = \frac{1}{\sqrt2}|(\phi_1\Phi_1)\Phi_2\rangle
+ \frac{1}{\sqrt2}|\Phi_1(\phi_2\Phi_2)\rangle
\end{equation}
or corresponding many-body density operators
\begin{equation}
\begin{split}
|\Psi\rangle\langle\Psi|
&=\frac12|(\phi_1\Phi_1)\Phi_2\rangle\langle(\phi_1\Phi_1)\Phi_2|
+\frac12|\Phi_1(\phi_2\Phi_2)\rangle\langle\Phi_1(\phi_2\Phi_2)|
\\
&\qquad
+\frac12|(\phi_1\Phi_1)\Phi_2\rangle\langle\Phi_1(\phi_2\Phi_2)|
+\frac12|\Phi_1(\phi_2\Phi_2)\rangle\langle(\phi_1\Phi_1)\Phi_2|,
\end{split}
\end{equation}
which include the channel $|(\phi_1\Phi_1)\Phi_2\rangle$ in which the fragments with mass numbers $A_1+1$ and $A_2$ are formed, and the other channel $|\Phi_1(\phi_2\Phi_2)\rangle$ in which the fragment mass numbers are $A_1$ and $A_2+1$.  If one is tempted to solve these channels independently by ignoring the interference terms, the mean field is $U[\hat{\rho}_{c=1}]$ or $U[\hat{\rho}_{c=2}]$ depending on the channel.  The one-body density operators for $|\Psi\rangle$ and for individual channels are
\begin{align}
&\hat{\rho}=\frac12\hat{\rho}_{c=1}+\frac12\hat{\rho}_{c=2}
+\frac12|\phi_1\rangle\langle\phi_2|+\frac12|\phi_2\rangle\langle\phi_1|
\label{eq:rho-coherent}
\\
&\hat{\rho}_{c=1}=|\phi_1\rangle\langle\phi_1|
+(A_1+A_2)\mathop{\textrm{Tr}}_{2,\ldots}|\Phi_1\Phi_2\rangle\langle\Phi_1\Phi_2|
\\
&\hat{\rho}_{c=2}=|\phi_2\rangle\langle\phi_2|
+(A_1+A_2)\mathop{\textrm{Tr}}_{2,\ldots}|\Phi_1\Phi_2\rangle\langle\Phi_1\Phi_2|.
\end{align}
As in the example of Fig.~\ref{fig:rios2011} by Rios et al., the elimination of the elements far from diagonal in the coordinate representation, $\hat{\rho}\rightarrow\hat{\rho}'=\tfrac12\hat{\rho}_{c=1}+\tfrac12\hat{\rho}_{c=2}$, does not introduce much change in the quantities of our interest.  However, this new density operator no longer corresponds to a Slater determinant, $\hat{\rho}^{\prime2}\ne\hat{\rho}^{\prime}$.  In the scheme of Eq.~(\ref{eq:stdhf}), this induces a branching into the two channels of different configurations of fragmentation.  We intuitively or empirically believe that the branching should be introduced in this case.  For example, due to the energy and momentum conservation, the fragment velocities should be different in different mass decompositions, which cannot be described by a model without branching.  In the true many-body dynamics, the coherence between $\phi_1$ and $\phi_2$ is destroyed by the interactions and correlations of the last nucleon with all the other nucleons, which is difficult to predict because it is most likely beyond the usual two-nucleon collision effects.

The above simple example gives us several hints for the extension to general cases.  In the first step, the usual single-particle dynamics in the mean field is considered (to predict $\psi$).  Then the single particle wave function is decomposed (into $\phi_1$ and $\phi_2$) in such a way that each component is localized in a nucleus.  After the wave function of the many-body system is decomposed, the mean field ($U[\hat{\rho}_{c=1}]$ or $U[\hat{\rho}_{c=2}]$) is different from channel to channel, and therefore the time evolution of channels should be solved independently.  We may still employ the idea of mean field in each channel.  The single-particle states in different channels do not interfere in a usual way because of the many-body correlations.

This idea was introduced into AMD by first calculating the single-particle dynamics without any restriction and then decomposing or splitting each evolved wave function into wave packets, which results in quantum branching \cite{ono1996amdv,ono1999,ono2002,ono2004ppnp}.  Let us first apply a mean field theory to the AMD wave function $|\Phi_{\text{AMD}}(Z)\rangle$, which describes one of the branches, at a time $t=t_0$.  In practice, we will utilize the Vlasov equation (\ref{eq:vlasov}) to solve the time evolution starting with a Gaussian wave packet, labeled by $k$, corresponding to each term of the approximate one-body distribution of Eq.~(\ref{eq:f-qmd-w}) with the physical coordinates.  This may be solved from $t=t_0$ to $t=t_1\equiv t_0+\tau_{\text{coh}}$, where $\tau_{\text{coh}}$ is a parameter of the model.  The meaning of $\tau_{\text{coh}}$ will be discussed below.  Note that the solution $|\psi_k(t_1)\rangle$ is no longer a Gaussian wave packet because no restriction is put on $|\psi_k(t)\rangle$.  Then at the time $t=t_1$, we assume that coherence is lost so that the state turns into an ensemble of Gaussian wave packets as
\begin{equation}
|\psi_k(t_1)\rangle\langle\psi_k(t_1)| \rightarrow \int
\frac{d\bm{R}'d\bm{P}'}{(2\pi\hbar)^3}
\mu_k(\bm{R}',\bm{P}')
|\phi(\bm{R}',\bm{P}')\rangle\langle\phi(\bm{R}',\bm{P}')|
\label{eq:SingleparticleBranching}
\end{equation}
where $|\phi(\bm{R}',\bm{P}')\rangle$ represents a (normalized) Gaussian wave packet with the centroid $(\bm{R}',\bm{P}')$ in phase space.  The weight function $\mu_k(\bm{R}',\bm{P}')$ may be determined by requiring that the both sides of Eq.\ (\ref{eq:SingleparticleBranching}) should be similar in some sense.  We require that the phase-space distribution for $|\psi_k(t_1)\rangle$ is reproduced by the right hand side of Eq.\ (\ref{eq:SingleparticleBranching}) only in the phase-space directions to which the distribution is spreading beyond the width of the original Gaussian wave packet \cite{ono2002}.  By applying the same procedure for all the nucleons $k$, the total state $|\Phi(Z)\rangle$ at the time $t=t_0$ branches into an ensemble of AMD wave functions at $t=t_1$.  This process is repeated from the initial state of the reaction to the final state.

In this approach, the parameter $\tau_{\text{coh}}$ controls how long the coherence of the single particle dynamics is kept before the independence of channels is respected by quantum branching.  In this sense, the parameter $\tau_{\text{coh}}$ may be called the coherence time.  Decoherence in the single-particle state is physically due to many-body correlations.  A suitable choice of $\tau_{\text{coh}}$ is an important element of the model.  When this kind of extension was first applied to AMD \cite{ono1996amdv,ono1999}, the limit of $\tau_{\text{coh}}\rightarrow0$ was studied by assuming that the coherence time is sufficiently short compared to the typical time scales of reactions.  It is also possible to assume that the coherence is lost for a nucleon when it collide with another nucleon (symbolically $\tau_{\text{coh}}=\tau_{\text{NN}}$), which may be a reasonable choice since the decoherence should be due to many-body correlations.

It is very important to ensure the total energy conservation when quantum branching is introduced, though it is not so straightforward because branching is introduced by considering only the single-particle dynamics, while the total energy should be conserved through many-body correlations.  In practice, the total energy is conserved by adding a frictional term to the equation of motion, with the conservation of the center-of-mass position, the
total momentum and the total angular momentum.  Furthermore, it seems important to constrain some global one-body quantities for the frictional term.  See Refs.~\cite{ono2002,ono2004ppnp} for more details.  It seems difficult to automatically derive how to conserve the energy.  The problem is easily seen also in the case of the one-body density operator of Eq.~(\ref{eq:rho-coherent}).  The elimination of the off-diagonal elements $|\phi_1\rangle\langle\phi_2|$ and $|\phi_2\rangle\langle\phi_1|$ does not much affect the energy conservation.  See the arguments for Fig.~(\ref{fig:rios2011}) and in Ref.~\cite{rios2011}.  However, the problem is that the energies do not necessarily agree among the three different Slater determinants corresponding to $\hat{\rho}$, $\hat{\rho}_{c=1}$ and $\hat{\rho}_{c=2}$.

\begin{figure}
\centering
\includegraphics[width=\textwidth]{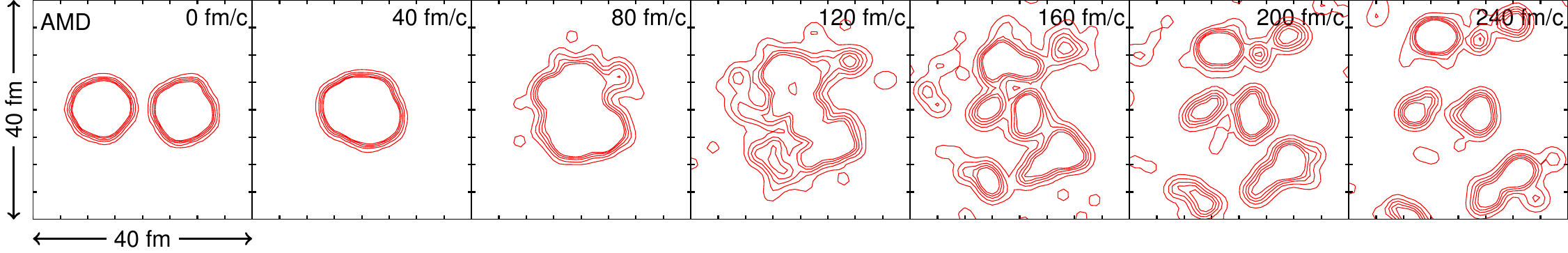}
\caption{\label{fig:denev-amd}
Result of AMD with wave-packet splitting ($\tau_{\text{coh}}=\tau_{\text{NN}}$) for the time evolution of the density distribution projected onto the reaction plane in an central event of $\nuc[112]{Sn}+\nuc[112]{Sn}$ at 50 MeV/nucleon.  Adapted from Ref.~\cite{colonna2010}.
}
\end{figure}

Figure \ref{fig:denev-amd} shows a typical example of the time evolution of ${}^{112}\mathrm{Sn}+{}^{112}\mathrm{Sn}$ central collisions at 50 MeV/nucleon \cite{rizzo2007,colonna2010}.  The AMD calculation was performed with the coherence time $\tau_{\text{coh}}=\tau_{\text{NN}}$.  This result can be directly compared with the SMF result of Fig.~\ref{fig:denev-smf} because they were calculated with similar effective interactions and free two-nucleon cross sections with a maximum cutoff at 150 mb.  As mentioned in Sec.~\ref{sec:buu-ext}, the variance $\overline{\Delta f^2}=\bar{f}(1-\bar{f})$ for the occupation probability assumed by SMF is equivalent to the fully occupied ($f=1$) or vacant ($f=0$) phase-space cells.  If a phase-space cell is identified with a Gaussian wave packet, the SMF and AMD models are conceptually similar, but the results can be different due to the different approximate treatments of fluctuations.  In fact, as quantitatively investigated in Ref.~\cite{rizzo2007}, the density fluctuation among different events is already developing in AMD at a relatively early stage of $50\lesssim t\lesssim 100$ fm/$c$, while the fluctuation develops in SMF only at a later stage $t\sim100$ fm/$c$ suggesting a fragmentation mechanism by spinodal decomposition \cite{chomaz2004}.  When the density fluctuation develops in AMD, the energy released by forming prefragments is converted to the kinetic energies of prefragments, which then facilitates the expansion of the system and the spatial separation of prefragments.   This is interpreted as an origin of the differences in the expansion velocity, the nucleon emission and so on predicted by AMD and SMF \cite{rizzo2007,colonna2010}.

\begin{figure}
\centering
\includegraphics[scale=0.7]{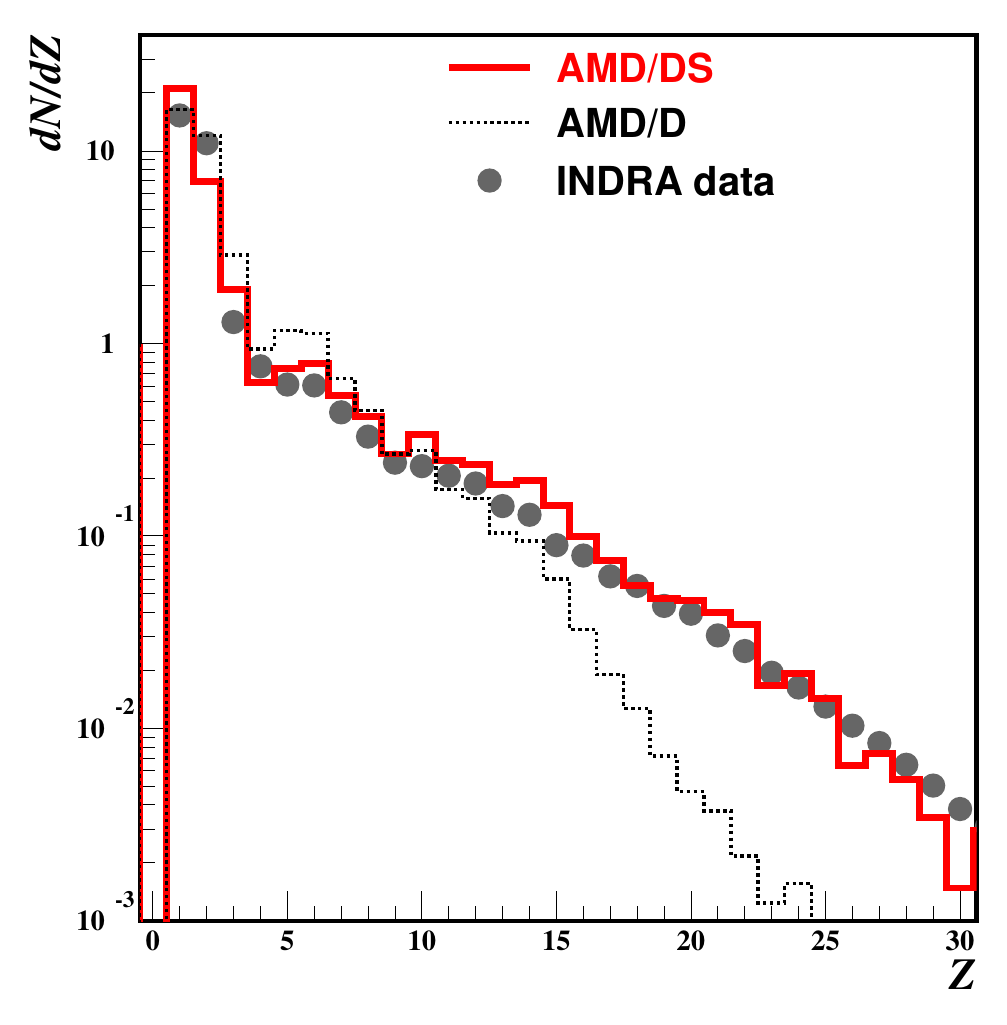}
\caption{\label{fig:XeSn-zmulti} The charge distribution of the produced clusters in ${}^{129}\mathrm{Xe}+\mathrm{Sn}$ collisions at 50 MeV/nucleon with the impact parameter $0<b<4$ fm, after calculating the secondary decay of excited clusters and applying the experimental filter for the detector setup.  Solid histogram (labeled AMD/DS) shows the result of AMD with the coherence time $\tau_{\text{coh}}=\tau_{\text{NN}}$, while the dotted histogram (labeled AMD/D) shows the result with the strongest decoherence $\tau_{\text{coh}}\rightarrow0$.  The INDRA experimental data are shown by solid points.  Taken from Ref.~\cite{ono2002}.}
\end{figure}

Figure \ref{fig:XeSn-zmulti} shows the AMD results of the fragment charge distribution compared with data for central $\nuc{Xe}+\nuc{Sn}$ collisions at 50 MeV/nucleon.  We find that the result depends very much on the choice of the coherence time for the wave-packet splitting.  With the coherence time $\tau_{\text{coh}}=\tau_{\text{NN}}$, the fragment yields for $Z\gtrsim 3$ are well reproduced.  However, a problem is found in the yields of light particles. The $\alpha$-particle multiplicity $M_\alpha\approx7$ is too small and the proton multiplicity $M_p\approx20$ is too large compared to the experimental data $M_\alpha\approx M_p\approx 10$.  In Ref.~\cite{ono1996amdv}, the AMD results for $\nuc[40]{Ca}+\nuc[40]{Ca}$ collisions at 35 MeV/nucleon are compared without wave-packet splitting and with wave-packet splitting in the limit of the small coherence time $\tau_{\text{coh}}\rightarrow0$ (i.e., the strongest decoherence).  The latter result is consistent with the experimental data.  Without wave-packet splitting, the two nuclei tend to go through each other with two big fragments at the end.  The wave-packet splitting allows the mixing of the two nuclei or the neck formation, so that more than two fragments can be formed from the system expanding in the beam direction.  The $\alpha$-particle multiplicity also depends on the wave-packet splitting very much.  The isotope distributions in projectile fragmentation reactions were  also reproduced reasonably well by the AMD with wave-packet splitting for $\nuc[40,48]{Ca}+\nuc{Be}$ and $\nuc[58,64]{Ni}+\nuc{Be}$ collisions at 140 MeV/nucleon \cite{mocko2008}.

In short, the strength of the wave-packet splitting (viewed from AMD) or the decoherence of the single particle states (viewed from the mean field theory) is one of the key ingredients for the description of fragmentation.  For the stronger splitting, the system tends to expand strongly and to break into small fragments and many $\alpha$ particles.  Unfortunately, the appropriate strength of decoherence seems to depend on the size of the system and/or the incident energy.  A more consistent understanding may be possible if the cluster correlations in dynamical systems are more explicitly treated (see Sec.~\ref{sec:clstmodels}).

\begin{figure}
\centering
\includegraphics[scale=2]{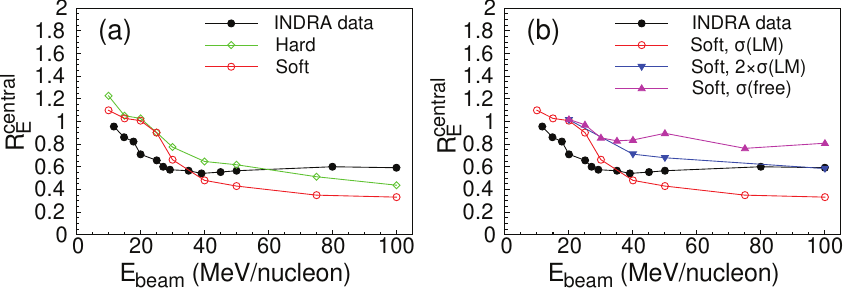}
\caption{\label{fig:zhao2014amdstopping}
Comparisons of the stopping observable $R_E$ between experimental data (black filled circles) \cite{lehaut2010} and the AMD calculation with wave-packet splitting ($\tau_{\text{coh}}\rightarrow0$) for central $\nuc[129]{Xe}+\nuc[120]{Sn}$ collisions. The default calculation shown by red open circles was done with a soft EOS and the in-medium cross section by Li and Machleidt.  (a)  Dependence on the EOS.  Green diamonds show the case with a hard EOS.  (b)  Dependence on the two-nucleon cross sections.  The purple triangles are with the free cross sections, and the blue inverted triangles are with the Li-Machleidt cross sections multiplied by a factor of two.  Adapted from Ref.~\cite{zhao2014amdstopping}.
}
\end{figure}

Nuclear stopping was studied by Zhao et al.~\cite{zhao2014amdstopping} employing the AMD with the strongest wave-packet splitting ($\tau_{\text{coh}}\rightarrow0$) for $\nuc{Xe}+\nuc{Sn}$ systems from 10 to 100 MeV/nucleon.  They find that, above 25 MeV/nucleon, the stopping observable $R_E$ is sensitive to both the two-nucleon collision cross sections and the effective interaction.  A stiff EOS results in a stronger stopping than a soft EOS as shown in Fig.~\ref{fig:zhao2014amdstopping}(a).  With the energy-dependent parametrization of the in-medium two-nucleon cross sections by Li and Machleidt \cite{li1993,li1994}, stopping is gradually underestimated when the incident energy is raised as shown in Fig.~\ref{fig:zhao2014amdstopping}(b), while the stopping is too strong with the free cross sections in this energy region.  This may be a useful information on the two-nucleon cross sections as well as the EOS.  At low energies, the full stopping ($R_E\approx 1$) is seen in the calculation even with the in-medium cross sections below 25 MeV/nucleon, while the full stopping is seen in the experimental data only at the lowest energy ($\approx 10$ MeV/nucleon).  A similar problem was also seen in Fig.~\ref{fig:bonnet2014}(a) for the SMF calculation \cite{bonnet2014}, though there is some difference in the incident energy of the transition from full to partial stopping.


In AMD and QMD models in their standard treatment described in Sec.~\ref{sec:basicmodels}, two nucleon collisions are considered as stochastic processes, which is also a source of branching into channels.  However, there is an important difference between AMD and QMD which is not necessarily related to the antisymmetrization.  In QMD, the momentum variables $\bm{P}_k$ in $f_{\text{QMD}}$ of Eq.~(\ref{eq:f-qmd}) also have a stochastic character because they are randomly sampled following some distribution in preparing the initial nuclei.  Compared to the case of AMD which may be approximately regarded as using Gaussian wave packets as in $f_{\text{W}}$ of Eq.~(\ref{eq:f-qmd-w}), the fluctuations contained in each wave packet in the momentum space is randomly realized in QMD.  Therefore the branching originating from the momentum width is naturally reflected in the dynamics through the time evolution of QMD.  This may be a reason why QMD results sometimes better agree with data than the basic version of AMD.

AMD handles wave functions which automatically satisfy Heisenberg's uncertainty principle, and therefore there is no way to accept $f_{\text{QMD}}$ of Eq.~(\ref{eq:f-qmd}) to treat the momentum fluctuation.  Instead, the problem has been overcome by considering the wave-packet splitting, with which the momentum fluctuations contained in the Gaussian wave packet are reflected in the dynamics.  A simpler prescription to incorporate momentum fluctuation was also proposed in Ref.~\cite{ono1996mflct}, in which a momentum fluctuation is given to a wave packet when it is emitted, i.e.~when it is isolated for the first time under some criterion.  The method can be generalized to the center-of-mass momentum of clusters and may be applied in the AMD with clusters in Sec.~\ref{sec:amd-cluster}, where the wave-packet splitting is not so straightforward.

Lin et al.~\cite{lin2016} introduced a new method into AMD with the strongest wave-packet splitting ($\tau_{\text{coh}}\rightarrow0$) to take into account the momentum fluctuation in the Gaussian wave packet also in the two-nucleon collision process.  When two nucleons collide, momentum fluctuations are first given to individual wave packets and the scattering is processed.  The momentum and energy conservation is restored by modifying other nucleons.  This additional momentum fluctuation process has drastically improved the high energy part of the proton energy spectra in $\nuc[40]{Ar}+\nuc[51]{V}$ at 44 MeV/nucleon and $\nuc[36]{Ar}+\nuc[181]{Ta}$ at 94 MeV/nucleon.

\subsection{\label{sec:comd}CoMD}

The Pauli principle requires that the phase-space density should not exceed one nucleon per a phase-space volume $(2\pi\hbar)^3$ for each spin-isospin state in semiclassical descriptions.  The exact treatment of the Pauli principle requires significant computational power for large systems.  In order to overcome the computational difficulty, an approximate implementation of the Pauli principle was proposed as constrained molecular dynamics (CoMD) \cite{papa2001,papa2005,papa2013}.  In this approach, a stochastic process is added to the usual QMD in order to prevent the violation of the Pauli principle due to the deterministic part of the equation of motion, in addition to the usual two-nucleon collisions with Pauli blocking.  The process is invoked when the phase-space density ${f}_i$ around a nucleon $i$ becomes greater than 1.  The momenta of the nucleon $i$ and other nucleon(s) are changed as in the two-nucleon scattering so that the Pauli principle ${f}_i\le1$ is finally satisfied after several trials.  This is one of the ways to satisfy the Pauli principle, though it is not derived from first principles.

For the mean field effect, the most reasonable formulation at the semiclassical level is the Vlasov equation (\ref{eq:vlasov}) which is consistent with the Pauli principle because it conserves the phase-space volume.  If the stochastic process for the Pauli principle is introduced into QMD so that the one-body distribution follows the Vlasov equation on average, CoMD can be similar to the AMD with wave-packet splitting.

CoMD can reproduce the multifragmentation data such as that at the incident energy of 35 MeV/nucleon \cite{papa2001}.  The effect of the stochastic process for the Pauli principle results in stronger stopping and expansion towards instability of multifragmentation.  The charge distribution of intermediate mass fragments are reasonably reproduced, except for the problems in the light particle multiplicities.  Another QMD code of Ref.~\cite{su2014}, which incorporates the same CoMD procedure for the Pauli principle, also successfully describes the anisotropic expansion of the system and the fragment formation in central $\nuc{Xe}+\nuc{Sn}$ collisions at 50 MeV/nucleon.

\section{\label{sec:clstmodels}Models with explicit cluster correlations}

As we have seen in the previous sections, practically all the transport models in Sec.~\ref{sec:basicmodels} and Sec.~\ref{sec:flctmodels} fail to consistently explain the large yield of $\alpha$ particles and the relatively small yield of protons observed in multifragmentation in heavy-ion collisions. 
This suggests that cluster correlations should be stronger in reality than these models can handle.  In this section, we will review approaches to overcome this problem by taking into account the cluster correlations more explicitly in transport models.

A cluster correlation means a correlation among several nucleons which propagate, at least for some time period, as if they form an internal bound or resonance state, which may be formulated as a quasiparticle corresponding to a peak in a spectral function in the Green's function formalism.  We may study the existence of such a cluster correlation in a similar way to the equilibrium case as in Sec.~\ref{sec:clusterinmed}, though idealistic equilibrium is not reached in heavy-ion collisions.  Once we know that cluster correlations can exist in medium, we may have a picture that the system is composed of clusters and non-clustered nucleons which interact and change chemical composition by reactions such as $X+p+n\leftrightarrow X+d$, where $X$ can be any particle.

When a many-body state $|\Psi\rangle$ is known at a time $t=t_0$, can one tell the number of clusters and their phase-space distribution without solving the many-body time evolution?  This seems to be a difficult, maybe ill-posed, question because a cluster is a composite particle.  For e.g.~a deuteron cluster, one may construct an operator
\begin{equation}
  \label{eq:adagger-d}
  a_d^\dagger(\bm{P}) = \int \frac{d\bm{p}}{(2\pi\hbar)^3}
  \psi_d(\bm{p})
a_n^\dagger(\tfrac12\bm{P}+\bm{p})a_p^\dagger(\tfrac12\bm{P}-\bm{p}),
\end{equation}
where $a_n^\dagger$ and $a_p^\dagger$ are the nucleon creation operators and $\psi_d(\bm{p})$ is the deuteron wave function in the momentum representation.  Spin indices should be implicitly understood. This operator certainly creates a deuteron with the momentum $\bm{P}$ when it is operated on the vacuum.  However, it does not satisfy the commutation relation of bosons (see e.g.\ Refs.~\cite{combescot2008,sahlin1965}), and therefore one cannot count the number of deuterons by $a_d^\dagger(\bm{P})a_d(\bm{P})$ for a general many-body state $|\Psi\rangle$.  Clusters can be safely treated as an elementary particle only in the dilute limit.

Although the aim of this section is to review transport models that handle the dynamics of cluster correlations during the time evolution of heavy-ion collisions, we here briefly mention the coalescence prescription which is often applied to calculate cluster observables, in particular in high-energy collisions (e.g.\ Refs.~\cite{sato1981,mattiello1997}).  Mean-field models such as BUU can only predict the single-nucleon distribution function.  At the time $t=t_0$ of freeze-out, the many-nucleon distribution is assumed to be given by the product of single-nucleon distributions, and then the cluster (e.g.~deuteron) distribution is calculated by the overlap of the product distribution function with the deuteron wave function \cite{lwchen2003,mattiello1997}.  With the above-mentioned caution in mind, we may get the deuteron momentum spectrum by
\begin{equation}
  \label{eq:coales}
  \sum_{\text{spin}}\langle a_d^\dagger(\bm{P})a_d(\bm{P})\rangle
  =\frac34\int\frac{d\bm{r}_1d\bm{r}_2d\bm{p}}{(2\pi\hbar)^3}
  \rho_d^{\text{W}}(\bm{r}_1-\bm{r}_2,\bm{p})
  f_n(\bm{r}_1,\tfrac12\bm{P}+\bm{p},t_0)
  f_p(\bm{r}_2,\tfrac12\bm{P}-\bm{p},t_0),
\end{equation}
where $\rho_d^{\text{W}}$ is the Wigner transform of the deuteron internal wave function, and $f_n$ and $f_p$ are the neutron and proton distributions at freeze-out summed over spins.  In Ref.~\cite{lwchen2003} for example, the prescription is applied to the nucleons which are assumed to be frozen out when their local densities are less than $\rho_0/8$.  Coalescence prescription is justified in case cluster correlations have only perturbative effects so that e.g.\ the violation of energy conservation does not significantly influence the dynamics.  Also a caution is again that it should be applied only to the cases of sufficiently low phase-space densities.  For example, Eq.~(\ref{eq:coales}) integrated over $\bm{P}$,
\begin{equation}
  \label{eq:coales-num}
  \frac34\int \frac{d\bm{r}_1d\bm{p}_1d\bm{r}_2d\bm{p}_2}{(2\pi\hbar)^6}
  \rho_d^{\text{W}}\bigl(\bm{r}_1-\bm{r}_2,\tfrac12(\bm{p}_1-\bm{p}_2)\bigr)
  f_n(\bm{r}_1,\bm{p}_1,t_0)
  f_p(\bm{r}_2,\bm{p}_2,t_0)
\end{equation}
cannot give a reasonable value as a number of deuterons if the neutron and/or proton phase-space densities are high.

Koonin's formula (\ref{eq:Koonin}) for the momentum correlations is similar to Eq.~(\ref{eq:coales}).  In fact, by replacing the deuteron Wigner function in Eq.~(\ref{eq:coales}) with that $\rho_{\bm{q}}^{\text{W}}(\bm{r},\bm{p})$ of the two-particle continuum state $\psi_{\bm{q}}(\bm{r})$ for the asymptotic relative momentum $\bm{q}$, and by performing the integration over $\bm{p}$ under an assumption that one-body distributions are almost constant in the momentum range where $\rho_{\bm{q}}^{\text{W}}(\bm{r},\bm{p})$ takes significant values, we arrive at Eq.~(\ref{eq:Koonin}) in the case of $D(\bm{r},m\bm{v},t)=f_p(\bm{r},\tfrac12\bm{P},t_0)\delta(t-t_0)$.

In the following, we will review the approaches to go beyond the coalescence prescription by treating the dynamics of cluster correlations during the time evolution of heavy-ion collisions.

\subsection{BUU with clusters}

Danielewicz et al.~\cite{danielewicz1991,danielewicz1992} extended the BUU model by incorporating light clusters based on the Green's function formalism.  In the extended BUU equation, light clusters are treated as new particle species so that the distribution functions $f_n(\bm{r},\bm{p},t)$, $f_p(\bm{r},\bm{p},t)$, $f_d(\bm{r},\bm{p},t)$, $f_t(\bm{r},\bm{p},t)$ and $f_h(\bm{r},\bm{p},t)$ are considered.  The set of equations is
\begin{equation}
\frac{\partial f_x}{\partial t}
+\frac{\partial H_x}{\partial\bm{p}}\cdot\frac{\partial f_x}{\partial\bm{r}}
-\frac{\partial H_x}{\partial\bm{r}}\cdot\frac{\partial f_x}{\partial\bm{p}}
=I^{\text{coll}}_x[f_n,f_p,f_d,f_t,f_h],
\end{equation}
where $H_x$ is the single-particle Hamiltonian with the mean field for the particle $x=n,p,d,t$ and $h$ ($h=\nuc[3]{He}$).  The equations are coupled by the collision terms $I_x^{\text{coll}}$ which take into account the gain and loss of particle $x$ by various reactions such as $ppn\leftrightarrow pd$ and $ppnn\leftrightarrow nh$, as well as by elastic collisions.  The contributions from different reaction channels are summed in $I_x^{\text{coll}}$.  The factor $(1\mp f_x)$ is included in the collision terms for the fermionic/bosonic particle $x$ in the final state.  A similar formulation was also given by R\"opke and Schulz \cite{roepke1988}.

The transition rates or cross sections in $I_x^{\text{coll}}$ are expressed by the matrix elements for various reaction channels such as $|\mathcal{M}_{ppn\rightarrow pd}|^2$ which are evaluated in Refs.~\cite{danielewicz1991,danielewicz1992} by the impulse approximation so that they are related to the two-nucleon elastic cross sections.  The matrix elements are, however, corrected by multiplying an energy-dependent factor to reproduce some experimental data such as the $pd\rightarrow ppn$ cross sections.  Another important fact to take into account is that a cluster cannot exist as a bound quasiparticle in some phase-space region as in the example of Fig.~\ref{fig:roepke2011} due to the medium effect or Pauli principle, i.e., the Mott effect.  Therefore a condition is imposed that a cluster can only be created when the nucleon occupation averaged over the phase-space corresponding to the internal cluster wave function is less than a cutoff value $\langle f\rangle <f_{\text{cut}}$.  The cutoff value $f_{\text{cut}}$ may be determined phenomenologically, guided by the solution of the in-medium Schr\"odinger equation, e.g.~Eq.~(\ref{eq:deuteron-inmed}) and Fig.~\ref{fig:roepke2011}.  The effect of the choice of $f_{\text{cut}}$ was studied in Ref.~\cite{kuhrts2001} as well as the medium effect of the reaction rates.

\begin{figure}
\centering
\includegraphics{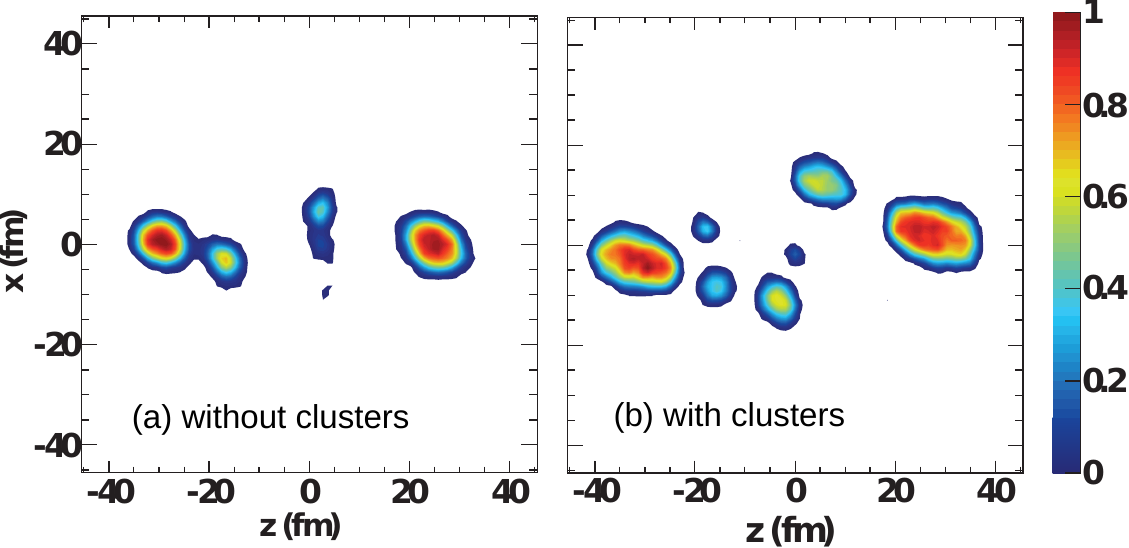}
\caption{\label{fig:coupland2011}
Density profile in the reaction plane at $t=270$ fm/$c$ in a $\nuc{Sn}+\nuc{Sn}$ collision at 50 MeV/nucleon, calculated by a BUU model with a momentum-dependent mean field.  The left and right panels show results without clusters and with clusters, respectively.  Adapted from Ref.~\cite{coupland2011}.
}
\end{figure}

Coupland et al.~\cite{coupland2011} studied the effect of clusters with this version of BUU model in $\nuc{Sn}+\nuc{Sn}$ collisions at 50 MeV/nucleon, as well as the effects of the density dependence of the symmetry energy and the in-medium two-nucleon cross sections.  They studied the isospin transport ratio \cite{tsang2004}, similar to Eq.~(\ref{eq:isospin-transport-ratio}), which characterizes the equilibration of the $N/Z$ ratio through the neck region between the projectile-like and target-like parts in e.g.~$\nuc[124]{Sn}+\nuc[112]{Sn}$ collisions of a neutron-rich projectile and a neutron-deficient target.  Inclusion of clusters had effects on the isospin transport ratio to weaken the isospin equilibration and to reduce the sensitivity to the density dependence of the symmetry energy.  These effects of clusters were found to be very similar to those of increasing the two-nucleon collision cross sections, which may be because the number of collisions also increases when the reaction channels of cluster production are added.  Furthermore, including clusters had effects on the fragment production.  As compared in Fig.~\ref{fig:coupland2011}, the neck region fragments into many smaller pieces in the right panel for the calculation with clusters.  The fragmentation process also continues longer, as seen in the shape of the residues, which have not yet reached a compact shape.  It was also found that the whole neck region tends to expel its asymmetry, when clusters are turned on, with less sensitivity to the symmetry energy.

The average kinetic energies of light clusters are well reproduced for $\nuc{Xe}+\nuc{Sn}$ central collisions at 50 MeV/nucleon \cite{kuhrts2001}.  However, $\alpha$ clusters have not been introduced in this BUU model, and the missing production of IMFs is a general problem of BUU models without fluctuation.  It should be desirable to have mean-field models with fluctuating collision terms with clusters, which may be achieved by adding collision-induced fluctuations in this BUU model with clusters or by incorporating clusters in a BUU model with fluctuations of Sec.~\ref{sec:buu-ext}.

\subsection{\label{sec:amd-cluster}AMD with clusters}

The AMD wave function (\ref{eq:amd-wf}) is suitable for describing cluster correlations in the ground state and low-lying excited states of nuclei in the studies of nuclear structure \cite{kanada2012}.  An important question is, however, whether such clusterized states are realized with correct probabilities during the time evolution of reactions.  Under the equation of motion (\ref{eq:amd-eq-of-motion}) with the usual stochastic two-nucleon collisions, the cluster correlation is governed by the classical phase space.  Let us consider here the density of states $\tilde{D}(\epsilon)$ (or more precisely the difference of the density of states from that of the non-interacting system) for the internal state of an $A_{\text{c}}$-nucleon cluster.  For a deuteron cluster, for example, $\tilde{D}(\epsilon)$ is a function starting continuously with $\tilde{D}(E_{\text{gs}})=0$ at the ground state.  Since the binding energy is small due to the cancellation of the potential and kinetic energies, the bound phase-space volume $\int_{E_{\text{gs}}}^0\tilde{D}(\epsilon)d\epsilon$ is not as large as to correspond to a single quantum state, and therefore cluster correlations do not emerge dynamically with reasonable probabilities.

It should be noted that irreversible stochastic processes, such as the wave-packet splitting, can change the phase-space weights and improve the cluster multiplicities as in the case of the strongest decoherence ($\tau_{\text{coh}}\rightarrow0$) in Fig.\ \ref{fig:XeSn-zmulti}, though a fully consistent reproduction of fragmentation is difficult by adjusting only the coherence time.

A way to overcome this problem is to add a quantum ground-state contribution $\delta(\epsilon-E_{\text{gs}})$ to $\tilde{D}(\epsilon)$ (and modify it to keep the total number of states) so that the new density of state $D(\epsilon)$ is similar to Eq.~(\ref{eq:den-of-states-deuteron}) for the quantum statistical treatment.  This is done in Refs.~\cite{ono2013nn,ikeno2016} by generalizing the two-nucleon collision process in AMD to allow the possibility that each colliding nucleon may form a cluster of mass numbers $A_{\text{c}}=2,3$ or 4 with some other wave packets.  Namely, when two nucleons $N_1$ and $N_2$ collide, we consider the process
\begin{equation}
N_1+N_2+B_1+B_2\rightarrow C_1+C_2
\end{equation}
in which each of the scattered nucleons $N_j$ ($j=1,2$) may form a cluster $C_j$ with a spectator particle $B_j$.  This process includes the collisions without cluster formation as the special case of $C_j=N_j$ with empty $B_j$.  The transition rate of the cluster-forming process is given by Fermi's golden rule, e.g~Eq.~(\ref{eq:ccrate}) with the suitable choice of the set of final states.  When a cluster is formed, the corresponding wave packets are placed at the same phase-space point, i.e., the cluster internal state is represented by the harmonic-oscillator $(0s)^n$ configuration.  Denoting the initial and final states of the $N_j+B_j$ system by $|\varphi_j\rangle$ and $|\varphi_j'\rangle$, respectively, the transition rate is
\begin{equation}
v_id\sigma = \frac{2\pi}{\hbar}
|\langle\varphi_1'|\varphi_1^{\bm{q}}\rangle|^2
|\langle\varphi_2'|\varphi_2^{-\bm{q}}\rangle|^2
|\mathcal{M}|^2\delta(E_f-E_i)
\frac{p_{\text{rel}}^2dp_{\text{rel}}d\Omega}{(2\pi\hbar)^3},
\label{eq:transitionrate-cluster}
\end{equation}
where $|\varphi_j^{\pm\bm{q}}\rangle = e^{\pm i\bm{q}\cdot\bm{r}_j}|\varphi_j\rangle$ are the states after the momentum transfer $\pm\bm{q}$ to the nucleons $N_j$ ($j=1,2$), and $(p_{\text{rel}},\Omega)$ is the relative momentum between $N_1$ and $N_2$ in these states.  The matrix element $|\mathcal{M}|^2$ is essentially the same as for the usual two-nucleon collisions.

The above equation for the rate should be generalized because there are many possible ways of forming a cluster for each $N$ of the scattered nucleons $N_1$ and $N_2$.  It should be done carefully for two main reasons.  One is the non-orthogonality of final states, which is the same problem as that Eq.~(\ref{eq:coales-num}) cannot give the right number of deuterons.  For example, when there are two neutron wave packets $n_{\text{a}}$ and $n_{\text{b}}$ with the same spin near the scattered proton, both $X+p+n_{\text{a}}+n_{\text{b}}\rightarrow X'+d_{\text{a}}+n_{\text{b}}$ and $X+p+n_{\text{a}}+n_{\text{b}}\rightarrow  X'+n_{\text{a}}+d_{\text{b}}$ may be possible as the final states of a deuteron production.  Let us distinguish these different channels by the formed clusters $C=d_{\text{a}}$ and $C=d_{\text{b}}$.  The probabilities cannot not be simply summed because different channels are not orthogonal.  Overcounting is avoided by calculating the overlap probability with the subspace spanned by non-orthogonal final states.  Another point is that e.g.~the probability of the usual two-nucleon collision $X+p+n\rightarrow X'+p+n$ ($C=p$) should be reduced when the cluster formation $X+p+n\rightarrow X'+d$ ($C=d$) is taken into account.  This corresponds to keeping the number of states in $D(\epsilon)$.  The method was extended in an iterative way to consider the formation of clusters $C_1$ and $C_2$ with mass numbers $A_{\text{c}}=1,2,3$ and 4.  The general formula is
\begin{equation}
v_id\sigma(C_1,C_2,p_{\text{rel}},\Omega)
 = \frac{2\pi}{\hbar}
P_1(C_1,p_{\text{rel}},\Omega)
P_2(C_2,p_{\text{rel}},\Omega)
|\mathcal{M}|^2\delta\bigl(E_f(C_1,C_2,p_{\text{rel}},\Omega)-E_i\bigr)
\frac{p_{\text{rel}}^2dp_{\text{rel}}d\Omega}{(2\pi\hbar)^3},
\label{eq:transitionrate-cluster-general}
\end{equation}
where the overlap probabilities in Eq.~(\ref{eq:transitionrate-cluster}) have been replaced by the probabilities of specific channels which satisfy $\sum_{C_1}P_1(C_1,p_{\text{rel}},\Omega) = 1$ and $\sum_{C_2}P_2(C_2,p_{\text{rel}},\Omega) = 1$.  The relative momentum $p_{\text{rel}}$ in the final state is adjusted for the energy conservation depending on the channel $(C_1, C_2)$.  The phase-space factor $p_{\text{rel}}^2/(\partial E_f/\partial p_{\text{rel}})$ also depends on $(C_1, C_2)$.

Even when the cluster formation is introduced, the many-body state is always represented by an AMD wave function which is a Slater determinant of nucleon wave packets.  The time evolution of the many-body state is solved just as usual without depending on whether some of the wave packets form clusters due to collisions in the past (except for the inter-cluster binding process below).  This is in contrast to the version of BUU by Danielewicz et al.~\cite{danielewicz1991} where clusters are treated practically as new particle species.  In the AMD with clusters, a nucleon in a formed cluster may collide with some other nucleon so that the cluster is broken.  It may be the case that the scattered nucleon forms the same cluster as before, so that an elastic scattering of the cluster is possible.  All of these kinds of processes are based on the nucleon-nucleon scattering matrix elements $|\mathcal{M}|^2$.  It is, of course, possible to modify the cluster formation probabilities.  For example, in the calculation of Ref.~\cite{ikeno2016}, the overall cluster production probability was suppressed when the momentum transfer is extremely small.

\begin{figure}
\centering
\includegraphics[scale=0.6]{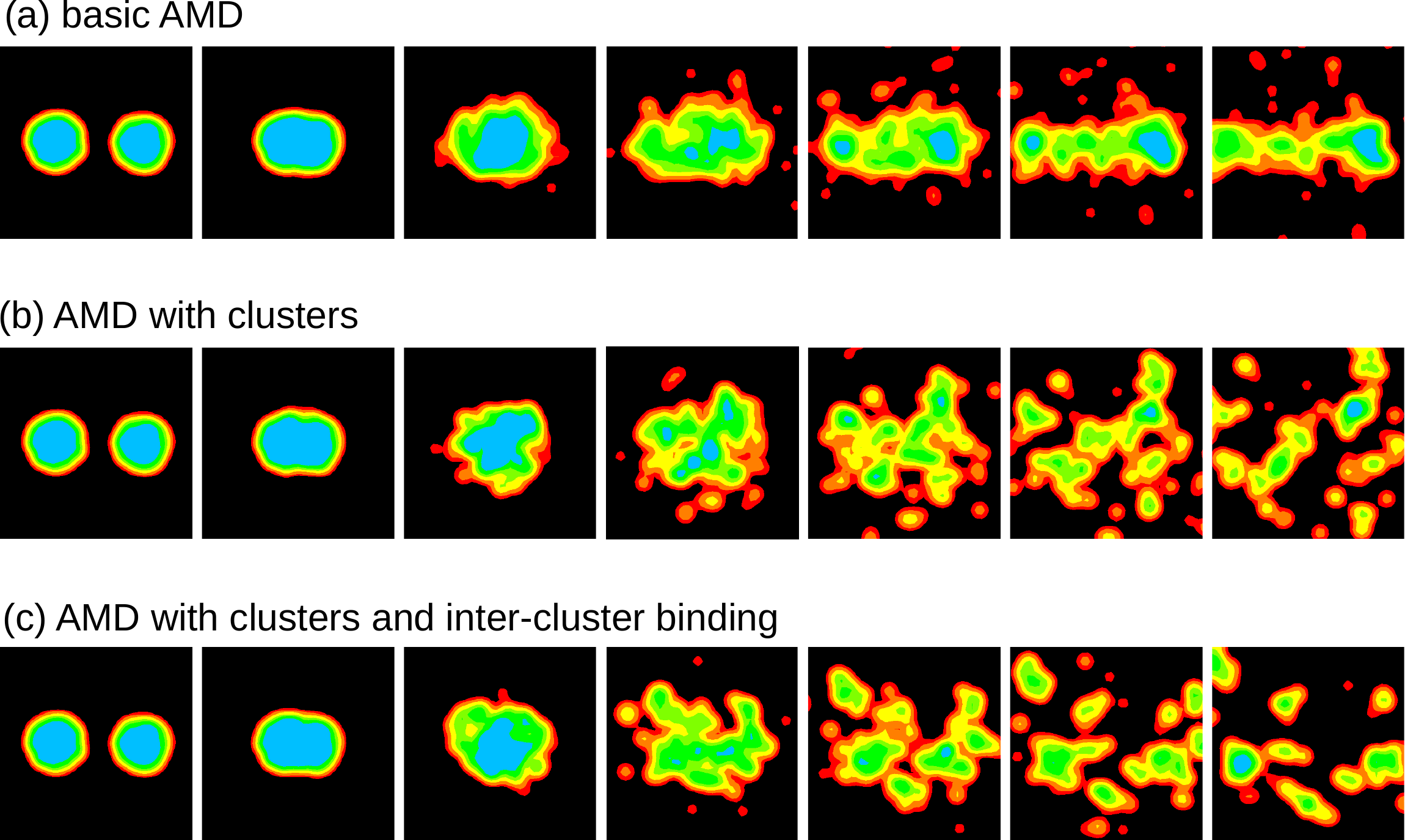}
\caption{\label{fig:ono2013nn-denev}
Time evolution of the density distribution projected onto the reaction plain in events of central $\nuc{Xe}+\nuc{Sn}$ collisions at 50 MeV/nucleon, calculated by (a) the basic AMD with usual two-nucleon collisions, (b) AMD with clusters in the final states of two-nucleon collisions, and (c) AMD with clusters and inter-cluster binding process.  The density distribution is shown at every 40 fm/$c$ from left to right.  The size of the shown area is $40\ \text{fm}\times40\ \text{fm}$.  Adapted from Ref.~\cite{ono2013nn}.
}
\end{figure}

As we saw in Sec.~\ref{sec:amd-basic}, when the basic version of AMD with usual two-nucleon collisions was applied to central heavy-ion collisions,  multifragmentation was not well described because two nuclei tend to penetrate each other so that there are two large nuclei and many emitted nucleons in the final state, which is clearly seen in the time evolution of the density distribution shown in Fig.~\ref{fig:ono2013nn-denev}(a).  When the two-nucleon collision process is extended to take into account the formation of light clusters ($A_{\text{c}}\le 4$), the system now expands strongly to disintegrate into small pieces, e.g.~in $\nuc{Xe}+\nuc{Sn}$ central collisions at 50 MeV/nucleon \cite{ono2013nn}, as shown in Fig.~\ref{fig:ono2013nn-denev}(b).  It is as if the system totally turns into a gas of light clusters and nucleons.  The yields of light clusters, including $\alpha$ particles, are too large compared with the experimental data, while the IMF yields are underestimated in particular for heavy IMFs.  Thus the cluster correlations can change the collision dynamics completely.  Because clustering effectively reduces the number of degrees of freedom by freezing the internal degrees of freedom in individual clusters, the energy distributed to each particle increases so that they do not have much chance to form bound heavier fragments.

This problem of producing too many light clusters and too few IMFs may be solved by taking into account the quantum correlations to bind light clusters to form heavier fragments \cite{ono2013nn}.  Many of light nuclei, such as Li and Be isotopes, have only one or a few bound states which may be regarded as bound states of internal clusters.  By the same reason for the issue of light clusters, the quantum-mechanical probability of forming such a nucleus in a discrete state is not consistent with the semiclassical phase space for the motions of clusters.  Therefore, for a better description, inter-cluster correlation is introduced as a stochastic process of inter-cluster binding.  The basic idea is to replace the relative momentum between clusters by zero if moderately separated clusters are moving away from each other with a small relative kinetic energy.  Since this prescription is introduced rather phenomenologically, the condition to invoke this method has to be carefully chosen.  Figure \ref{fig:ono2013nn-denev}(c) is an example of the result with inter-cluster binding process, with which the experimental data of fragment charge distribution and the yields of light charged particles are now well reproduced in a preliminary study in Ref.~\cite{ono2013nn}.  In recent studies, it seems sufficient to consider the formation of light nuclei with mass numbers $A\le9$ or 10 \cite{ikeno2016,tian2018}.

\begin{figure}
\centering
\includegraphics{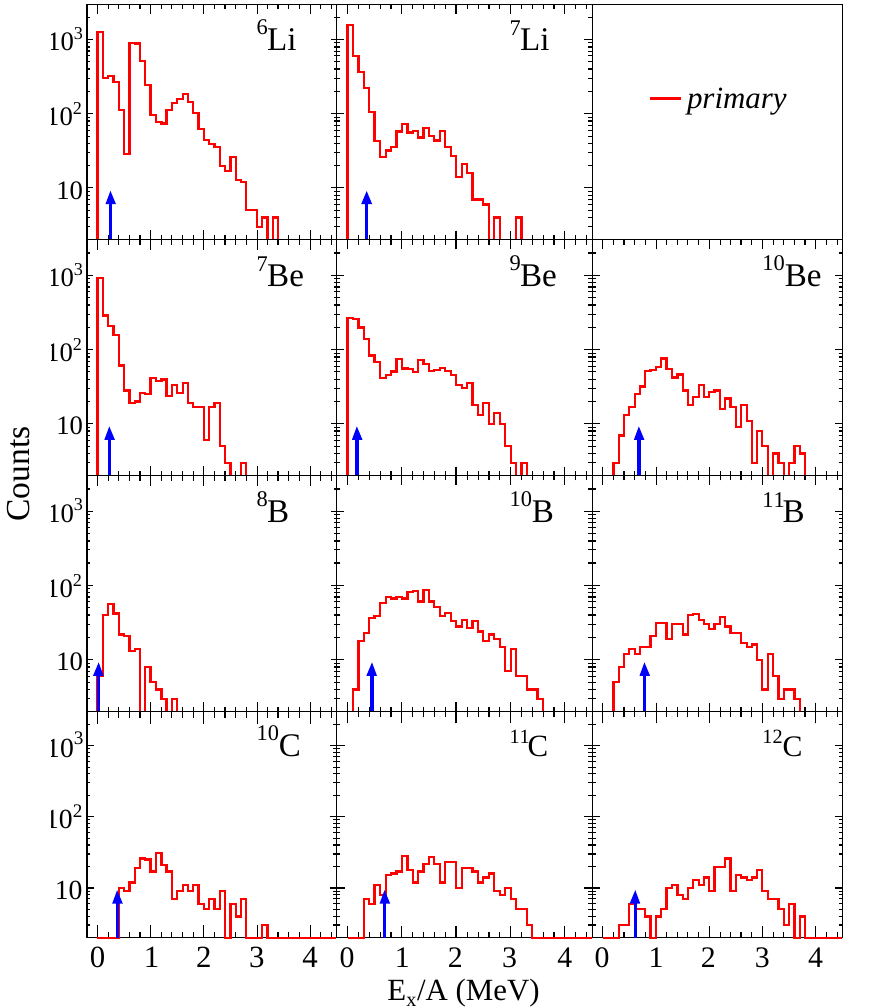}
\caption{\label{fig:tian2018-exdstiv} Excitation energy distributions of the primary fragments produced by $t=300$ fm/$c$ in the AMD calculations for $\nuc[12]{C}+\nuc[12]C$ at 95 MeV/nucleon.  The histograms show the distribution of fragments in the intermediate velocity source, i.e.~those emitted at $\theta_{\text{lab}} > 20^\circ$ and with kinetic energies above experimental energy thresholds.  The particle decay thresholds are shown by arrows. Adapted from Ref.~\cite{tian2018}.}
\end{figure}

Tian et al.~\cite{tian2018} studied fragmentation in $\nuc[12]{C}+\nuc[12]{C}$ collisions at 95 MeV/nucleon by the AMD with cluster correlations and with the inter-cluster binding process.  Reasonable agreements with the experimental data were obtained for the yields of light particles and fragments, angular distributions and energy spectra.  Besides the projectile-like (and undetected target-like) component, there is also an intermediate velocity component in the particle spectra at large angles in the laboratory frame.  Such fragments in the latter component can be produced only through violent processes.  The fragment excitation energies at the end of the dynamical calculation ($t=300$ fm/$c$) were analyzed and it was found in Fig.~\ref{fig:tian2018-exdstiv} that many IMFs with $A\le9$ in the intermediate velocity component are produced with very low excitation energies, which is most probably due to the inter-cluster binding process.  Many of them are below the particle decay threshold and survive through the statistical decay that is applied to the primary fragments produced by the dynamical calculation.  This is how the present calculation is consistent with data, while other AMD calculations \cite{tian2017}, which do not consider inter-cluster binding process, underestimated the IMF production in the intermediate velocity component.  It should be remarked, however, that when the excitation energy is low and close to the particle decay threshold, a subtle uncertainty in the estimated excitation energy may result in a large difference in the final fragment yield.
Thus this study suggests that cluster correlations and inter-cluster correlations have a strong impact on the IMF production.

\begin{figure}
\centering
\includegraphics[scale=0.45]{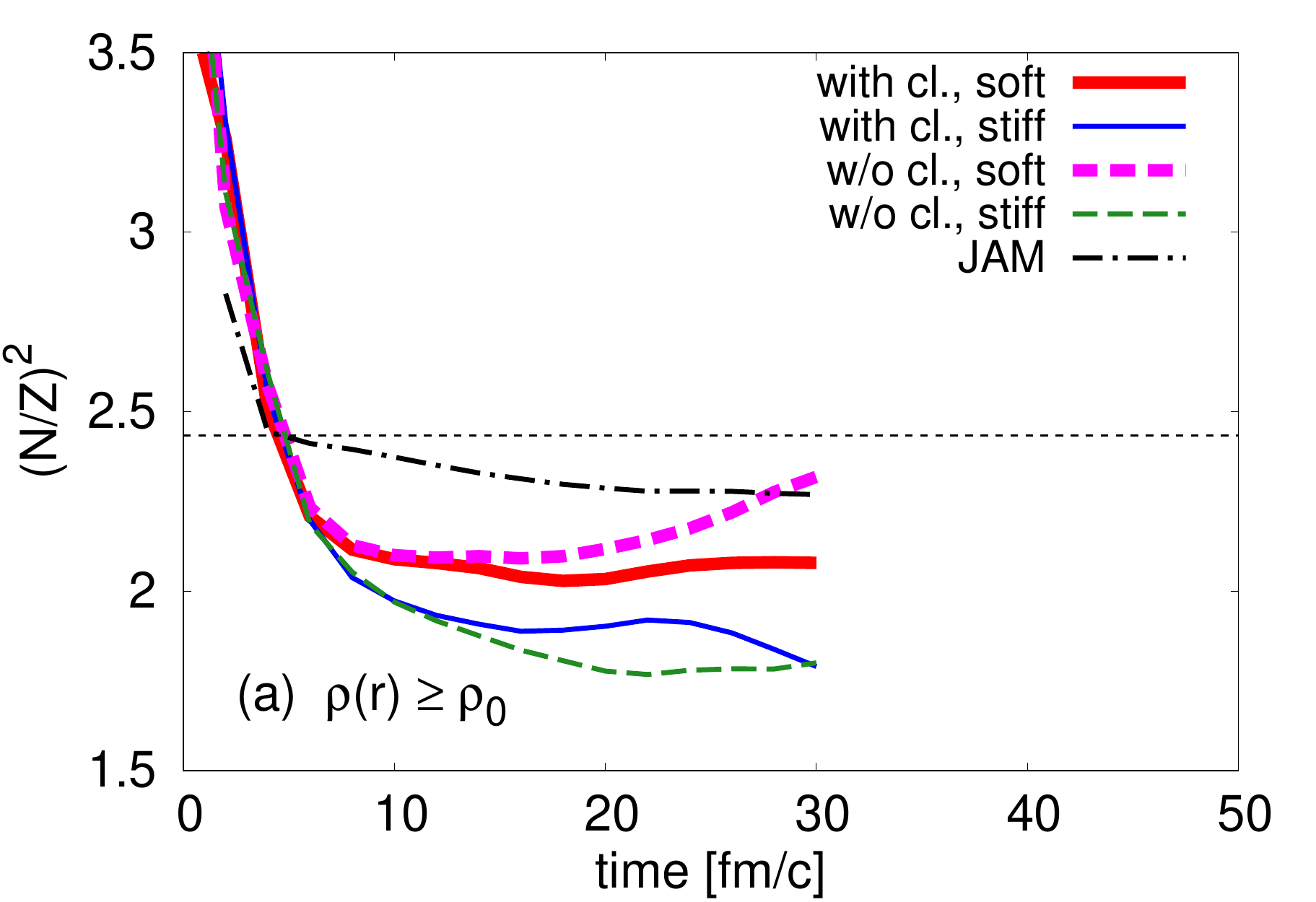}
\includegraphics[scale=0.45]{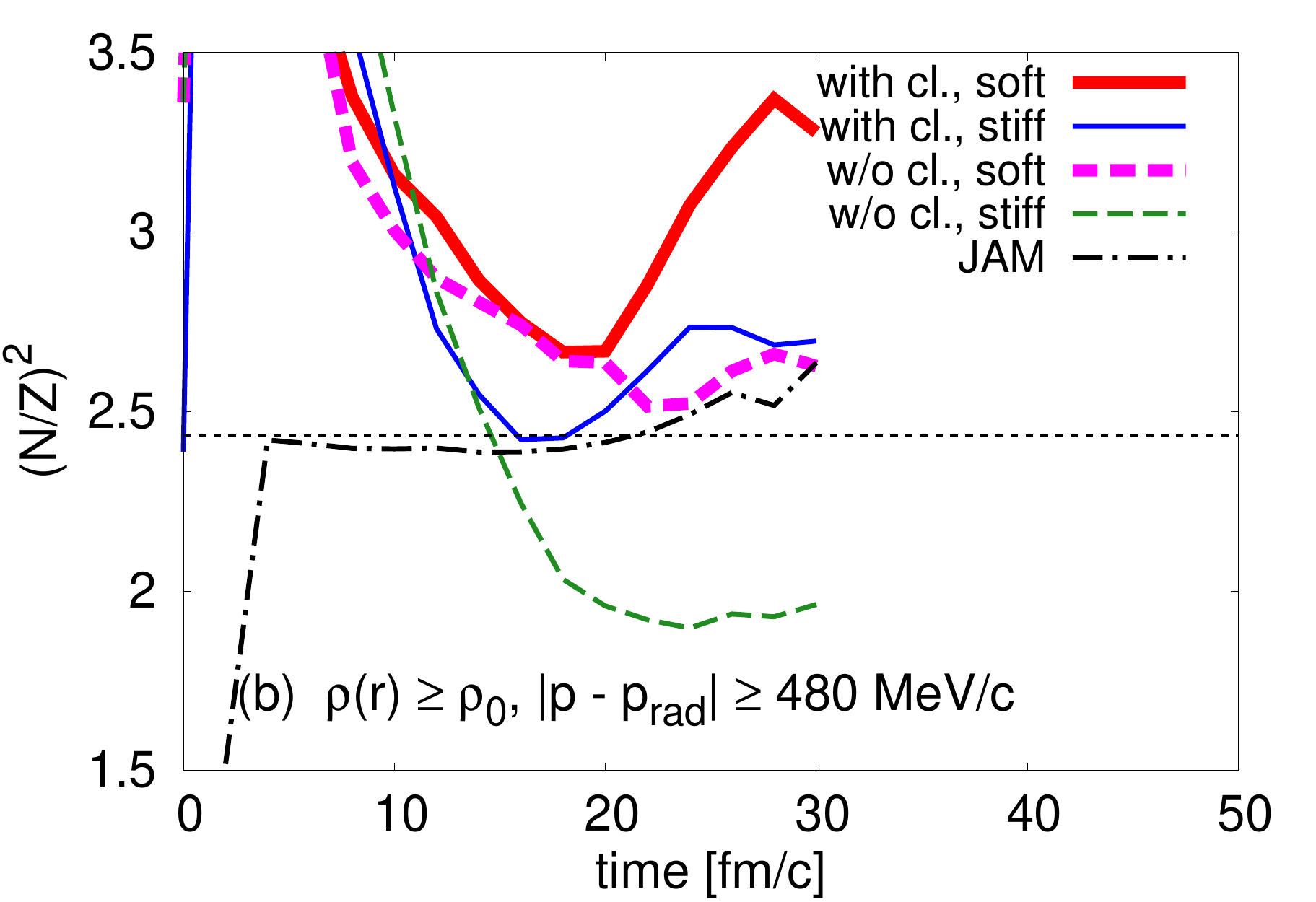}
\caption{\label{fig:nzsqratio} The time evolution of the squared ratio of neutron and proton $(N/Z)^2$ in $\nuc[132]{Sn}+\nuc[124]{Sn}$ central collisions at 300 MeV/nucleon, calculated (a) for the nucleons in the high-density region of $\rho(r)\ge\rho_0$, and (b) for those in the high-density region $\rho(r)\ge\rho_0$ and with high momenta $|\bm{p}-\bm{p}_{\text{rad}}|\ge 480$ MeV/$c$.  The results of AMD with (solid lines) and without (dashed lines) cluster correlations are shown for a soft (thick lines) and a stiff (thin lines) density-dependence of the symmetry energy.  The result of the JAM model \cite{nara1999} without mean field is shown by a dot-dashed line.  The horizontal thin dashed line shows $(N/Z)^2$ of the total system.  Taken from Ref.~\cite{ikeno2016erratum}.}
\end{figure}

\begin{figure}
\centering
\includegraphics[scale=0.7]{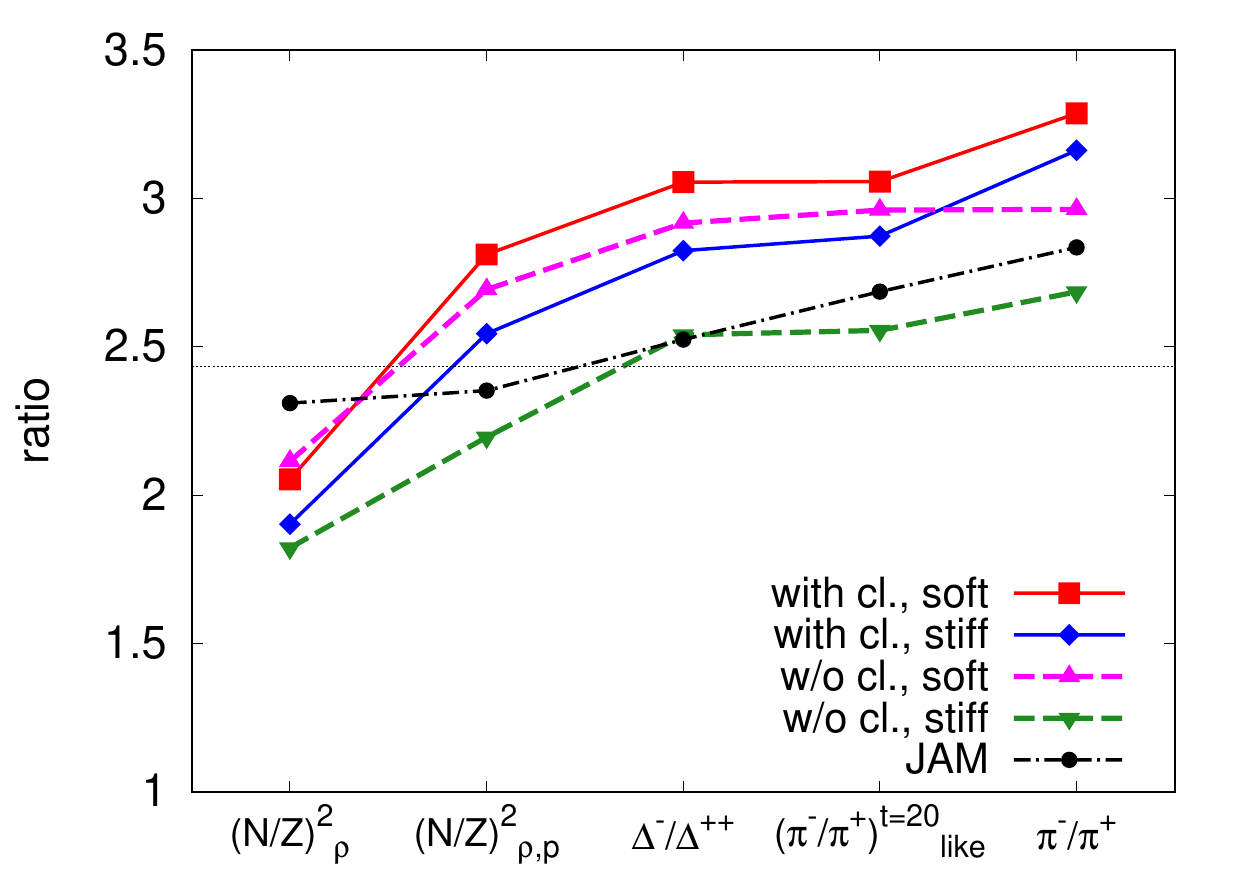}
\caption{\label{fig:ratiosummary}
Relation of various isospin ratios in $\nuc[132]{Sn}+\nuc[124]{Sn}$ central collisions at 300 MeV/nucleon calculated with AMD+JAM approach, from the $(N/Z)^2$ ratio in the high-density region to the final $\pi^-/\pi^+$ yield ratio.  The meaning of line styles is the same as in Fig.~\ref{fig:nzsqratio}.  Taken from Ref.~\cite{ikeno2016erratum}.
}
\end{figure}

Ikeno el al.~\cite{ikeno2016,ikeno2016erratum} studied $\nuc[132]{Sn}+\nuc[124]{Sn}$ central collisions at 300 MeV/nucleon by employing AMD with cluster correlations.  The effect of cluster correlations was studied as well as that of symmetry energy by comparing the results of calculations with different options.  In Fig.~\ref{fig:nzsqratio}(a), the time evolution of $(N/Z)^2$ of the nucleons in the high-density region $\rho\ge\rho_0$ is shown.  The solid lines show the results with clusters and the dashed lines without clusters.  The effect of the density dependence of the symmetry energy is seen by comparing thick and thin lines.  In general, due to the symmetry energy effect, the high-density region becomes less neutron rich than the total system for which the $(N/Z)^2$ value is shown in the figure by the dotted horizontal line.  In the right panel [Fig.~\ref{fig:nzsqratio}(b)], the same $(N/Z)^2$ ratio is shown but by further selecting the nucleons which have high momenta $|\bm{p}-\bm{p}_{\text{rad}}|^2$ where $\bm{p}_{\text{rad}}$ is the momentum of collective radial expansion.  In general, high-momentum nucleons are more neutron rich and this tendency is strong with cluster correlations in particular after the maximum compression at $t\approx 20$ fm/$c$.  This can be understood because producing e.g.~$\alpha$ clusters, which have relatively low velocities and contain the same number of neutrons and protons, makes the rest of high-momentum nucleons neutron rich.  We also see that the symmetry energy effect, i.e.~the difference between thick and thin lines, is weaker with cluster correlations, which may be because forming a cluster tends to force neutrons and protons move together.  Thus this example shows that the cluster correlations, as well as the density dependence of the symmetry energy, have strong impacts on the dynamics of nucleons in the compression and expansion stages.  This effect may be observable in spectra of neutrons, protons and light clusters if the particle momenta are not much influenced by the interactions in late stages \cite{ono2017nn}.

It was suggested by Li \cite{bali2002,bali2008} that the dynamics at high densities may be reflected in the emitted pions.  For the AMD calculation of nucleon dynamics discussed above, another transport code JAM \cite{nara1999} was employed to calculate the dynamics of $\Delta$ resonances and pions.  As shown in Fig.~\ref{fig:ratiosummary}, the $(N/Z)^2$ ratio of the high-density and high-momentum part, labeled as $(N/Z)_{\rho,p}^2$, is strongly correlated to the ratio $\Delta^-/\Delta^{++}$ of the production rates of $\Delta$ resonances, and then it is almost identical to the ratio $(\pi^-/\pi^+)_{\text{like}}$ at $t=20$ fm/$c$, where the pion-like particles include the contributions of existing $\Delta$ resonances with their decay branching ratios.  From $t=20$ fm/$c$ to the final state, the symmetry energy effect in $\pi^-/\pi^+$ is slightly reduced and the value of $\pi^-/\pi^+$ increases when there are cluster correlations.  This may be understood from the difference in the exterior region of the expanding system which pions have to go through \cite{ikeno2016}.  Currently the transport code results for the pion production in this energy domain do not necessarily agree with each other \cite{reisdorf2007,xiao2009,zqfeng2010,cozma2013,hong2014,song2015,bali2015,cozma2016,tsang2017,zhenzhang2017}.  It will be important, as in this analysis of Fig.~\ref{fig:ratiosummary}, to find out how the pion ratio is related to the nucleon dynamics and how deviations start to appear among different transport codes.

\section{\label{sec:summary}Summary and perspectives}

This article reviewed our understandings on how light clusters and heavier fragments are so copiously produced in heavy-ion collisions at incident energies between several tens of MeV/nucleon and several hundred MeV/nucleon.  For central collisions, accumulated experimental data indicate that clusters and fragments are formed in the radially expanding system with some anisotropy, i.e.~with incomplete stopping, which quantitatively depends on the incident energy and the size of the colliding nuclei.  Although we focused on experimental data for central collisions in this article, clusters and fragments are also produced in semi-peripheral collisions, often copiously, from the projectile-like and target-like parts as well as from the participant region which sometimes forms a neck.  A well-known but surprising fact is that, even in a central collision which is supposed to be a violent phenomenon, only a minor fraction protons in the whole system are emitted as free protons and all the other protons are bound in clusters and fragment nuclei in the final state.  Thus we confront strong many-body correlations in excited nuclear many-body systems at densities lower than the saturation density $\rho_0$, and in a very wide range of the excitation energy.

The excited low-density states realized during heavy-ion collisions should have some relation to the equilibrium properties of warm low-density nuclear matter which are also directly important in some astrophysical phenomena such as core-collapse supernovae.  Statistical ensembles of nucleons, light clusters and heavier nuclei are typically considered in order to construct the EOS and calculate the composition of matter.  Except in the lowest densities, the properties of clusters as quasiparticles may be affected by the medium and the interactions between particles should be properly treated in principle.  Recent developments and improvements of the treatment of clusters in nuclear matter can be linked to improving the understanding of clusters in heavy-ion collisions.

We need to understand the mechanisms how the dynamical evolution in heavy-ion collisions arrives at such highly correlated states with many clusters and fragments.  This is important not only for the interest in correlations at low densities but also for extracting information of nuclear matter properties, such as the EOS of isospin-symmetric and asymmetric nuclear matter, at high densities which are realized at early reaction times.  One might assume that the information on the high-density stage is mainly contained in the global one-body dynamics so that one does not need to worry about correlations at first.  However, the real situation may not be so simple.  Many experimental observables are inevitably of the clusters and fragments in the final state.  Furthermore, it is probable that strong cluster correlations, which develop  during the collision, have impacts on the global one-body dynamics through the conservation and distribution of energy.  In any case, transport models have been indispensable to understand the dynamics of heavy-ion collisions and to extract information on the nuclear EOS.

We reviewed various transport models for heavy-ion collisions putting emphasis both on their theoretical features and on their practical performances in describing formation of clusters and fragments.  All the practical models, maybe unfortunately, start with the consideration on the single-particle distribution function $f(\bm{r},\bm{p})$.  The BUU equation describes a deterministic time evolution of $f$.  Practical BUU codes employ a finite number ($N_{\text{tp}}$) of test particles per nucleon.  In principle, the precise solution should be obtained in the limit of $N_{\text{tp}}\rightarrow\infty$.  The collaboration of transport code comparison \cite{xu2016,yxzhang2018} is supposed to solve the deviations among different BUU codes.  While the BUU models do not describe clusters and fragments, QMD models can describe fragmentation, by representing each nucleon by a wave packet and by treating  two-nucleon collisions as a stochastic process.  Many events are generated corresponding to different fragmentation configurations.  The treatment of the momentum of each nucleon, without considering any width, is also an important feature of QMD.  Essential aspects of fragmentation data are often explained by QMD models.  However, typical deviations from data seem to be related to a possible underestimation of degeneracy pressure due to the violation of Pauli principle.  The approach by CoMD is one of the possible ways to improve the situation.  The collaboration of transport code comparison may be able to reduce the difference among QMD codes to some extent but some differences will remain as real differences in model assumptions, such as the methods and parameters for Pauli blocking.  In both BUU and QMD models, there are some issues in the treatment of two-nucleon collisions when the mean field is momentum dependent.

There are extended approaches to go beyond the above-mentioned standard transport models of BUU and QMD types.  A key concept to consistently understand these extensions is how to handle single-nucleon motions in the mean field, fluctuation or branching induced by two-nucleon collisions, and localization of nucleons to form fragments and clusters.  There can be two routes starting with BUU and QMD (or AMD) models approaching to a similar goal.  Based on the BUU models, fluctuations are introduced on $f$ as an effect of two-nucleon collisions in addition to the average effect by the usual collision term.  For example, in the recently developed BLOB model, two entire nucleons, each of which consists of $N_{\text{tp}}$ nucleons, are scattered by a two-nucleon collision, which induces fluctuations in a similar way to a two-nucleon collision in QMD and AMD.  On the other hand, from AMD which represents nucleons as antisymmetrized Gaussian wave packets with fixed width in both coordinate and momentum spaces, the effect of the change of the distribution shape was introduced as fluctuations to the wave-packet centroids, so that the time evolution of the one-body distribution is similar to mean-field models on average.  Thus these extended models developed from different origins are now conceptually similar.  However, in AMD with wave-packet splitting, branching may occur even without two-nucleon collisions.  A consideration on this point was given in this article, in relation to decoherence due to many-body correlations.  By these models, the characteristics of multifragmentation is often well explained, depending on the situation, except for the yields of light clusters and free nucleons.  It seems that these extended models show stronger stopping in low-energy collisions than usual QMD models.

All the above-mentioned transport models, based more or less on single-nucleon motions, seriously overestimate the proton multiplicity and underestimate the $\alpha$-particle multiplicity in heavy-ion collisions.  This may be requiring us to change the picture drastically by putting more emphasis on cluster correlations.  Clusters should be explicitly treated because the classical phase space for nucleons is not consistent with the quantum bound state of a cluster.  If we already know that cluster correlations are strong, it would be more natural to describe the system as a mixture of nucleons and clusters, in a similar way to statistical ensembles of nucleons, clusters and nuclei.  In the BUU model by Danielewicz, light clusters with $A_{\text{c}}\le 3$ are treated as new particle species and the BUU equations for them are coupled by various reactions to produce or break clusters.  The properties and the existence of a cluster, in principle, depend on the medium and on its momentum in the same way as in the statistical calculations.  We reviewed an example which shows the effect of clusters in fragmentation dynamics, as well as in the sensitivity to the symmetry energy.  Another model with explicit cluster correlations has been developed based on AMD.  Clusters  with $A_{\text{c}}\le 4$ are allowed to be formed in the final states of two-nucleon collisions.  A formed cluster is still described in an AMD wave function by placing the corresponding nucleon wave packets at the same phase-space point.  Calculations show that cluster correlations drastically change the global collision dynamics so that the whole system turns into an expanding gas of clusters and nucleons, without producing heavier fragments sufficiently.  Further investigations were made by considering inter-cluster correlations to form nuclei, such as Li and Be isotopes, in a phenomenological way.  With a reasonable set of parameters for inter-cluster correlations, the yeilds of protons, light clusters and heavier fragments can be well explained e.g.~in $\nuc{Xe}+\nuc{Sn}$ central collisions at 50 MeV/nucleon.  We also reviewed an example in which cluster correlations have impacts on the dynamics of neutron-to-proton ratio in compressed and expanding neutron-rich system and on the sensitivity to the symmetry energy, which consequently influences the charged pion ratio in collisions at $\sim300$ MeV/nucleon.  It is an important question at which stage cluster correlations start to play roles during the collision dynamics.  Investigations by changing the condition for the existence of clusters may be able to answer this question.

Presently, the data of heavy-ion collisions are analyzed mostly with standard transport models, i.e.~the usual BUU and QMD models, in order to extract physical information such as the EOS of symmetric and asymmetric nuclear matter and the effective masses of neutrons and protons.  We may expect that the deviations among different standard codes will be better understood in near future by the international collaboration of transport code comparison, which will enable us to have `standard' conclusions on extracted physical information.  However, it is still indispensable to improve extended transport models for better descriptions of fragments and clusters, not only for the interesting physics of correlations but also for confirmations or revisions of `standard' conclusions which may be affected by strong many-body correlations in principle.

Assumption of strong cluster correlations sounds to be a promising way to understand the particle yields observed in heavy-ion collisions.  It then seems that correlations between light clusters are also important to form heavier fragments.  However, this is a rather drastic change from traditional understanding based on independent motions of nucleons in the mean field.  Model calculations and experimental data should be carefully compared in order to establish a picture of strongly correlated and highly excited system realized in heavy-ion collisions.  Such studies are in progress.

\section*{Acknowledgments}
The author acknowledges support from Japan Society for the Promotion of Science KAKENHI Grant No.~17K05432.

\bibliography{ono_nucl}
\end{document}